\documentclass[12pt]{article}

\usepackage{graphicx} %
\usepackage{amsmath}
\usepackage{amssymb}
\usepackage[dvipsnames]{xcolor} %
\usepackage{ytableau} %
\usepackage{mathtools} %
\usepackage{dsfont} %
\usepackage{commath} %
\usepackage{BOONDOX-uprscr} %
\usepackage{amsthm} %
\usepackage[numbers, square, sort&compress]{natbib} %
\usepackage{hyperref} %

\makeatletter
\newcommand{\mathgraphics}[3]{%
  \vcenter{\hbox{\includegraphics[height=#1, width=#2]{#3}}}%
}
\makeatother

\newtheoremstyle{mydef}
  {10pt}{10pt}      %
  {\itshape}   %
  {0 em}              %
  {\bfseries\itshape}     %
  {.}             %
  {0.5em}         %
  {\thmname{#1}\thmnumber{ #2}\thmnote{\hspace{0.1em}#3}}

\theoremstyle{mydef}
\newtheorem{theorem}{Theorem}[section]
\newtheorem{lemma}[theorem]{Lemma}
\newtheorem{proposition}[theorem]{Proposition}
\newtheorem{corollary}[theorem]{Corollary}
\newtheorem{definition}[theorem]{Definition}

\newcommand{\offbig}{\mathfrak{hsgl}(d - 1 , 2 | 2)}
\newcommand{\onbig}{\mathfrak{hso}(d-1, 2 | 2)}
\newcommand{\off}{\mathfrak{hsgl}_{p}(d - 1 , 2 | 2)}
\newcommand{\onsh}{\mathfrak{hso}_p(d-1, 2 | 2)}
\newcommand{\id}{\mathcal{I}_{\mathfrak{sp}(2)}}
\newcommand{\so}{\mathfrak{so}(d-1,2)}

\renewcommand{\thesubsection}{\arabic{section}.\arabic{subsection}}

\makeatletter 
\@addtoreset{equation}{section} 
\makeatother

\begin{document}

    \vspace{0.5cm}
\begin{center}
{\large\bf An explicit product for the on-shell higher-spin algebra in $AdS_d$-space}

\vspace{1 cm}

\textbf{M.A.~Povarnin$^{1,2}$}\\

\vspace{1 cm}

\textbf{}\textbf{}\\
 \vspace{0.5cm}
 \textit{$^{1}$I.E. Tamm Department of Theoretical Physics,
Lebedev Physical Institute,}\\
 \textit{ Leninsky prospect 53, 119991, Moscow, Russia }\\
 \vspace{0.5cm}
 {\it
			$^2$Moscow Institute of Physics and Technology,\\
			Institutsky lane 9, 141700, Dolgoprudny, Moscow region, Russia}

\par\end{center}

\begin{center}
\vspace{0.6cm}
e-mail: povarnin.ma@gmail.com \\
\par\end{center}

\vspace{0.4cm}

\begin{abstract}

\noindent 
    For $d>3$ we construct the product of the bosonic on-shell higher-spin algebra directly on the AdS-traceless representatives. We find the traceless projector in closed form, determine all linear relations among the trace-producing operators, and prove that every class modulo the $\mathfrak{sp}(2)$ ideal has a unique traceless representative. The resulting product is an explicit differential operator, which we test on the $\mathfrak{so}(d-1,2)$ generators. A toy model shows how the factorisation lifts from bilinear products to trilinear ones in minimal $A_\infty$-algebras.

\end{abstract}
\newpage

    \tableofcontents
    \newpage
   
    \section{Introduction}
Higher-spin (HS) gauge theories are a class of theories describing interaction between massless fields of all spins\cite{Fradkin:1987ks,Vasiliev:1990en,Vasiliev:1992av,vasiliev1999higher,Vasiliev:2003ev}. Such theories have been actively researched in recent years\cite{Didenko:2019xzz,Didenko:2020bxd,Didenko:2022qga,Didenko:2023txr,Didenko:2023vna,Neiman:2023orj,Didenko:2024zpd,Ponomarev:2024jyg,Tatarenko:2024csa,Didenko:2025xca,Kirakosiants:2025gpd,Metsaev:2025wcv,Tarusov:2025sre,Cho:2026mjp,Didenko:2026apw,Didenko:2026nag,Faliakhov:2026yum,King:2026vch,Kirakosiants:2026spd,Metsaev:2026tym,Sundborg:2026cut,Tarusov:2026ich,Zaigraev:2026xzr}, but the interest in the topic dates back to the first half of the 20th century\cite{fierz1939relativistic}.
Fronsdal’s works during the 1970s provided a comprehensive treatment of linear HS theory, including the quadratic action for massless fields of arbitrary spin on flat or constant-curvature backgrounds and the double-traceless constraints that ensure its consistency and gauge invariance\cite{Fronsdal:1978rb,Fang:1978wz,Fang:1979hq} (see \cite{Bittencourt:2025roa} for a recent unconstrained formulation).
Around the same time, several no-go theorems\cite{coleman1967all,weinberg1964photons,weinberg1980limits} were proven, seemingly prohibiting interaction of HS fields at all, while the Haag--{\L}opusza\'nski--Sohnius classification of the supersymmetries of the $S$-matrix\cite{haag1975all} admits fermionic generators of spin $\frac{1}{2}$ only.
However, these theorems assume a Minkowski background. This obstruction was circumvented by Fradkin and Vasiliev\cite{fradkin1987cubic,Fradkin:1987ks} and by Vasiliev\cite{Vasiliev:1990en,Vasiliev:1992av}, who constructed nontrivial interactions of massless HS fields with gravity and themselves on an anti-de Sitter (AdS)
background.  

One of the most substantial results in this field is the formulation of generating equations by Vasiliev\cite{Vasiliev:1990en,Vasiliev:2003ev} (see \cite{Didenko:2022qga,Vasiliev:2025hfh,Tarusov:2025sre,Korybut:2025vdn,Didenko:2026nag,Diaz:2024kpr,Diaz:2024iuz} for recent developments). Such systems are manifestly gauge and diffeomorphism invariant due to the unfolded formulation\cite{vasiliev2002conformal} (see \cite{Misuna:2024ccj,Misuna:2024dlx,Misuna:2026bhy} for recent progress). The generating system approach describes HS theory at the classical level via equations of motion, which schematically read
\begin{subequations}
    \label{intro}
    \begin{align}
        &d_x \omega + \omega \ast \omega = \mathcal{V}(\omega, \omega, C) + \mathcal{V}(\omega, \omega, C, C) + \dots \, , \\
        &d_x C + [\omega, C]_{\pi} = \Upsilon(\omega, C, C) + \Upsilon(\omega, C, C, C) + \dots \, ,
    \end{align}
\end{subequations}
where $\ast$ is the associative product in the HS algebra \cite{Vasiliev:1990en,vasiliev2004higher}, while $\mathcal{V}$ and $\Upsilon$ represent interaction vertices. Such  a formulation has led to substantial progress in lower
dimensions ($d=3,4$), where the HS algebra admits a simple oscillator realisation\cite{Vasiliev:1990en, Vasiliev:1989qh}, while in a space-time of generic dimension a convenient realisation of the underlying
HS algebra is lacking. Instead, the oscillator formalism leads to a so-called off-shell HS algebra\cite{vasiliev2004higher}. The
term off-shell implies that the system based on this algebra describes no dynamics; instead it provides an infinite set of
differential identities, which can be deemed an HS version of Bianchi identities in General Relativity or Yang-Mills theory. In this context, the off-shell HS theory was extensively studied in \cite{Didenko:2023vna,Didenko:2024zpd}; in particular, all-order local vertices $\mathcal{V}$ and $\Upsilon$ were reconstructed within the generating equations approach.

Despite being remote from the original problem of HS interactions, the off-shell system actually can
be used for the on-shell reduction. As demonstrated by Vasiliev, the on-shell HS algebra, which describes the actual dynamics of HS fields, can be obtained as a factor of the off-shell one with respect to a certain ideal\cite{Vasiliev:2003ev,vasiliev2004higher}. This algebra is isomorphic to the algebra of higher symmetries of the Laplacian\cite{eastwood2005higher,vasiliev2004higher}. However, the factorisation at the level of algebra applies only to the deformation of the $\ast$-product on the l.h.s. of \eqref{intro}. To obtain interaction vertices for the on-shell theory, it is possible to factor out the deformation of the aforementioned ideal within generating equations\cite{Vasiliev:2003ev} (see \cite{Vasiliev:2025hfh} for recent developments). A related approach imposes a strong $\mathfrak{sp}(2)$ projection on the zero-form master field and derives the resulting linearised on-shell constraints\cite{Sagnotti:2004pod}.

While the described factorisation is well-defined as a concept, it has so far rarely been implemented in practice.
Invariant bilinear and trilinear forms of the algebra, and hence its structure constants, were obtained in \cite{joung2014notes} using the trace defined by the quasiprojector of \cite{vasiliev2004higher}. Here we give the product itself as a differential operator acting on traceless generating functions.
The main aim of the present paper is to fill this gap for $d>3$ by constructing an explicit realisation of the on-shell HS algebra on AdS-traceless $\mathfrak{sp}(2)$-singlet polynomials.

In addition to the generating equations approach, the system \eqref{intro} can be described as a minimal $A_\infty$-algebra\cite{stasheff1963homotopy,keller2001introduction}. Within this framework, we show in a toy model, whose ideal is simpler than the $\mathfrak{sp}(2)$-ideal, how the factorisation can be lifted from the product $m_2$ to the lowest higher product $m_3$, and we list what such a lift requires for the $\mathfrak{sp}(2)$-ideal. 

The paper is organised as follows. Section~\ref{sec:hs-algebras} reviews the oscillator realisation of higher-spin algebras, the $\mathfrak{sp}(2)$-ideal and the realisation of a factor-algebra on a space of representatives. In Section~\ref{sec:projector} we analyse the reduction of $\mathfrak{gl}(d+1)$-representations to $\mathfrak{so}(d+1)$-representations and the associated trace structure, and construct the horizontal projector onto AdS-traceless $\mathfrak{sp}(2)$-singlets. Section~\ref{sec:factoring} contains the main results of the paper: the proof of the uniqueness of the corresponding decomposition, the splitting map, and the resulting realisation of the higher-spin factor-algebra and its induced product, together with a check on the $\so$-generators, the Lorentz-traceless representatives and the case $d=3$. Section~\ref{sec:ainfty} reformulates HS interactions in terms of minimal $A_\infty$-algebras and lifts the factorisation to the lowest higher product in a toy model. Appendix~\ref{app:lift-proof} contains the proof of Theorem~\ref{thm:lift}, Appendix~\ref{app:B} the derivation of the projector, including the (in)dependence of the trace-producing operators, Appendix~\ref{app:C} the properties of the quasiprojector used for the uniqueness and for the splitting map, and Appendix~\ref{app:MM} the evaluation of $M_{AB}\circ M_{CD}$.

\newpage

\section{Bosonic HS algebras in $d$ dimensions}
\label{sec:hs-algebras}

This section reviews the HS algebras underlying the unfolded description of interacting symmetric bosonic HS fields in $d$ dimensions.
\subsection{Frame-like formulation of $AdS_d$ space}

The unfolded approach\cite{vasiliev2002conformal} to HS fields is closely related to the frame-like formulation of field dynamics; both are reviewed in \cite{Didenko:2014dwa,bekaert2005nonlinear} and recalled here briefly.

The relation between unconstrained frame-like and metric-like formulations in the limit of vanishing cosmological constant is examined in \cite{Bekaert:2008sa}, including the gauge transformations to first order in a weak-field expansion.

The $AdS_d$ background is described by the $\so$-valued one-form
\begin{equation}
    \Omega^{AB} = -\Omega^{BA} = \Omega^{AB}_{\underline{m}}(x) \, dx^{\underline{m}} \, ,
\end{equation}
satisfying the zero-curvature condition
\begin{equation}
    \label{AdS}
    d_x \Omega^{AB} + \Omega^{AC}\Omega^{DB} \eta_{CD} = 0 \, .
\end{equation}
Here $d_x = dx^{\underline{m}} \, \partial_{\underline{m}}$, and capital Latin indices $A, B, C, D \in \{0, \, \dots \, , d \}$ are raised and lowered by the $\so$-invariant metric
\begin{equation}
    \eta_{AB}=\text{diag}\{- \, + \dots + \, - \} \, , \quad \eta^{AB}\eta_{AB}=d+1 \, .
\end{equation}
The vielbein $e^a_{\underline{m}}(x) dx^{\underline{m}}$ and the spin connection $\varpi^{ab}_{\underline{m}}(x)dx^{\underline{m}}$, which determine the metric $g_{\underline{m}\underline{n}}(x)$, are identified with components of $\Omega^{AB}$ as\footnote{A more systematic approach based on the \textit{compensator field} is presented in \cite{vasiliev2004higher}.}\textsuperscript{,}\footnote{In general, this identification involves a factor fixed by the cosmological constant. Throughout the paper, the cosmological constant is set to $-1$.}
\begin{equation}
    e^{a}:= \Omega^{a \, d} \, , \quad \varpi^{ab} := \Omega^{ab} \, ,
\end{equation}
with $a, b \in \{0, \, \dots \, , d - 1 \}$.

\subsection{Field spectrum within the unfolded approach}
As shown by Vasiliev\cite{vasiliev2004higher}, the unfolded form of a spin-$s$ gauge field\footnote{Throughout the text we use the condensed notation standard in the HS literature for symmetrisation of indices: $\psi_{\underline{m(s)}}(x): = \psi_{\underline{m_1 \, \dots \, m_s}}(x) = \psi_{\underline{(m_1 \, \dots \, m_s)}}(x)$.} $\psi_{\underline{m(s)}}(x)$ in $AdS_d$ consists of two irreducible $\so$-modules: the adjoint and the twisted-adjoint module. In the linear theory, the former describes gauge-dependent degrees of freedom, while the latter contains the gauge-invariant linearised Weyl tensor and its derivatives. They are therefore called the \textit{gauge module} and the \textit{Weyl module}, respectively.

The gauge module of $\psi_{\underline{m(s)}}(x)$ is a one-form $\omega^{A(s-1) \, , \, B(s-1)} = \omega^{A(s-1) \, , \, B(s-1)}_{\underline{m}}(x) \, dx^{\underline{m}}$
valued in the irreducible $\so$-module described by the traceless two-row rectangular Young diagram of length $s-1$,
\begin{equation}
    \label{Young}
    \omega^{(A_1 \dots A_{s-1} \, , \, A_s ) B_2 \dots B_{s-1} } = 0 \, , \quad \eta_{CD} \, \omega^{A_1 \dots A_{s-3} \, CD \, , \, B_1 \dots B_{s-1}} = 0 \, .
\end{equation}
Such Young diagrams are conveniently handled with generating variables $Y^\alpha_A$. Here $A$ is an $\so$ index, i.e. a vector index of the ambient space of $AdS_d$, while $\alpha \in \{ 1,2\}$ is an $\mathfrak{sp}(2)$ index, raised and lowered by the $\mathfrak{sp}(2)$-invariant metric $\epsilon_{\alpha\beta}$:
    \begin{equation}
        \label{epsilon}
        \begin{split}
            &\epsilon_{\alpha\beta} = - \epsilon_{\beta\alpha} \, , \quad \epsilon_{12} = +1 \, , \\
            T^\alpha = \epsilon^{\alpha\beta} \, T_\beta \, & , \quad  T_\beta = \epsilon_{\alpha\beta} \, T^\alpha \, , \quad \epsilon^{\alpha \gamma}\epsilon_{\beta\gamma} = \delta^\alpha_\beta \, .
        \end{split}
    \end{equation}

In terms of $Y^\alpha_A$, the spin-$s$ gauge module is encoded in a one-form $\omega^{(s)}:=\omega^{(s)}_{\underline{m}}(x|Y) \, dx^{\underline{m}}$ satisfying
\begin{subequations}
    \label{gauge_mdl}
    \begin{align}
         \label{gauge_mdl:1}Y^\alpha \cdot  \frac{\partial}{\partial Y^\alpha}  \omega^{(s)} = 2(s-1) &\omega^{(s)} \, , \\
          \label{gauge_mdl:2}\Big(Y_\alpha  \cdot \frac{\partial}{\partial Y^\beta}  + (\alpha \leftrightarrow \beta) \Big) \omega^{(s)}&= 0 \, , \\
          \label{gauge_mdl:3}\frac{\partial^2}{\partial Y^\alpha \cdot \partial Y^\beta} \omega^{(s)} = 0& \, ,
    \end{align}
\end{subequations}
where $\cdot$ denotes contraction of $\so$ indices with $\eta_{AB}$. These conditions imply
\begin{equation}
    \label{gauge_mdl:cnsq}
    \begin{split}
    \omega^{(s)}_{\underline{m}}(x|Y) &= \Big(\frac{1}{2}\Big)^{s-1}\omega^{A(s-1) \, , \, B(s-1)}_{\underline{m}}(x) \big( M_{AB} \big)^{(s-1)} \\
    &:= \Big(\frac{1}{2}\Big)^{s-1}\omega^{A_1 \dots A_{s-1} \, , \, B_1 \dots B_{s-1}}_{\underline{m}}(x) M_{A_1 B_1} \dots M_{A_{s-1}B_{s-1}} \, ,
    \end{split}
\end{equation}
where $\omega^{A(s-1) \, , \, B(s-1)}_{\underline{m}}(x)$ satisfies \eqref{Young}, and $M_{AB} = -M_{BA}$ is defined as
\begin{equation}
    M_{AB} := \frac{i}{2}\, \epsilon_{\alpha\beta} \, Y^\alpha_A Y^\beta_B \, .
\end{equation}
The derivation of \eqref{gauge_mdl:cnsq} from \eqref{gauge_mdl} is given in \cite{vasiliev2004higher}.

The conditions \eqref{gauge_mdl} define an irreducible $\so$-module only for $d>3$: for $d=3$, the isometry algebra is $\mathfrak{so}(2,2)$, whose tensors admit (anti)self-duality conditions via $\epsilon_{ABCD}$. For $d = 3$ the isometry algebra $\mathfrak{so}(2,2) \cong \mathfrak{sl}(2) \oplus \mathfrak{sl}(2)$ admits a direct oscillator realisation \cite{Vasiliev:1989qh}, in which the one-parameter family of three-dimensional HS algebras is generated by deformed oscillators and no factorisation is required. We assume $d > 3$ unless stated otherwise; the $d = 3$ case of the present construction is analysed in Section~\ref{sec:d3}.

The Weyl module has a significantly different structure. To describe it, the generating variables are split in accordance with the identification of $e^a$ above\cite{vasiliev2004higher}:
\begin{equation}
    \label{small_y}
    \begin{split}
        &y^\alpha := Y^\alpha_d \, , \\
        \forall a\in \{0, \, \dots , \, & d-1 \}: \mathbf{y}^\alpha_a := Y^\alpha_a \, .
    \end{split}
\end{equation}
In these terms, the Weyl module $C^{(s)}(x|Y)$ is defined by
\begin{subequations}
    \begin{align}
         \bigg(\mathbf{y}^\alpha_a & \frac{\partial}{\partial \mathbf{y}^\alpha_a} - y^\alpha \frac{\partial}{\partial y^\alpha}\bigg) C^{(s)}(x|Y) = 2s \, C^{(s)}(x|Y) \, , \\
        \label{Weyl_mdl:2}&  \Big(Y_\alpha  \cdot \frac{\partial}{\partial Y^\beta}  + (\alpha \leftrightarrow \beta) \Big) C^{(s)}(x|Y)= 0 \, , \\
        \label{Weyl_mdl:3}&\bigg(\eta_{ab} \, \frac{\partial^2}{\partial \mathbf{y}^\alpha_a \partial \mathbf{y}^\beta_b} + y_\alpha y_\beta \bigg)C^{(s)}(x|Y) = 0 \, ,
    \end{align}
\end{subequations}
where $\eta_{ab} = \text{diag}\{- + \dots + \}$. This module is studied in detail in \cite{TatarenkoPREP}.

\subsection{Star product and the free level equations}
\label{sec:star}
The last ingredient of the linearised dynamics is the \textit{Moyal star product}\cite{vasiliev1999higher,vasiliev2004higher}
\begin{equation}
    \label{Moyal}
    f(Y) \ast g(Y) = \int \frac{dUdV}{\big(2\pi \big)^{2(d+1)}} \, e^{iU_\alpha^A V^\alpha_A} \, f(Y+U) \, g(Y+V) = : \int_{U,V} e^{iUV} f(Y+U) g(Y+V) \, .
\end{equation}
It is formally associative and possesses the automorphism (recall \eqref{small_y})
\begin{equation}
    \label{twist}
    \pi \big( f(y, \mathbf{y}) \big) = f(-y , \mathbf{y}) \, : \quad \pi \big( f \ast g \big) = \pi(f) \ast \pi(g) \, .
\end{equation}
It also admits the antiautomorphism $\dagger$ defined by
\begin{equation}
    \label{antiauto}
    \begin{split}
    \forall \lambda &\in \mathds{C}: (\lambda)^\dagger = \bar{\lambda} \, , \\
    &\Big( Y^\alpha_A \Big)^\dagger = Y^\alpha_A \, .
    \end{split}
\end{equation}
With respect to the $\ast$-commutator, the bilinears $M_{AB}$ form a representation of $\so$:
\begin{equation}
    \label{so-alg}
    \big[M_{AB}\, , \, M_{CD}\big]_\ast = \eta_{BC}M_{AD} - \eta_{AC}M_{BD} - \eta_{BD}M_{AC} + \eta_{AD}M_{BC} \, .
\end{equation}

In terms of $\ast$, the linear HS equations take the simple form
\begin{subequations}
    \label{linear}
    \begin{align}
        & d_x \Omega + \Omega \ast \Omega = 0 \, , \\
        &d_x \omega + \omega \ast \Omega + \Omega \ast \omega = \mathcal{V}(\Omega, \Omega, C) \, , \\
        &d_x C + \Omega \ast C - C \ast \pi(\Omega) = 0 \, ,
    \end{align}
\end{subequations}
where the one-form $\Omega = \frac{1}{2}\,\Omega^{AB} M_{AB}$ describes the $AdS_d$ background (by \eqref{so-alg}, the first equation is equivalent to \eqref{AdS}), the one-form $\omega = \omega_{\underline{m}}(x|Y)\,dx^{_{\underline{m}}}$ satisfies \eqref{gauge_mdl:2} and \eqref{gauge_mdl:3}, and the zero-form $C = C(x|Y)$ satisfies \eqref{Weyl_mdl:2} and \eqref{Weyl_mdl:3}. The algebraic content is encoded in the associative product $\ast$, the automorphism $\pi$ and the cocycle
\begin{equation}
    \mathcal{V}(\Omega, \Omega,C) = \frac{\epsilon^{\alpha\beta}}{4}\, e_a e_b \, \frac{\partial^2 C}{\partial \mathbf{y}^\alpha_a \partial \mathbf{y}^\beta_b}(0, \mathbf{y}) \, .
\end{equation}

Vasiliev's nonlinear equations\cite{Vasiliev:1990en,Vasiliev:1992av,Vasiliev:2003ev} describe interacting HS fields, and \eqref{linear} is expected to be their linearisation.
Their naive nonlinear completion $d_x \omega + \omega \ast \omega = \mathcal{V}(\omega, \omega, C) + \dots$ in the gauge sector is, however, inconsistent with the traceless condition \eqref{gauge_mdl:3}, which is not preserved by the Moyal star product\cite{vasiliev2004higher}. Dropping the traceless condition leads to the \textit{off-shell HS equations}\cite{vasiliev2004higher}.

The \textit{on-shell HS system}, in turn, can be formulated by factoring out the $\mathfrak{sp}(2)$-ideal in the generating equations\cite{Vasiliev:2003ev} (see \cite{Vasiliev:2025hfh} for the BRST formulation). Resolving this factorisation should lead to the gauge-sector equation
\begin{equation}
    d_x \omega + \omega \circ \omega = \mathscr{V}(\omega, \omega, C) + \dots \ ,
\end{equation}
where $\omega$ satisfies both \eqref{gauge_mdl:2} and \eqref{gauge_mdl:3}, and $\circ$ is the product in the factor-algebra.

\subsection{Off-shell HS algebra and $\mathfrak{sp}(2)$-ideal}
To analyse the factorisation, we define the ``big'' off-shell HS algebra $\offbig$ and its subalgebra $\off$.

\begin{definition}[$\big(\offbig \, \text{algebra} \big)$]
The algebra $\offbig$ is the associative algebra with product \eqref{Moyal}, spanned by formal series $f(Y)$ in the real variables $Y^\alpha_A$ satisfying the $\mathfrak{sp}(2)$-invariance condition
\begin{equation}
    \label{sp2-inv}
    \Big(Y_\alpha  \cdot \frac{\partial}{\partial Y^\beta}  + (\alpha \leftrightarrow \beta) \Big) f(Y) = 0 \, .
\end{equation}
\end{definition}

To show that \eqref{Moyal} preserves \eqref{sp2-inv}, we introduce
\begin{equation}
    t^{\alpha\beta} := \eta^{AB} \, Y^\alpha_A Y^\beta_B \, .
\end{equation}
Although $t^{\alpha\beta}$ do not belong to $\offbig$, they play a crucial role in its analysis. Direct calculation gives
    \begin{subequations}
        \label{t-star}
        \begin{align}
            \label{t-star:left}
            &t^{\alpha\beta} \ast f(Y) = t^{\alpha\beta} \cdot f(Y) - i \Big( Y^\alpha \cdot \frac{\partial f}{\partial Y_\beta} + (\alpha \leftrightarrow \beta) \Big) - \frac{\partial^2 f(Y)}{\partial Y_\alpha \cdot \partial Y_\beta} \, , \\
            &f(Y) \ast t^{\alpha\beta} = t^{\alpha\beta} \cdot f(Y) + i \Big( Y^\alpha \cdot \frac{\partial f}{\partial Y_\beta} + (\alpha \leftrightarrow \beta) \Big) - \frac{\partial^2 f(Y)}{\partial Y_\alpha \cdot \partial Y_\beta} \, ,
        \end{align}
    \end{subequations}
which implies
\begin{subequations}
    \begin{align}
        \label{sp2-alg}&\big[t^{\alpha\beta}\, , \, t^{\mu\nu} \big]_{\ast} = 2i \big(t^{\alpha\mu}\epsilon^{\beta\nu} + t^{\beta\mu}\epsilon^{\alpha\nu} + t^{\alpha\nu}\epsilon^{\beta\mu} + t^{\beta\nu}\epsilon^{\alpha\mu} \big) \, ,\\
        \label{sp2-inv:ast}\Big(&Y_\alpha   \cdot\,\frac{\partial}{\partial Y^\beta}  + (\alpha \leftrightarrow \beta) \Big) f(Y) = 0 \, \Longleftrightarrow \, \big[t_{\alpha\beta}\, , \, f(Y) \big]_{\ast} = 0 \, .
    \end{align}
\end{subequations}
By \eqref{sp2-alg}, $t^{\alpha\beta}$ form a representation of $\mathfrak{sp}(2)$ with respect to $\big[\bullet \, , \, \bullet \big]_{\ast}$, while \eqref{sp2-inv:ast} explains the name of condition \eqref{sp2-inv}. It also proves that \eqref{sp2-inv} is compatible with the $\ast$-product: if $f(Y)$ and $g(Y)$ satisfy \eqref{sp2-inv}, then
\begin{equation}
    \big[t_{\alpha\beta}\, , \, f(Y) \ast g(Y) \big]_{\ast} = f(Y) \ast \big[t_{\alpha\beta}\, , \, g(Y) \big]_{\ast} + \big[t_{\alpha\beta}\, , \, f(Y) \big]_{\ast}   \ast g(Y) = 0 \, .
\end{equation}

The label $\mathfrak{gl}$ in $\off$ reflects that, by \eqref{sp2-inv}, $\offbig$ is spanned by tensors corresponding to two-row $\mathfrak{gl}(d+1)$ Young diagrams, whereas HS dynamics requires two-row $\so$ Young diagrams. These are obtained by passing to a factor-algebra, as described below.

Strictly speaking, $\offbig$ is well defined only once a functional class preserved by \eqref{Moyal} is specified. A well-known function outside $\offbig$ is the \textit{quasiprojector}\cite{vasiliev2004higher}
\begin{equation}
    \label{2.31}
    \sum_{m=0}^{+\infty} \frac{\Gamma\big( \frac{d}{2} \big)}{\Gamma\big(m+ \frac{d}{2} \big)} \cdot \frac{1}{8^m} \cdot \frac{\big(t_{\mu\nu}t^{\mu\nu} \big)^m}{m!} \, .
\end{equation}
The series itself and its $\ast$-product with any polynomial are well defined, but its $\ast$-product with itself diverges\cite{vasiliev2004higher}. For the on-shell HS algebra, this functional-class problem is avoided by restricting to polynomials.

\begin{definition}[$\big(\off \, \text{algebra} \big)$]
The algebra $\off$ is the subalgebra of $\offbig$ consisting of polynomials, i.e.
\begin{equation}
    \off := \bigl\{ f(Y) \in \offbig \bigm| \exists n \in \mathbb{N}
    \ \forall (\alpha_k, A_k): \
    \frac{\partial^n f(Y)}{\partial Y^{\alpha_1}_{A_1} \dots \partial Y^{\alpha_n}_{A_n}} = 0 \bigr\} .
\end{equation}
\end{definition}
This is indeed a subalgebra, since the $\ast$-product of two polynomials is a polynomial\cite{vasiliev2004higher}.

Moreover, $t^{\alpha\beta}$ generate the $\mathfrak{sp}(2)$-ideal.
\begin{definition}[$\big(2\text{-sided ideal }\id \big)$]
The ideal $\id$ is the subspace of $\off$ spanned by elements of the form
\begin{equation}
    i = t^{\alpha\beta} \ast A_{\alpha\beta}(Y) \, ,
\end{equation}
where $A_{\alpha\beta}(Y) = A_{\beta\alpha}(Y)$ is a polynomial in $Y^\alpha_A$ satisfying
\begin{equation}
    \label{sp2-covar}
    \big[t^{\alpha\beta}, A^{\mu\nu} \big]_{\ast} = 2i \cdot \big( A^{\alpha\mu}\,\epsilon^{\beta\nu} + A^{\beta\mu}\,\epsilon^{\alpha\nu} +  A^{\alpha\nu}\,\epsilon^{\beta\mu} + A^{\beta\nu}\,\epsilon^{\alpha\mu}\big) \, .
\end{equation}
\end{definition}
Since $A_{\alpha\beta}(Y)$ is polynomial, so is $t^{\alpha\beta} \ast A_{\alpha\beta}(Y)$. Condition \eqref{sp2-covar} ensures that it is also an $\mathfrak{sp}(2)$-singlet (recall \eqref{sp2-inv:ast}):
\begin{equation}
    \big[t^{\alpha\beta}, A^{\mu\nu} \big]_{\ast} = 2i \cdot \big( A^{\alpha\mu}\,\epsilon^{\beta\nu} + \text{3 terms}) \implies \big[ t^{\mu\nu}, t^{\alpha\beta} \ast A_{\alpha\beta} \big]_{\ast} = 0
\end{equation}
Hence $\id \subset \off$. That $\id$ is two-sided also follows from \eqref{sp2-covar}:
\begin{equation}
    \label{tA}
    \big[t^{\alpha\beta}, A_{\alpha\beta} \big]_{\ast} = 0 \implies t^{\alpha\beta} \ast A_{\alpha\beta} = A_{\alpha\beta}\ast t^{\alpha\beta} = t^{\alpha\beta} \cdot A_{\alpha\beta} - \frac{\partial^2 A_{\alpha\beta}}{\partial Y_\alpha \cdot \partial Y_\beta} \, .
\end{equation}
For $f \in \off$, both $f \ast A^{\mu\nu}$ and $A^{\mu\nu} \ast f$ are polynomials satisfying \eqref{sp2-covar}:
\begin{subequations}
    \begin{align}
        &\big[t^{\alpha\beta}, f\ast  A^{\mu\nu} \big]_{\ast} = 2i \cdot \big( f \ast A^{\alpha\mu}\, \cdot  \,\epsilon^{\beta\nu} + \text{3 terms}) \, ,\\
        &\big[t^{\alpha\beta},   A^{\mu\nu} \ast f \big]_{\ast} = 2i \cdot \big( A^{\alpha\mu} \ast f \, \cdot  \,\epsilon^{\beta\nu} + \text{3 terms}) \, .
    \end{align}
\end{subequations}

By \eqref{sp2-covar}, the $\mathfrak{sp}(2)$ indices of $A_{\alpha\beta}(Y)$ can be carried only by $Y$ factors, while the indices of all other $Y$'s are contracted via $\epsilon_{\bullet \bullet}$:
\begin{equation}
    \label{sp2-covar:cnsq}
    A_{\alpha\beta}(Y) = Y^A_{\alpha} \, Y^B_{\beta} \, \mathcal{B}_{AB}(Y) \hookrightarrow \begin{cases}
        & \mathcal{B}_{AB}(Y) = \mathcal{B}_{BA}(Y) \, , \\
        & \big[t_{\alpha\beta}, \mathcal{B}_{AB}(Y) \big]_{\ast} =0 \, .
    \end{cases}
\end{equation}
This is analogous to \eqref{gauge_mdl:cnsq}, where \eqref{sp2-inv} forces all $\mathfrak{sp}(2)$ indices within elements of $\offbig$ to be contracted via $\epsilon_{\bullet\bullet}$.

Finally, even among polynomials, different $A_{\alpha\beta}(Y) \neq A'_{\alpha\beta}(Y)$ satisfying \eqref{sp2-covar} may give the same element of $\id$:
\begin{equation}
    \label{divisor-of-0}
    t^{\alpha\beta} \ast A_{\alpha\beta}(Y) = t^{\alpha\beta} \ast A'_{\alpha\beta}(Y) \, .
\end{equation}
Indeed, if $B_{\alpha\beta}(Y)$ satisfies \eqref{sp2-covar}, then
\begin{equation}
    t^{\mu\nu} \ast \epsilon^{\alpha\beta} \big\{ t_{\alpha(\mu}, B_{\nu)\beta} \big\}_{\ast} \equiv 0 \, .
\end{equation}
This is irrelevant for the factorisation itself, which involves only $t^{\alpha\beta} \ast A_{\alpha\beta}$, but it may affect how the factorisation is lifted to the interaction level (see Section~\ref{sec:ainfty}).

A similar $\mathfrak{sp}(2)$-ideal can formally be defined in $\offbig$, but this again requires a precise definition of the underlying vector space.

\subsection{On-shell HS algebra}

The polynomial on-shell HS algebra $\onsh$ is defined as a factor-algebra\cite{Vasiliev:2003ev,vasiliev2004higher,eastwood2005higher}.
\begin{definition}[$\big(\onsh \, \text{algebra} \big)$]
The algebra $\onsh$ is the factor-algebra
\begin{equation}
    \onsh := \off \, / \, \id \, .
\end{equation}
\end{definition}

Related constructions using spinor oscillators were developed in five and seven dimensions. In five dimensions, a central generator is eliminated by a coset construction\cite{Sezgin:2001zs}; in seven dimensions, tracelessness is obtained by factoring out an ideal containing the traces\cite{Sezgin:2001ij}.

Among the two-sided ideals of $\off$, $\id$ is singled out by the traceless condition \eqref{gauge_mdl:3}. As shown in \cite{vasiliev2004higher}, there exists the splitting
\begin{equation}
    \label{known-split}
    \forall f \in \off \, \,  \exists \big(t^{\alpha\beta} \ast \mathcal{A}_{\alpha\beta}\big) \in \id: \frac{\partial^2}{\partial Y^\alpha \cdot \partial Y^\beta} \big( f + t^{\mu\nu} \ast \mathcal{A}_{\mu\nu}  \big) = 0 \,,
\end{equation}
i.e., each element of $\off$ can be represented as
\begin{equation}
    f = \underbrace{\big( f + t^{\alpha\beta} \ast \mathcal{A}_{\alpha\beta} \big)}_{\text{AdS traceless}} -  \underbrace{t^{\alpha\beta} \ast \mathcal{A}_{\alpha\beta}}_{\in \, \id} \, .
\end{equation}
This suggests that $\onsh$ can be realised as an algebra of two-row $\so$ Young diagrams, which is what should appear at zeroth order of the interacting on-shell HS equations.

The splitting \eqref{known-split} is necessary but not sufficient for this: it must also be unique, i.e., exactly one $t^{\alpha\beta} \ast \mathcal{A}_{\alpha\beta}(f)$ must make $f+t^{\alpha\beta} \ast \mathcal{A}_{\alpha\beta}(f)$ AdS-traceless. Uniqueness is proved in the following sections, where $t^{\alpha\beta} \ast \mathcal{A}_{\alpha\beta}(f)$ is also derived explicitly.

As shown below, $\onsh$ can be modelled on the space of AdS-traceless $\mathfrak{sp}(2)$-singlet polynomials in $Y^\alpha_A$ (recall \eqref{gauge_mdl:3} and \eqref{gauge_mdl:2}). However, nontrivial interaction equations cannot be formulated with polynomials alone\cite{vasiliev2004higher}. In the gauge-sector equation
\begin{equation}
    d_x \omega + \omega \circ \omega = \dots \, ,
\end{equation}
the connection $\omega = \omega(Y)$ cannot in general be polynomial: $\onsh$ has no finite-dimensional subalgebra containing generators of spin greater than two, so a gauge sector containing such a generator involves all spins.

Hence there is a larger algebra $\onbig \supset \onsh$, spanned by power series in $Y^\alpha_A$, with the product induced from $\onsh$. Its explicit construction requires a functional class closed under the $\circ$-product, which is beyond the scope of this paper.

\subsection{Factor-algebra and representatives}
\label{sec:factor-algebra}

This subsection recalls the standard realisation of a factor-algebra on a space of representatives of equivalence classes. Parts of this scheme are needed later to lift the factorisation of an associative algebra to the level of a minimal $A_\infty$-algebra.

Let $\mathfrak{A}=(A, \ast)$ be an associative algebra over a field $\mathds{F}$ with a two-sided ideal $\mathcal{I}$. The equivalence relation
\begin{equation}
    u \thicksim v: (u-v) \in \mathcal{I} \, ,
\end{equation}
defines the factor-algebra $\mathfrak{A}/\mathcal{I}$, which is associative since $\mathcal{I}$ is two-sided.

\begin{definition}[$\big(\text{Space of representatives} \big)$] A linear subspace $\mathcal{R}(A) \subset A$ is a \textbf{space of representatives} if
\begin{equation}
    A = \mathcal{R}(A) \,  \oplus  \, \mathcal{I} \, ,
\end{equation}
in the sense of linear spaces, i.e.
    \begin{equation}
    \label{AA:decomp}
    \forall v \in A: \exists! \, \tilde{v} \in \mathcal{R}(A)\hookrightarrow  \tilde{v} - v =:  i(v) \in \mathcal{I}\, .
\end{equation}
\end{definition}
A linear complement of $\mathcal{I}$ always exists; the nontrivial requirement, relevant below, is that $\mathcal{R}(A)$ be singled out by additional properties such as $\so$-covariance.

As vector spaces, $\mathcal{R}(A) \cong A / \mathcal{I}$. The problem is to endow $\mathcal{R}(A)$ with a product $\circ$ such that $\big( \mathcal{R}(A), \circ\big) \cong \mathfrak{A}/\mathcal{I}$. To this end, we regard $i( \, \bullet \, )$ from \eqref{AA:decomp} as a map $i( \, \bullet \, ): A \rightarrow \mathcal{I}$. Its definition immediately implies the following lemma.
\begin{lemma} $i( \, \bullet \, ): A \rightarrow \mathcal{I}$ is a linear map, $i( \, \bullet \, )\Big|_{\mathcal{R}(A)} = 0$ and $i( \, \bullet \, )\Big|_{\mathcal{I}} = -\mathds{1}$, i.e.:
\begin{equation}
    \label{AA:lemma}
    \begin{split}
        \forall u, v \in A  \, \,  \forall \mu, \nu \in \mathds{F}&: \,i( \mu \cdot u + \nu \cdot v) = \mu \cdot i(u) + \nu \cdot i(v) \, \\
        \forall& \tilde{v} \in \mathcal{R}(A): \,  i(\tilde{v}) = 0 \, , \\
        &\forall u \in \mathcal{I}: \, i(u) = - u \, .
    \end{split}
\end{equation}
\end{lemma}

This lemma yields the realisation of $\mathfrak{A}/\mathcal{I}$ on $\mathcal{R}(A)$.

\begin{lemma}
The map $\, \circ \, :\mathcal{R}(A) \times \mathcal{R}(A) \rightarrow \mathcal{R}(A)$ defined as
    \begin{equation}
        \label{AA:circ-prod}
        \tilde{f} \circ \tilde{g} = \tilde{f} \ast \tilde{g} + i(\tilde{f} \ast \tilde{g}) \, ,
    \end{equation}
    is a bilinear associative product on $\mathcal{R}(A)$, and $\big( \mathcal{R}(A), \circ\big) \cong \mathfrak{A}/\mathcal{I}$.
\end{lemma}
The key point of the proof is the associativity of \eqref{AA:circ-prod}. Direct application of \eqref{AA:circ-prod} gives
    \begin{equation}
        \begin{split}
            &\tilde{f} \circ (\tilde{g} \circ \tilde{h}) = \tilde{f} \ast \tilde{g} \ast \tilde{h} + \tilde{f} \ast i(\tilde{g} \ast \tilde{h}) + i\Big( \tilde{f} \ast \tilde{g} \ast \tilde{h} + \tilde{f} \ast i(\tilde{g} \ast \tilde{h}) \Big) \, , \\
            & (\tilde{f} \circ \tilde{g}) \circ \tilde{h} = \tilde{f} \ast \tilde{g} \ast \tilde{h} + i(\tilde{f} \ast \tilde{g}) \ast \tilde{h} + i\Big(  \tilde{f} \ast \tilde{g} \ast \tilde{h} + i(\tilde{f} \ast \tilde{g}) \ast \tilde{h}\Big) \, .
        \end{split}
    \end{equation}
These expressions look different, but \eqref{AA:lemma} implies
    \begin{equation}
        \begin{split}
            &i\Big( \tilde{f} \ast \tilde{g} \ast \tilde{h} + \tilde{f} \ast i(\tilde{g} \ast \tilde{h}) \Big) =i(\tilde{f} \ast \tilde{g} \ast \tilde{h}) - \tilde{f} \ast i(\tilde{g} \ast \tilde{h}) \,,  \\
            & i\Big(  \tilde{f} \ast \tilde{g} \ast \tilde{h} + i(\tilde{f} \ast \tilde{g}) \ast \tilde{h}\Big) = i(\tilde{f} \ast \tilde{g} \ast \tilde{h}) - i(\tilde{f} \ast \tilde{g}) \ast \tilde{h} \, ,
        \end{split}
    \end{equation}
so that
    \begin{equation}
        \tilde{f} \circ (\tilde{g} \circ \tilde{h}) = \tilde{f} \ast \tilde{g} \ast \tilde{h} + i(\tilde{f} \ast \tilde{g} \ast \tilde{h}) = (\tilde{f} \circ \tilde{g}) \circ \tilde{h} \, ,
    \end{equation}
    which proves the associativity of $\, \circ \,$.

\subsection{Traceless conditions}
In HS algebras, a space of representatives is naturally chosen by imposing a traceless\linebreak{} condition\cite{vasiliev2004higher}. Such conditions are linear differential equations in $Y^\alpha_A$, so their solutions form linear subspaces of $\off$ or $\offbig$.
\begin{itemize}
    \item \textbf{The AdS-traceless condition} reads
    \begin{equation}
        \label{AdS-traceless}
        \frac{\partial^2}{\partial Y^\alpha \cdot \partial Y^\beta} f(Y) = 0 \, .
    \end{equation}
    This condition is natural in the gauge sector because the differential operator in \eqref{AdS-traceless} commutes with the background adjoint covariant derivative
    \[
    D_\Omega := d_x + [\Omega,\bullet]_\ast .
    \]
    AdS-traceless functions form a linear subspace of $\offbig$; since the operator in \eqref{AdS-traceless} is homogeneous in $Y^\alpha_A$, AdS-traceless polynomials form a linear subspace of $\off$.

    This condition is the main focus of the paper: we prove that \eqref{AdS-traceless} defines a space of representatives in $\off$ with respect to $\id$, and derive the map $i(\bullet): \off \rightarrow \id$ explicitly.

    \item \textbf{The twisted-traceless condition} reads (recall \eqref{small_y})
    \begin{equation}
        \label{Twist-traceless}
        \bigg(\eta_{ab} \, \frac{\partial^2}{\partial \mathbf{y}^\alpha_a \partial \mathbf{y}^\beta_b} + y_\alpha y_\beta \bigg) f(Y) = 0 \, .
    \end{equation}
    This condition is natural in the Weyl sector because the differential operator in \eqref{Twist-traceless} commutes with the background twisted-adjoint covariant derivative
    \[
    \widetilde{D}_\Omega := d_x + \Omega\ast\bullet - \bullet\ast\pi(\Omega) .
    \]
    Its solutions form a linear subspace of $\offbig$, but, unlike in the AdS case, the only polynomial solution is trivial. Hence twisted-traceless functions do not belong to $\off$ and are not analysed here; see \cite{TatarenkoPREP} for details.

    \item \textbf{The Lorentz-traceless condition} reads
    \begin{equation}
        \label{Lorentz-traceless}
        \eta_{ab} \, \frac{\partial^2}{\partial \mathbf{y}^\alpha_a \partial \mathbf{y}^\beta_b} f(Y) = 0 \, .
    \end{equation}
    Although this condition is less directly adapted to the linear dynamics\cite{vasiliev2004higher}, it offers a common traceless condition for the gauge and Weyl sectors. This choice is motivated by the explicit form of the vertices obtained in \cite{Didenko:2024zpd}, in which AdS covariance is not manifest. Manifest AdS covariance can be restored by introducing a compensator, as mentioned above. Lorentz-traceless polynomials form a linear subspace of $\off$, which, as proven later, is also a space of representatives with respect to $\id$.
\end{itemize}
\newpage

\section{Projecting onto $AdS$-traceless polynomials}
\label{sec:projector}
This section constructs the horizontal projector onto $AdS$-traceless polynomials in terms of the generating variables $Y^\alpha_A$. \textit{Horizontal} means that the projector maps homogeneous polynomials of degree $N$ to homogeneous polynomials of the same degree. In terms of Young diagrams, it extracts the $AdS$-traceless part of a two-row $\mathfrak{gl}(d+1)$ diagram, so we first recall how $\mathfrak{gl}(d+1)$-irreps restrict to $\mathfrak{so}(d+1)$. These finite-dimensional statements do not depend on the signature and hence apply to $\so$ as well. Throughout this section, $s$ denotes the row length of a diagram, so that\footnote{Not to be confused with the spin, which corresponds to row length $s-1$.} $\hat{N} = 2s$.

\subsection{Restriction of $\mathfrak{gl}(d+1)$ to $\mathfrak{so}(d+1)$}
\label{sec:gl-so}
Homogeneous $\mathfrak{sp}(2)$-singlet polynomials of degree $2s$ span the $\mathfrak{gl}(d+1)$-irrep described by the two-row rectangular diagram of length $s$. As an $\mathfrak{so}(d+1)$-module, a $\mathfrak{gl}(d+1)$-irrep is, in general, a direct sum of $\mathfrak{so}(d+1)$-irreps; see \cite{bekaert2021unitary} for a review. For two-row diagrams, the branching rule is\footnote{We assume $d>3$. For $d=3$, the $\mathfrak{so}(4)$-modules on the r.h.s. split further into (anti)self-dual parts.}
\begin{equation}
        \label{GLtoSO}
        \Bigg(\,\raisebox{0.6ex}{$\overbrace{\ytableausetup{mathmode, boxsize=1.25em}
        {\begin{ytableau}
        {} & {\bullet} & {\bullet} & {\bullet} & {\bullet} & {}\\
        {} & {\bullet} & {\bullet} & {\bullet} & {\bullet} & {}
        \end{ytableau}}}^{s}$}\, \Bigg)_{\mathfrak{gl}(d+1)} = \raisebox{-2.3ex}{\scalebox{4.8}{$\oplus$}} \Bigg( \mathgraphics{5em}{7.5em}{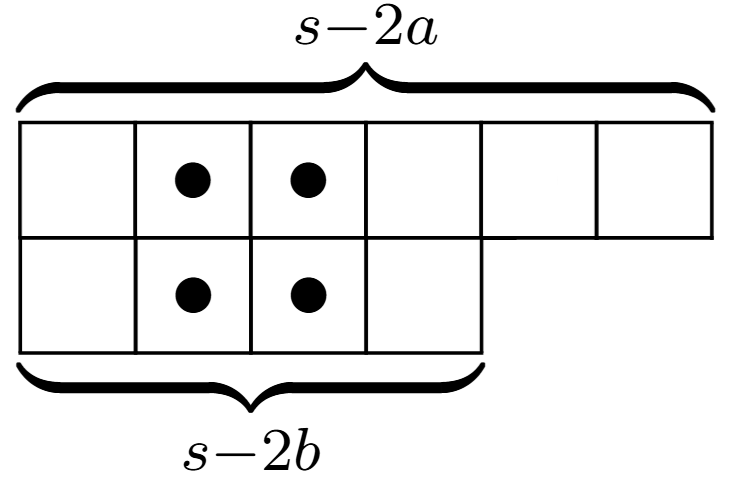}\Bigg)_{\mathfrak{so}(d+1)}  \, ,
\end{equation}
where the sum runs over all $a, b$ with $0 \leqslant 2a \leqslant 2b \leqslant s$, i.e., over all legitimate Young diagrams obtained by removing an even number of cells. This reflects the extraction of traces: all tensor powers of the metric, viewed as $\mathfrak{gl}(d+1)$ diagrams\footnote{Known as plethysms\cite{bekaert2021unitary}.}, contain an even number of cells. A diagram with $2s-2p$ cells on the r.h.s. of \eqref{GLtoSO} is the traceless part of the $p$-fold trace of the original diagram. Each diagram appears at most once, i.e., the decomposition is multiplicity-free. Consequently,
\begin{itemize}
    \item the r.h.s. of \eqref{GLtoSO} contains $\frac{(n+1)(n+2)}{2}$ diagrams, where\footnote{$\lfloor x \rfloor$ and $\lceil x \rceil$ denote rounding $x$ down and up, respectively.} $n = \lfloor s/2 \rfloor$;
    \item the number of diagrams corresponding to the traceless part of the $p$-fold trace is
    \begin{equation}
        \label{reduction}
        \# \Big(\text{diagrams with $2(s-p)$ cells} \Big)= \min(p,n) - \lceil p/2 \rceil +1 \, .
    \end{equation}
\end{itemize}
Both properties shape the structure of the projector, as shown below. 
For example, for the box diagram ($s = 2$, $n = 1$), \eqref{GLtoSO} reads
    \begin{equation}
        \label{box_diag}
        \ytableausetup{mathmode, boxsize=1em}
        \Bigg(\raisebox{0.6ex}{\begin{ytableau}
        {} & {} \\
        {} & {}
        \end{ytableau}}\Bigg)_{ \, \mathfrak{gl}(d+1)} =  \, \, \underbrace{\Bigg(\raisebox{0.6ex}{\begin{ytableau}
        {} & {} \\
        {} & {}
        \end{ytableau}}\Bigg)_{ \, \mathfrak{so}(d+1)}}_{\text{traceless part}} \raisebox{-0.3ex}{\scalebox{1.5}{$\oplus$}}  \underbrace{\Bigg(\raisebox{-0.5ex}{\begin{ytableau}
        {} & {}
        \end{ytableau}}\Bigg)_{ \, \mathfrak{so}(d+1)}}_{\text{ t-s part of single trace} } \raisebox{-0.3ex}{\scalebox{1.5}{$\oplus$}}  \underbrace{\bigg(\, \,\raisebox{-0.3ex}{\scalebox{1.4}{$\bullet$}}\, \, \bigg)_{ \, \mathfrak{so}(d+1)}}_{\text{t-s part of double trace}} \, ,
    \end{equation}
where the scalar $\bullet$ is traceless by definition.
In agreement with \eqref{reduction}, each of the traces $p = 0, 1, 2$ contributes one diagram. The 8-cell diagram ($s = 4$, $n = 2$) is the smallest one in which the traceless part of the double trace consists of two distinct diagrams.
By \eqref{reduction}, the number of diagrams for a fixed number of traces depends on the size of the initial diagram. This leads to a certain redundancy in the ansatz for the horizontal projector.

\subsection{Constructing the horizontal projector $\mathcal{P}\big( \bullet \big)$}
\label{sec:horproj}
We introduce the Euler operator $\hat{N}$, which induces a $\mathds{Z}_{\geq0}$-grading on $\off$, and the trace operator $\Gamma_{\alpha\beta}$:
\begin{equation}
    \hat{N}:= Y^\alpha_A \frac{\partial}{\partial Y^\alpha_A} \, , \quad \Gamma_{\alpha\beta} := \eta_{AB}\frac{\partial^2}{\partial Y^\alpha_A \partial Y^\beta_B}.
\end{equation}
The AdS-traceless condition \eqref{AdS-traceless} then reads
\begin{equation}
    \Gamma_{\alpha\beta}f = 0 \, .
\end{equation}
To solve it explicitly, we construct a projector onto $\ker \Gamma_{\alpha\beta}$. Since such projectors are not unique\footnote{For example, if $\mathcal{P}\big( \bullet \big)$ is a projector onto $\ker \Gamma_{\alpha\beta}$, then so is $\mathcal{P}\big( \bullet \big) + \mathcal{P}\Big( \Gamma^{\alpha\beta}\Gamma_{\alpha\beta}\big(\bullet\big) \Big)$, which, however, violates condition 3 below.}, we require that:
\begin{enumerate}
    \item $\mathcal{P}\big( \bullet \big)$ is $\mathfrak{sp}(2)$-invariant, which ensures
    \begin{equation}
        \mathrm{Im}\,\mathcal{P}\big( \bullet \big) \subset \off \, ;
    \end{equation}
    \item $\mathcal{P}\big( \bullet \big)$ is $\so$-covariant, i.e., it involves only $AdS$-tensors present in $f$ or their traces;
    \item $\mathcal{P}\big( \bullet \big)$ is grade preserving (\textit{horizontal}), i.e., $\mathcal{P}\big(\hat{N}f\big) = \hat{N} \,\mathcal{P}\big(f\big)$;
    \item $\mathcal{P}(\tilde{f}) = \tilde{f}$ for all $\tilde{f}\in \ker \Gamma_{\alpha\beta}$ (normalisation).
\end{enumerate}
We also write $\mathcal{P}\big( \bullet \big)$ as a differential operator, which may be convenient within the sources formalism widely used in the spin-locality analysis (see \cite{Didenko:2019xzz,Didenko:2020bxd,Didenko:2024zpd} for examples).

Conditions 1--4 fix $\mathcal{P}\big( \bullet \big)$ uniquely. Indeed, in degree $2s$ the space of $\mathfrak{sp}(2)$-singlets is a single $\mathfrak{gl}(d+1)$-irrep, whose restriction \eqref{GLtoSO} is multiplicity-free. By Schur's lemma, the grade-preserving, $\so$-covariant projector onto its traceless component is unique, and its kernel is the sum of the remaining (traceful) components, i.e., the image of multiplication by $t^{\alpha\beta}$ (Lemma~\ref{lem:pos}(a)):
\begin{equation}
    \forall f_s \in \off: \, \hat{N} f_s = 2s \cdot f_s \, \, \, \exists! \tilde{f}_s \in \ker \Gamma_{\alpha\beta}: \, \hat{N} \tilde{f}_s = 2s \cdot \tilde{f}_s \hookrightarrow f_s = \tilde{f}_s + t^{\alpha\beta} \cdot A_{\alpha\beta} \, ,
\end{equation}
with some $\mathfrak{sp}(2)$-covariant $A_{\alpha\beta}$ (only $t^{\alpha\beta} \cdot A_{\alpha\beta}$, not $A_{\alpha\beta}$, is unique). Hence, in the splitting
\begin{equation}
    f_s = \mathcal{P}\big(f_s \big) + f_s - \mathcal{P}\big(f_s \big) \, ,
\end{equation}
$\mathcal{P}\big(f_s \big)$ is exactly $\tilde{f}_s$.

$\mathfrak{sp}(2)$-invariance and $\so$-covariance of a differential operator dictate the ansatz
\begin{equation}
    \label{ansatz}
    \mathcal{P}\big( f \big) = \sum_{r=0}^{+\infty}\#_r(\hat{N}) (\epsilon_{\bullet \bullet} \eta^{\bullet \bullet})^r \, Y^{\alpha_1}_{A_1}  \, \dots \, Y^{\alpha_r}_{A_r} \frac{\partial}{\partial Y^{\beta_1}_{B_1}} \, \dots \frac{\partial}{\partial Y^{\beta_r}_{B_r}} f \, ,
\end{equation}
where $(\epsilon_{\bullet \bullet} \eta^{\bullet \bullet})^r$ denotes a product of $r$ factors $\epsilon_{\alpha\beta}$ and $r$ factors $\eta^{AB}$. Thus all $\mathfrak{sp}(2)$ indices of $Y$ and $\partial/\partial Y$ are contracted via $\mathfrak{sp}(2)$-invariant tensors, and likewise for the $\so$ indices\footnote{Here $d>3$ is essential; the case $d=3$ is discussed in Section~\ref{sec:d3}.}. Functions of $\hat{N}$ act on everything to their right and are defined by
\begin{equation}
    \forall f_s \in \off: \hat{N} f_s = 2s \cdot f_s \hookrightarrow \mathcal{F}(\hat{N}) f_s : = \mathcal{F}(2s) \cdot f_s \, .
\end{equation}
The ansatz simplifies further once the $\so$ contractions are specified. Two $Y$'s contracted via $\eta^{AB}$ form $t^{\alpha\beta}$, and two $\partial/\partial Y$'s form $\Gamma_{\alpha\beta}$. A contracted pair $Y$, $\partial/\partial Y$ can be eliminated using the differential consequences of the $\mathfrak{sp}(2)$-singlet condition \eqref{sp2-inv}:
\begin{equation}
    \begin{split}
       &Y^\alpha_A \frac{\partial }{\partial Y^\beta_A}\bigg( \frac{\partial^nf}{\partial Y^{\gamma_1}_{C_1} \dots \partial Y^{\gamma_n}_{C_n}} \bigg) = \\
       =\delta^\alpha_\beta \,  \frac{\hat{N}+n}{2} \,  \frac{\partial^nf}{\partial Y^{\gamma_1}_{C_1} \dots \partial Y^{\gamma_n}_{C_n}}  &- \sum_{k=1}^n \delta^\alpha_{\gamma_k}  \frac{\partial^n f}{\partial Y^{\gamma_1}_{C_1} \dots \partial Y^{\gamma_{k-1}}_{C_{k-1}} \, \partial Y^\beta_{C_k} \, \partial Y^{\gamma_{k+1}}_{C_{k+1}} \dots \partial Y^{\gamma_n}_{C_n}} \, .
    \end{split}
\end{equation}
Therefore, \eqref{ansatz} reduces to
\begin{equation}
    \mathcal{P}\big( f \big)  = \sum_{r=0}^{+\infty} \Big(\#(\hat{N}) \Big)_{\alpha_1...\alpha_{2r} } ^{\beta_1...\beta_{2r}} t^{\alpha_1 \alpha_2} \dots t^{\alpha_{2r-1} \alpha_{2r}} \Gamma_{\beta_1 \beta_2} \dots \Gamma_{\beta_{2r-1} \beta_{2r}}  \, f \, ,
\end{equation}
where each $\mathfrak{sp}(2)$ index of $\Big(\#(\hat{N}) \Big)_{\alpha_1...\alpha_{2r} } ^{\beta_1...\beta_{2r}}$ is carried by $\epsilon_{\alpha \beta}$, $\epsilon^{\alpha\beta}$ or $\delta^\alpha_\beta$. The Schouten identities \eqref{B1} then yield

\begin{equation}
    \label{HorProj}
    \boxed{\mathcal{P}\big( f \big)  = \sum_{k=0}^{+\infty}\sum_{m=0}^{+\infty}c_{(k,m)}(\hat{N}) \, \mathcal{T}_{(k,m)}\, f \, ,}
\end{equation}
The terms of \eqref{HorProj} are \textit{trace-producing operators} (TPOs)
    \begin{equation}
        \label{TPO}
            \boxed{
            \mathcal{T}_{(k,m)} := \big(t^{\mu\nu}\big)^k \big(t^{\sigma\lambda}t_{\sigma\lambda}\big)^m\big(\Gamma_{\mu\nu}\big)^k \big(\Gamma^{\sigma\lambda}\Gamma_{\sigma\lambda}\big)^m : \off \rightarrow \off } \, ,
    \end{equation}
with the shorthand
\begin{equation}
    \label{TPO-shorthand}
    (t^{\mu\nu})^k\dots (\Gamma_{\mu\nu})^k = \bigg(\prod_{i=1}^{k} t^{\mu_i \nu_i}\bigg) \dots \bigg(\prod_{j=1}^{k} \Gamma_{\mu_j \nu_j}\bigg) \, .
\end{equation}

Since all $\Gamma_{\mu\nu}$ commute, mixed expressions such as $t_{\alpha\beta}\,\mathcal{T}_{(k,m)}\big(\Gamma^{\sigma\lambda}\Gamma_{\sigma\lambda} f\big)$ in \eqref{Proj-eq-1} are well defined. On any polynomial, only finitely many of them act nontrivially, since for large $k+2m$ the number of derivatives exceeds the degree. Their key properties are:
\begin{enumerate}
    \item All TPOs are grade preserving:
    \begin{equation}
        \forall \, k ,  m \in \mathds{Z}_{\geqslant 0}: \quad \Big[ \mathcal{T}_{(k,m)} \, , \,\hat{N}\Big] = 0 \, .
    \end{equation}
    \item The TPO $ \mathcal{T}_{(k,m)} f$ produces precisely the $(k+2m)$-fold trace\footnote{Single trace for $k+2m=1$, double trace for $k+2m=2$, etc.} of $f$.
    \item TPOs with $k+2m \leqslant n = \lfloor s/2 \rfloor$ are linearly independent on homogeneous $\mathfrak{sp}(2)$-singlets of degree $2s$:
    \begin{equation}
        \label{indpnd}
        \forall f_s: \hat{N} f_s = 2s \cdot f_s \hookrightarrow \sum_{0 \leqslant k+2m\leqslant n} \zeta_{(k, m)} \, \mathcal{T}_{(k,m)}  f_s  = 0 \Leftrightarrow \forall k, m: 0\leqslant k+2m\leqslant n \hookrightarrow \zeta_{(k,m)} = 0 \, .
    \end{equation}
    \item For $k+2m > n$, TPOs are linearly dependent. Writing the number of traces as $2n-q$, the dependence is captured by
    \begin{equation}
        \label{reductionA}
        \mathcal{T}_{(2n-2m-q,\,m)} \, f_{s}  =\sum_{j=0}^{l}B_j(m;n,l) \, \mathcal{T}_{(2n-2j-q,\,j)} \, f_{s} \, ,
    \end{equation}
    where $l = \lfloor q/2 \rfloor$ and
    \begin{equation}
    \label{B-coef}
    B_j(m ; n, l) = \begin{cases}
        &\delta^m_j \, , \, 0 \leqslant m \leqslant l \, , \\
        & \frac{m!}{j!} \cdot  \frac{(-1)^{j+l}}{(m-j) \cdot \Gamma(l-j+1) \cdot \Gamma(m-l)}  \cdot \frac{\Gamma(n+3/2-j)}{\Gamma(n+3/2-m)} \cdot \frac{\Gamma(n-l-m+1/2)}{\Gamma(n-l-j+1/2)} \, , \, m > l \, .
    \end{cases}
    \end{equation}
    For odd $s$, the remaining case $k+2m=s$ is not covered by this parametrisation with $q\geqslant0$; all TPOs in this case vanish on $V_s$, as shown in Appendix~\ref{app:B2}.
\end{enumerate}

Properties 1 and 2 follow directly from \eqref{TPO}. Properties 3 and 4 are proved in Appendix~\ref{app:B2}: property 3 is Corollary~\ref{cor:indep} ($d \geqslant 4$), and the relations of property 4 are Proposition~\ref{prop:reduction}(a), valid for every $d$.
To illustrate property 2: schematically, $\Gamma_{\mu\nu}$ acts on a homogeneous element of $\off$ as
\begin{equation}
    \label{gamma-action}
    \Gamma_{\mu\nu}\Big( T_{A(s) \, , \, B(s)} \, \big( M^{AB} \big)^{(s)} \Big) \thicksim Y^{A'}_\mu Y^{B'}_\nu \, \big(T_{A'A'' A(s-2)\, , \, B'B''B(s-2)} \eta^{A'' B''}\big) \big( M^{AB} \big)^{(s-2)} \, ,
\end{equation}
i.e., it produces a single trace of $T_{A(s)\, , \, B(s)}$. The result is not an element of $\off$, but becomes one after contraction with $t^{\mu\nu}$. This contraction involves only $\mathfrak{sp}(2)$ indices and takes no further traces, so $t^{\mu\nu}\Gamma_{\mu\nu} f$ is the single trace of $f$. Similarly,
\begin{equation}
    \begin{split}
        &\Gamma_{\mu_1 \nu_1}\Gamma_{\mu_2 \nu_2}\Big( T_{A(s) \, , \, B(s)} \, \big( M^{AB} \big)^{(s)}  \Big) \thicksim \\
        &\thicksim Y^{A'}_{\mu_1} Y^{B'}_{\nu_1} \, Y^{A'}_{\mu_2} Y^{B'}_{\nu_2} \, \big(T_{A'(2)A''(2) A(s-4)\, , \, B'(2)B''(2) B(s-4)} (\eta^{A''B''})^{(2)}\big) \big( M^{AB} \big)^{(s-4)} + \\
        &+ \big( Y^{A'}_{\mu_1}Y^{B'}_{\mu_2} \,\epsilon_{\nu_1\nu_2} + \dots \big) \big(T_{A'A''(2) A(s-3)\, , \, B'B''(2)B(s-3)} (\eta^{A''B''})^{(2)}\big) \big( M^{AB} \big)^{(s-3)} + \\
        &+ \big(\epsilon_{\mu_1\mu_2}\epsilon_{\nu_1\nu_2}+\epsilon_{\mu_1\nu_2}\epsilon_{\nu_1\mu_2}\big)\big(T_{A''(2) A(s-2)\, , \, B''(2)B(s-2)} (\eta^{A''B''})^{(2)}\big) \big( M^{AB} \big)^{(s-2)}
    \end{split}
\end{equation}
is a double trace of $T_{A(s)\, , \, B(s)}$. Unlike in \eqref{gamma-action}, there are now two distinct $\mathfrak{sp}(2)$-singlets, $\mathcal{T}_{(2,0)} f$ and $\mathcal{T}_{(0,1)} f$, and again no further traces are taken. The generalisation to higher orders is straightforward and relates the grade of a TPO to the trace multiplicity.

This relation reduces the linear (in)dependence of TPOs to the branching \eqref{GLtoSO}. For example, the decomposition \eqref{box_diag} of the box diagram is the Ricci decomposition
\begin{equation}
        \ytableausetup{mathmode, boxsize=1em}
        \underbrace{\Bigg(\raisebox{0.6ex}{\begin{ytableau}
        {} & {} \\
        {} & {}
        \end{ytableau}}\Bigg)_{ \, \mathfrak{gl}(d+1)}}_{\text{Riemann tensor}} =  \, \, \underbrace{\Bigg(\raisebox{0.6ex}{\begin{ytableau}
        {} & {} \\
        {} & {}
        \end{ytableau}}\Bigg)_{ \, \mathfrak{so}(d+1)}}_{\text{Weyl tensor}} \raisebox{-0.3ex}{\scalebox{1.5}{$\oplus$}}  \underbrace{\Bigg(\raisebox{-0.5ex}{\begin{ytableau}
        {} & {}
        \end{ytableau}}\Bigg)_{ \, \mathfrak{so}(d+1)}}_{\text{t-s part of Ricci tensor} } \raisebox{-0.3ex}{\scalebox{1.5}{$\oplus$}}  \underbrace{\bigg(\, \,\raisebox{-0.3ex}{\scalebox{1.4}{$\bullet$}}\, \, \bigg)_{ \, \mathfrak{so}(d+1)}}_{\text{scalar curvature}} \, ,
\end{equation}
i.e., the Weyl tensor is the Riemann tensor plus its single and double traces with suitable coefficients. The same holds for any two-row diagram. Each diagram has a unique single-trace component, which is captured by the single TPO
\begin{equation}
    t^{\mu\nu}\Gamma_{\mu\nu}: \off \rightarrow \off
\end{equation}
and corresponds in \eqref{GLtoSO} to removing one pair of cells. For double traces, there are two TPOs:
\begin{equation}
    \label{3.31}
        \begin{split}
            \mathcal{T}_{(2,0)}: \off & \rightarrow \off \, , \\
            & \\
            &\text{and} \\ 
            & \\
            \mathcal{T}_{(0,1)}: \off & \rightarrow \off \, ,
        \end{split}
\end{equation}
in agreement with \eqref{GLtoSO}: two pairs of cells can be removed either as a box or as a row of four cells. For the box and the 6-cell diagrams, the second option is unavailable, so the two TPOs in \eqref{3.31} are linearly dependent there.

In general, for $p \leqslant n$ the number of TPOs with $k+2m = p$ equals $\lfloor p/2 \rfloor + 1$, which coincides with \eqref{reduction}, whereas for $p > n$ there are fewer diagrams than TPOs. This count is consistent with properties 3 and 4.

\subsection{Explicit form of $\mathcal{P}\big(  \bullet  \big)$}
On a homogeneous polynomial $\hat{N} f_s = 2s \cdot f_s$, the series \eqref{HorProj} terminates:
\begin{equation}
    \label{3.32}
    \mathcal{P} \big( f_s \big) = \sum_{k+2m\leqslant s} c_{(k, m)}(2s) \cdot \mathcal{T}_{(k,m)}  f_s  \, ,
\end{equation}
so finding $c_{(k,m)}(\hat{N})$ amounts to fixing $c_{(k, m)}(2s)$ for all $s \in \mathds{Z}_{\geqslant 0}$. By properties 3 and 4, we split \eqref{3.32} as
\begin{equation}
    \mathcal{P}\big( f_s \big) = \underbrace{\sum_{0\leqslant k+2m\leqslant n} c_{(k, m)}(2s) \cdot \mathcal{T}_{(k,m)}  f_s}_{\text{all terms are independent}} +\underbrace{\sum_{n<k+2m\leqslant s} c_{(k, m)}(2s) \cdot \mathcal{T}_{(k,m)}  f_s}_{\text{linear dependence of terms}} \, ,
\end{equation}
with $n = \lfloor s/2 \rfloor$. Hence the coefficients $c_{(k,m)}(2s)$ are
\begin{enumerate}
    \item fixed \textbf{unambiguously} for $0 \leqslant k+2m \leqslant n$;
    \item fixed only \textbf{up to} the relations \eqref{reduction}, \eqref{reductionA} for $n < k+2m \leqslant s$;
    \item irrelevant (can be set to $0$) for $k+2m > s$, since they do not enter \eqref{3.32}.
\end{enumerate}
Thus $c_{(k, m)}(\hat{N})$ are inherently ambiguous, and \textbf{any} choice satisfying
\begin{equation}
    \label{Proj-eq}
    \boxed{
        \begin{split}
        &\forall f \in \off: \Gamma_{\alpha\beta} \Big( \mathcal{P}\big( f \big) \Big) = 0 \, , \\
        &\forall \Tilde{f} \in \ker \Gamma_{\alpha\beta}: \, \mathcal{P}\big( \Tilde{f} \big) = \Tilde{f} \, ,
        \end{split}
    }
\end{equation}
defines the same projector. The derivation is given in Appendix~\ref{app:B3}; its main steps are:
\begin{enumerate}
    \item Using \eqref{B4}, the first equation in \eqref{Proj-eq} takes the form
    \begin{equation}
        \label{Proj-eq-1}
        \sum_{k=0}^{+\infty}\sum_{m=0}^{+\infty} \mathscr{A}_{(k, m)}(\hat{N}) \cdot \mathcal{T}_{(k,m)}\big(\Gamma_{\alpha\beta} f\big) + \sum_{k=0}^{+\infty}\sum_{m=0}^{+\infty} \mathscr{B}_{(k, m)}(\hat{N}) \cdot t_{\alpha\beta} \, \mathcal{T}_{(k,m)}\big(\Gamma^{\sigma\lambda} \Gamma_{\sigma\lambda} f\big) = 0 \, ,
    \end{equation}

    where $\mathscr{A}_{(k, m)}(\hat{N})$ and $\mathscr{B}_{(k, m)}(\hat{N})$, given in \eqref{B30} and \eqref{B31}, are linear in the unknown $c_{(k,m)}(\hat{N})$.
    \item On $f_s$ with $\hat{N} f_s = 2s \cdot f_s$, this becomes
    \begin{equation}
        \label{Proj-eq-cons}
        \sum_{k+2m\leqslant s-1}\mathscr{A}_{(k, m)}(2s-2) \cdot \mathcal{T}_{(k,m)}\big(\Gamma_{\alpha\beta} \, f_s\big) +\sum_{k+2m\leqslant s-2} \mathscr{B}_{(k, m)}(2s-2) \cdot t_{\alpha\beta} \, \mathcal{T}_{(k,m)}\big(\Gamma^{\sigma\lambda} \Gamma_{\sigma\lambda} f_s\big) = 0 \, .
    \end{equation}
    \item Because of the linear dependences, it is sufficient, though not necessary, to require that each term of \eqref{Proj-eq-cons} vanish separately:
        \begin{equation}
            \label{rqrmnt}
            \begin{split}
                &\forall (k, m): 0\leqslant k+2m \leqslant s - 1 \hookrightarrow \mathscr{A}_{(k, m)}(2s-2) \overset{!}{=} 0 \, , \\
                &\forall (k, m): 0\leqslant k+2m \leqslant s - 2 \hookrightarrow \mathscr{B}_{(k, m)}(2s-2) \overset{!}{=} 0 \, . \\
            \end{split}
        \end{equation}
    This is a convenient choice for the ambiguous coefficients. \textit{A priori}, \eqref{rqrmnt} might have no solution; in fact, it has exactly one, up to the initial condition.
    \item Together with the initial condition from $\mathcal{P}\big( \Tilde{f} \big) = \Tilde{f}$, \eqref{rqrmnt} yields the recurrence system
        \begin{equation}
            \scalebox{0.85}{ $
            \begin{split}
                & c_{(0, 0)}(2s) = 1 \, , \\
                &k+2m\leqslant s-1:  2(k+1)(d+2s-2k-4m-5) \cdot c_{(k+1, m)}(2s) + c_{(k, m)}(2s) = 0 \, , \\
                &k+2m \leqslant s -2:  (k+1)(k+2) c_{(k+2, m)}(2s)+2(m+1)(d+2s-2m-4) c_{(k, m+1)}(2s) = 0 \, .
            \end{split}  $}
        \end{equation}
    For $d>3$ and the stated ranges of $k, m$, the factor $d+2s-2k-4m-5 \geqslant d-3$ does not vanish, and the unique solution is
    \begin{equation}
         c_{(k, m)}(2s) = \frac{\Gamma\big( \frac{d+2s-3}{2} - k -2m \big) \Gamma\big(\frac{d+2s-2}{2} -m \big)}{\Gamma\big( \frac{d+2s-3}{2}\big) \Gamma\big(\frac{d+2s-2}{2}\big)} \cdot \frac{(-)^{k+m}}{k! \cdot m! \cdot 2^{2k+6m}} \, .
    \end{equation}
\end{enumerate}
The horizontal projector onto $\ker \Gamma_{\alpha\beta}$ is therefore
\begin{equation}
    \label{c}
    \boxed{
    \begin{split}
        \mathcal{P}\big( f \big) = \sum_{k=0}^{+\infty}\sum_{m=0}^{+\infty}c_{(k, m)}(\hat{N}) \, \mathcal{T}_{(k,m)} f& \, ,\\
        & \\
        c_{(k, m)}(\hat{N}) =\frac{\Gamma\big( \frac{d+\hat{N}-3}{2} - k-2m \big) \Gamma\big( \frac{d+\hat{N}-2}{2} - m\big)}{\Gamma\big( \frac{d+\hat{N}-3}{2}  \big)  \Gamma\big( \frac{d+\hat{N}-2}{2}\big)}& \cdot \frac{(-)^{k+m}}{k! \cdot m! \cdot 2^{2k+6m}} \, .
    \end{split}}
\end{equation}

\subsection{Properties of $\mathcal{P}\big( \bullet\big) $}

\subsubsection*{Preliminaries of the factorisation}
The structure of $\mathcal{P}\big( \bullet \big)$ implies, for any $f \in \off$,
\begin{equation}
    \label{split-start}
    f = \mathcal{P}(f) - t^{\alpha\beta} \cdot A_{\alpha\beta}(f) \, ,
\end{equation}
where $A_{\alpha\beta}(f)$ can be chosen as
\begin{equation}
    A_{\alpha\beta}(f)= \sum_{k=1}^{+\infty}\sum_{m=0}^{+\infty} a_{(k,m)}(\hat{N}) \, \mathcal{T}_{(k-1,m)}\big(\Gamma_{\alpha\beta} f\big) +\sum_{k=0}^{+\infty}\sum_{m=1}^{+\infty} b_{(k,m)}(\hat{N}) \, t_{\alpha\beta \, }  \mathcal{T}_{(k,m-1)}\big(\Gamma^{\sigma\lambda}\Gamma_{\sigma\lambda} f\big) \, ,
\end{equation}
with any coefficients satisfying\footnote{With the convention $a_{(0, m)}(\hat{N}) = b_{(k, 0)}(\hat{N}) = 0$.}
\begin{equation}
    a_{(k,m)}(\hat{N}-2) + b_{(k,m)}(\hat{N}-2) = c_{(k, m)}(\hat{N}) \, .
\end{equation}
This representation is not unique, but for any choice $A_{\alpha\beta}(f)$ is an $\mathfrak{sp}(2)$-covariant rank-2 tensor satisfying \eqref{sp2-covar}: since $f \in \off$, its $\mathfrak{sp}(2)$ indices are contracted via $\epsilon_{\bullet\bullet}$, so $A_{\alpha\beta}(f)$ has the form \eqref{sp2-covar:cnsq}. This opens a path to the existence of the splitting \eqref{known-split}.

\subsubsection*{Technical property}
Consider, for $m \in \mathds{N}$,
\begin{equation}
    \big( t^{\alpha\beta}t_{\alpha\beta} \big)^m \cdot \tilde{f}_s: \quad \tilde{f}_s \in \ker \Gamma_{\alpha\beta} \, , \ \hat{N} \tilde{f}_s = 2s \cdot \tilde{f}_s \, .
\end{equation}
All its traces come from the factor $\big( t^{\alpha\beta}t_{\alpha\beta} \big)^m$:
\begin{equation}
    \label{property-0}
    \Gamma_{\alpha\beta} \Big( \big(t^{\mu\nu}t_{\mu\nu}\big)^m \cdot \tilde{f}_s \Big) = 4m(d+2m+2s-2) \cdot t_{\alpha\beta} \cdot  \big(t^{\mu\nu}t_{\mu\nu}\big)^{m-1} \cdot \tilde{f}_s \, .
\end{equation}
A further $\Gamma_{\alpha'\beta'}$ yields an expression whose indices $\alpha'$, $\beta'$ are carried by $t_{\bullet\bullet}$ or $\epsilon_{\bullet\bullet}$: $\Gamma_{\alpha'\beta'}\tilde{f}_s = 0$, and a single derivative acting on $\tilde{f}_s$ reduces to $\hat{N}$ (see \eqref{B2} at $n=0$). Hence
\begin{equation}
    \label{3.46}
    \mathcal{P}\Big( \big( t^{\alpha\beta}t_{\alpha\beta} \big)^m \cdot \tilde{f}_s  \Big) = \kappa_{(m, s)} \cdot \big( t^{\alpha\beta}t_{\alpha\beta} \big)^m \cdot \tilde{f}_s \, .
\end{equation}
Since $\mathrm{Im}\,\mathcal{P}\big( \bullet \big) = \ker \Gamma_{\alpha\beta}$, \eqref{property-0} and \eqref{3.46} force $\kappa_{(m, s)} = 0$, so
\begin{equation}
    \label{property-1}
    \boxed{
    \forall m\in \mathds{N} \quad \forall \tilde{f} \in \ker \Gamma_{\alpha\beta} : \quad \mathcal{P}\Big( \big( t^{\alpha\beta}t_{\alpha\beta} \big)^m \cdot \tilde{f}  \Big) = 0 .}
\end{equation}
Equivalently, $\big( t^{\alpha\beta}t_{\alpha\beta} \big)^m \cdot \tilde{f}_s$ lies in an $\so$-irrep different from the traceless component of its degree, so \eqref{property-1} also follows from Schur's lemma.
\newpage

\section{Factoring-out ideal}
\label{sec:factoring}
In this section the factor-algebra $\onsh$ is constructed explicitly. We prove the uniqueness of the splitting \eqref{known-split}, derive the splitting map and the induced product, evaluate the induced product on the $\so$-generators, show that the Lorentz-traceless polynomials also form a space of representatives, and analyse the case $d=3$. Unless stated otherwise, $d>3$.

\subsection{Existence of the splitting}
\label{sec:existence}
The existence of the splitting \eqref{known-split} was established by Vasiliev\cite{vasiliev2004higher}; the explicit form of $\mathcal{P}\big( \bullet \big)$ makes each step of the argument explicit. The initial step is \eqref{split-start}: by \eqref{c} there exists $A_{\alpha\beta}(Y)$ satisfying \eqref{sp2-covar} such that
\begin{equation}
    f = \mathcal{P}(f) + \big(f - \mathcal{P}(f) \big) = \mathcal{P}(f) - t^{\alpha\beta} A_{\alpha\beta}(f)\, .
\end{equation}
Using \eqref{tA}, this is rewritten as
\begin{equation}
    f = \underbrace{\mathcal{P}(f)}_{\in \ker \Gamma_{\alpha\beta}} -\, \Gamma^{\alpha\beta} A_{\alpha\beta}(f) - \underbrace{t^{\alpha\beta} \ast A_{\alpha\beta}(f)}_{\in \id} \, .
\end{equation}
Applying the same transformation to $\Gamma^{\alpha\beta}A_{\alpha\beta}(f)$ gives
\begin{equation}
    \begin{split}
      f = &\underbrace{\mathcal{P}\big( f \big) - \mathcal{P}\Big( \Gamma^{\alpha\beta}A_{\alpha\beta}(f) \Big)}_{\in \ker \Gamma_{\alpha\beta}} -\, \Gamma^{\alpha\beta}A_{\alpha\beta}\Big( \Gamma^{\mu\nu}A_{\mu\nu}(f)\Big) - \\
      &-\underbrace{\Big( t^{\alpha\beta} \ast A_{\alpha\beta}(f) - t^{\alpha\beta} \ast A_{\alpha\beta}\big( \Gamma^{\mu\nu}A_{\mu\nu}(f) \big) \Big)}_{\in \id} \, ,
    \end{split}
\end{equation}
and so on. For polynomial $f$ the procedure terminates, since
\begin{equation}
    \mathcal{P}\big( \hat{N} f \big) = \hat{N} \, \mathcal{P}\big( f \big) \implies A_{\alpha\beta}\big( \hat{N} f \big) = \big(\hat{N} - 2 \big) \, A_{\alpha\beta}\big(f \big) \, .
\end{equation}
Hence for every $f \in \off$ there exists an $\mathfrak{sp}(2)$-covariant $\mathcal{A}_{\alpha\beta}\big(f \big)$ with
\begin{equation}
    \Gamma_{\alpha\beta} \Big( f +  t^{\alpha\beta} \ast\mathcal{A}_{\alpha\beta}\big(f \big) \Big) = 0 \, ,
\end{equation}
which proves the splitting
\begin{equation}
    \label{4.6}
    \forall f \in \off \, \exists \Tilde{f} \in \ker \Gamma_{\alpha\beta} \hookrightarrow  \big(\Tilde{f} - f \big) \in \id \, .
\end{equation}
Two remarks are in order. First, the algorithm never raises the degree, so the representative of $f$ has degree not exceeding that of $f$; this is used repeatedly below.
Second, without the uniqueness of \eqref{4.6}, $\ker \Gamma_{\alpha\beta}$ need not carry the algebraic structure of $\off / \id$. We are not aware of a direct proof of uniqueness in \cite{vasiliev2004higher}; here it follows from \eqref{qsprjctr:nice}.
At $d=3$ the existence of the splitting \eqref{4.6} follows from \cite{vasiliev2004higher}.

\subsection{Uniqueness of the splitting}
\label{sec:uniqueness}
Uniqueness of \eqref{4.6} amounts to
\begin{equation}
    \forall f \in \off: \exists! \,  \tilde{f} \in \ker \Gamma_{\alpha\beta} \, : \big(f -\tilde{f} \,\big) \in \id \, ,
\end{equation}
which, given existence, is equivalent to
\begin{equation}
    \label{4.8}
    \ker \Gamma_{\alpha\beta} \, \cap  \, \id = \{0 \} \, .
\end{equation}
To prove it we use the family of functions of the $\mathfrak{sp}(2)$- and $\so$-invariant $t^{\mu\nu}t_{\mu\nu}$
\begin{equation}
    \label{quasiprojector}
    \boxed{
    \Delta(t^{\mu\nu}t_{\mu\nu}; s) = \frac{\Gamma\big( \frac{d}{2} \big) \Gamma\big(s+\frac{d-3}{2} \big) }{\Gamma\big(s+\frac{d}{2})\Gamma\big(\frac{d-3}{2} \big)} \sum_{m=0}^{+\infty} \frac{\Gamma\big(s+\frac{d}{2}\big)}{\Gamma\big(m+s+\frac{d}{2}\big)} \cdot \frac{1}{8^m} \cdot \frac{\big(t^{\mu\nu}t_{\mu\nu} \big)^m}{m!} \, ,
    }
\end{equation}
with $s \in \mathds{Z}_{\geqslant 0}$. This family is defined for $d > 3$; the degenerate case $d = 3$ is treated in Section~\ref{sec:d3}. These are confluent hypergeometric functions
\begin{equation}
    \Delta(t^{\mu\nu}t_{\mu\nu}; s)  = \frac{\Gamma\big( \frac{d}{2} \big) \Gamma\big(s+\frac{d-3}{2} \big) }{\Gamma\big(s+\frac{d}{2})\Gamma\big(\frac{d-3}{2} \big)} \, {}_0F_{1} \bigg( \begin{array}{c}
        {} \\
        {s+d/2}\\
    \end{array} \, \Bigg| \, z\bigg) \, , \quad z = \frac{t^{\mu\nu}t_{\mu\nu} }{8} \, ,
\end{equation}
which converge absolutely for $z \in \mathds{C}$ when $s+d/2>0$. They are not elements of $\off$, but their $\ast$-products with polynomials are well defined and associative, provided at most one factor is a series of this class\cite{vasiliev2004higher}.

The function
\begin{equation}
    \label{qsprjctr}
    \Delta(t^{\mu\nu}t_{\mu\nu}; 0) = \sum_{m=0}^{+\infty} \frac{\Gamma\big(\frac{d}{2}\big)}{\Gamma\big(m+\frac{d}{2}\big)} \cdot \frac{1}{8^m} \cdot \frac{\big(t^{\mu\nu}t_{\mu\nu} \big)^m}{m!}
\end{equation}
is the quasiprojector \eqref{2.31} of \cite{vasiliev2004higher}, with the defining property
\begin{equation}
    \label{qsprjctr:central}
    \begin{split}
    t&_{\alpha\beta} \ast \Delta(t^{\mu\nu}t_{\mu\nu}; 0) = \Delta(t^{\mu\nu}t_{\mu\nu}; 0) \ast t_{\alpha\beta} = 0 \, \\\implies &\Delta(t^{\mu\nu}t_{\mu\nu}; 0) \ast \id = \id \ast \Delta(t^{\mu\nu}t_{\mu\nu}; 0) = 0 \, ,
    \end{split}
\end{equation}
the prefix \textbf{quasi} referring to $\Delta(t^{\mu\nu}t_{\mu\nu}; 0) \ast \Delta(t^{\mu\nu}t_{\mu\nu}; 0)$ being ill-defined. By \eqref{sp2-inv:ast}, $t^{\alpha\beta}\ast t_{\alpha\beta}$ is central in $\off$, hence so is any function of $t^{\mu\nu}t_{\mu\nu}$.

The $\ast$-product of an $AdS$-traceless function with the quasiprojector takes a simple form:
\begin{equation}
    \label{qsprjctr:nice}
    \boxed{
    \begin{split}
        &\forall \tilde{f}_s \in \ker \Gamma_{\alpha\beta}: \hat{N} \tilde{f}_s = 2s \cdot \tilde{f}_s \hookrightarrow \\
         \hookrightarrow \Tilde{f}_s \ast \Delta(t^{\mu\nu}&t_{\mu\nu}; 0) =\, \Delta(t^{\mu\nu}t_{\mu\nu}; 0) \ast \Tilde{f}_s =    \Delta(t^{\mu\nu}t_{\mu\nu}; s) \cdot \Tilde{f}_s \, .
    \end{split}
    }
\end{equation}
Both are derived in Appendix~\ref{app:C} and together give a direct proof of \eqref{4.8}. Appendix~\ref{app:C} derives $t_{\alpha\beta}\ast\Delta(t^{\mu\nu}t_{\mu\nu};0)$ and $\Delta(t^{\mu\nu}t_{\mu\nu};0)\ast\tilde{f}_s$; the other orderings follow: $\Delta(t^{\mu\nu}t_{\mu\nu};0)\ast t_{\alpha\beta}$ from \eqref{t-star} on singlets, and $\tilde{f}_s\ast\Delta(t^{\mu\nu}t_{\mu\nu};0)$ from the centrality of the powers of $t^{\mu\nu}t_{\mu\nu}$.
We have not found \eqref{qsprjctr:nice} in \cite{vasiliev2004higher,joung2014notes}, where the quasiprojector is used through $f \ast \Delta(t^{\mu\nu}t_{\mu\nu}; 0) = \Delta(t^{\mu\nu}t_{\mu\nu}; 0) \ast f$ and the trace $\big(\Delta(t^{\mu\nu}t_{\mu\nu}; 0) \ast f\big)\big|_{Y=0}$.

\begin{proof}
    Let $g \in \ker \Gamma_{\alpha\beta} \, \cap \, \id$. Since $\Big[ \Gamma_{\alpha\beta} \, , \hat{N} \Big] = 2 \cdot \Gamma_{\alpha\beta}$ and $\off$ is spanned by polynomials, one can choose $g_0, \, g_1, \, \dots \,  , g_s$ with $s<+\infty$ such that
    \begin{equation}
        \forall j \in [ 0 \, , \, s]: \hat{N} g_j = \, 2j \cdot g_j \, , \, \,   g_j \in  \ker \Gamma_{\alpha\beta}:\, g = \sum_{j=0}^{s} g_j \, \, .
    \end{equation}
Note that $g \in \id$ does not imply $g_j \in \id$, since $\id$ is not $\hat{N}$-graded; this is not needed below.

Because $g \in \id$ and the $g_j$ are homogeneous and traceless, \eqref{qsprjctr:central} and \eqref{qsprjctr:nice} give
\begin{equation}
    \Delta(t^{\mu\nu}t_{\mu\nu}; 0) \cdot g_0 + \Delta(t^{\mu\nu}t_{\mu\nu}; 1) \cdot g_1 + \dots + \Delta(t^{\mu\nu}t_{\mu\nu}; s) \cdot g_s = 0 \, .
\end{equation}
Let $j_0$ be the smallest index with $g_{j_0} \neq 0$, if any. Each term $\Delta(t^{\mu\nu}t_{\mu\nu}; j) \cdot g_j$ is a series whose lowest degree is $2j$, with coefficient
\begin{equation}
    \Delta(t^{\mu\nu}t_{\mu\nu}; j)\Big|_{z=0} = \frac{\Gamma\big( \frac{d}{2} \big) \Gamma\big(j+\frac{d-3}{2} \big) }{\Gamma\big(j+\frac{d}{2}\big)\Gamma\big(\frac{d-3}{2} \big)} \neq 0 \quad \text{for } d>3 \, .
\end{equation}
At degree $2j_0$ only the term with $j=j_0$ contributes, so $g_{j_0}=0$, a contradiction. Hence $g=0$.
\end{proof}
\begin{proposition}
\label{prop:K}
Let $d>3$ and $K := \big\{ f \in \off : \Delta(t^{\mu\nu}t_{\mu\nu}; 0) \ast f = 0 \big\}$. Then $K = \id$.
\end{proposition}
\begin{proof}
By \eqref{qsprjctr:central}, $\id \subset K$. Conversely, let $f \in K$ and let $\tilde{f} \in \ker \Gamma_{\alpha\beta}$ be as in \eqref{4.6}, so that $\tilde{f} - f \in \id \subset K$ and hence $\tilde{f} \in K$. The argument of the preceding proof, applied to $\tilde{f}$ in place of $g$, gives $\tilde{f} = 0$, so $f = -(\tilde{f} - f) \in \id$.
\end{proof}

\subsection{Constructing the splitting}
\label{sec:constructing}
We now determine
\begin{equation}
    \big(t^{\alpha\beta} \ast \mathcal{A}_{\alpha\beta}(\bullet)\big): \off \rightarrow \id
\end{equation}
such that
\begin{equation}
    \forall f \in \off: \, \Gamma_{\alpha\beta}\Big( f+ t^{\mu\nu} \ast \mathcal{A}_{\mu\nu}(f) \Big) = 0 \, .
\end{equation}
Denote by $R(f) := f+ t^{\mu\nu} \ast \mathcal{A}_{\mu\nu}(f)$ the representative of $f$. Its form is fixed by representation theory.

\begin{lemma}
\label{lem:universality}
Let $d>3$. For every $f \in \off$ homogeneous of degree $2s$,
\begin{equation}
    \label{4.19}
    R(f) = f + t^{\mu\nu} \ast \mathcal{A}_{\mu\nu}(f) = \sum_{m=0}^{+\infty} r_m(\hat{N}) \cdot \mathcal{P} \Big( \big( \Gamma_{\alpha\beta} \Gamma^{\alpha\beta} \big)^m f  \Big) \, ,
\end{equation}
with coefficients $r_m$ depending only on the degree, i.e. universal for all $f \in \off$.
\end{lemma}
\begin{proof}
Both $\ker \Gamma_{\alpha\beta}$ and $\id$ are invariant subspaces: the latter because $\big[M_{AB}, t^{\alpha\beta}\big]_\ast = 0$ and \eqref{sp2-covar} is preserved by the $\so$-action. Uniqueness \eqref{4.8} then implies that $R$ commutes with $\mathfrak{sp}(2)$ and $\so$, and by Section~\ref{sec:existence} it does not raise the degree. Hence $R$ decomposes into covariant maps from the degree-$2s$ component into $\ker \Gamma_{\alpha\beta}$ at degree $2s-2j$. The former is a single $\mathfrak{gl}(d+1)$-irrep and the latter is the $\so$-irrep described by the two-row diagram of length $s-j$; by \eqref{GLtoSO} the latter occurs in the former only for even $j=2m$, and then exactly once. By Schur's lemma each such map is unique up to normalisation, and $\mathcal{P} \Big( \big( \Gamma_{\alpha\beta}\Gamma^{\alpha\beta} \big)^m \Big)$ realises it, being non-zero by \eqref{beta_coef}. This gives \eqref{4.19} with $r_m$ a function of the degree alone.
\end{proof}
Functions of $\hat{N}$ act on everything standing to their right, so the argument of $r_m$ is the degree of $\mathcal{P} \big( ( \Gamma\Gamma )^m f \big)$; the series terminates because $\off$ consists of polynomials. By universality the $r_m$ may be extracted from a single family of test functions, so $t^{\mu\nu} \ast \mathcal{A}_{\mu\nu}(f)$ is determined without knowing $\mathcal{A}_{\mu\nu}(f)$ itself.

Two technical properties are needed. For $\tilde{f}_s \in \ker \Gamma_{\alpha\beta}$ with $\hat{N}\tilde{f}_s = 2s \cdot \tilde{f}_s$, formula \eqref{qsprjctr:nice} generalises to
\begin{equation}
        \label{qsprjctr:wow}
        \forall m \in \mathds{Z}_{\geqslant 0}: \Delta\big(t^{\mu\nu}t_{\mu\nu};0 \big) \ast \Big[ (t^{\alpha\beta}t_{\alpha\beta})^m \,  \tilde{f}_s \Big] = h_m(s) \cdot \Big[ \Delta\big(t^{\mu\nu}t_{\mu\nu};s \big) \cdot \tilde{f}_s \Big] \, ,
\end{equation}
where
\begin{equation}
        h_m(s) = 8^m \cdot \frac{\Gamma\Big(m+\frac{3}{2}\Big)}{\Gamma\Big(\frac{3}{2}\Big)} \cdot \frac{\Gamma\Big(\frac{d+2s+1}{2}+m\Big)}{\Gamma\Big(\frac{d+2s+1}{2}\Big)} \, ,
\end{equation}
and
\begin{equation}
        \label{beta-action}
        \forall m, k \in \mathds{Z}_{\geqslant 0}: \Big( \Gamma_{\alpha\beta}\Gamma^{\alpha\beta}\Big)^m \Big[ \big(t^{\mu\nu}t_{\mu\nu}\big)^k \cdot \tilde{f}_s \Big] = \beta_{(m, k)}(s) \cdot \big(t^{\mu\nu}t_{\mu\nu}\big)^{k-m} \cdot \tilde{f}_s \, ,
\end{equation}
where
    \begin{equation}
        \label{beta_coef}
        \beta_{(m, k)}(s) = 2^{6m} \cdot \frac{\Gamma(k+1)\Gamma(k+\frac{3}{2})}{\Gamma(k-m+1)\Gamma(k-m+\frac{3}{2})} \cdot \frac{\Gamma(\frac{d+2s}{2}+k) \Gamma(\frac{d+2s+1}{2}+k)}{\Gamma(\frac{d+2s}{2}+k-m)\Gamma(\frac{d+2s+1}{2}+k-m)} \, .
\end{equation}
Both are proven in Appendix~\ref{app:C}. Multiplying \eqref{4.19} by the quasiprojector gives
    \begin{equation}
        \forall f: \, \Delta\big(t^{\mu\nu}t_{\mu\nu}; 0 \big) \ast f = \Delta\big(t^{\mu\nu}t_{\mu\nu}; 0 \big) \ast \bigg[ \sum_{m=0}^{+\infty} r_m(\hat{N}) \cdot \mathcal{P} \Big( \big( \Gamma_{\alpha\beta} \Gamma^{\alpha\beta} \big)^m f  \Big) \bigg] \, ,
    \end{equation}
and choosing
    \begin{equation}
        f = \big(t_{\alpha\beta}t^{\alpha\beta} \big)^m \cdot \tilde{f}_s
    \end{equation}
\eqref{qsprjctr:wow}, \eqref{qsprjctr:nice}, \eqref{beta-action} and \eqref{property-1} give
\begin{equation}
    h_m(s) \cdot \Big[\Delta\big(t^{\mu\nu}t_{\mu\nu}; s \big) \cdot \tilde{f}_s \Big] = r_m(2s) \cdot \beta_{(m,m)}(s) \cdot \Big[\Delta\big(t^{\mu\nu}t_{\mu\nu}; s \big) \cdot \tilde{f}_s \Big] \, .
\end{equation}
By Lemma \ref{lem:universality} this fixes $r_m$ for all arguments:
\begin{equation}
    \label{r-coef}
    \boxed{
     r_m(2s) = \frac{h_m(s)}{\beta_{(m,m)}(s)} = \frac{1}{8^m \cdot m!} \cdot \frac{\Gamma\Big( \frac{d+2s}{2} \Big)}{\Gamma\Big( \frac{d+2s}{2} + m\Big)} \, .}
\end{equation}
Hence the map $t^{\alpha\beta} \ast \mathcal{A}_{\alpha\beta}(\bullet)$ is
\begin{equation}
    \label{tA-map}
    t^{\alpha\beta} \ast \mathcal{A}_{\alpha\beta}(f) = -f + \sum_{m=0}^{+\infty} \frac{\Gamma\big( \frac{d+\hat{N}}{2} \big)}{\Gamma\big( \frac{d+\hat{N}}{2}+m \big)} \, \frac{1}{m! \cdot 8^m} \, \mathcal{P}\Big( \big(\Gamma_{\alpha\beta}\Gamma^{\alpha\beta} \big)^m f \Big) \, .
\end{equation}

\subsection{The $\onsh$ algebra}
By \eqref{4.6} and \eqref{4.8}, $\ker \Gamma_{\alpha\beta} \subset \off$ is a space of representatives with respect to $\id$, which turns the definition $\onsh = \off / \id$ into the following theorem.
\begin{theorem}
For $d>3$, the algebra $\onsh$ is isomorphic to the algebra of polynomials $f(Y)$ satisfying
    \begin{equation}
        \Big(Y_\alpha \cdot \frac{\partial}{\partial Y^\beta}  + (\alpha \leftrightarrow \beta) \Big) f(Y) = 0 \, , \quad \Gamma_{\alpha\beta} \, f(Y) = 0 \, ,
    \end{equation}
    with the product $\circ$ defined as
    \begin{equation}
        \label{circ-prod}
        \boxed{
        f \circ g := \mathcal{P} \Bigg(\sum_{m=0}^{+\infty} \frac{\Gamma\big( \frac{d+\hat{N}}{2} \big)}{\Gamma\big( \frac{d+\hat{N}}{2}+m \big)} \,  \frac{\big(\Gamma_{\alpha\beta}\Gamma^{\alpha\beta} / 8 \big)^m }{m!}\big( f \ast g \big) \Bigg) \, .}
    \end{equation}
\end{theorem}
The product \eqref{circ-prod} is compatible with the structures of Section~\ref{sec:hs-algebras}.
\begin{proposition}
The antiautomorphism $\dagger$ of \eqref{antiauto} and the automorphism $\pi$ of \eqref{twist} descend to $\big( \ker \Gamma_{\alpha\beta}, \circ \big)$:
\begin{equation}
    \big( f \circ g \big)^\dagger = g^\dagger \circ f^\dagger \, , \quad \pi\big( f \circ g \big) = \pi(f) \circ \pi(g) \, .
\end{equation}
\end{proposition}
\begin{proof}
On the Weyl symbols of real $Y^\alpha_A$ the map $\dagger$ acts as complex conjugation combined with the reversal of the order of $\ast$-multipliers. The operators $t^{\mu\nu}$, $\Gamma_{\mu\nu}$ and $\hat{N}$ are real, and so are the coefficients $c_{(k, m)}(\hat{N})$ in \eqref{c} and the coefficients of the series in \eqref{circ-prod}; hence $\mathcal{P}$ and the whole operator in \eqref{circ-prod} commute with conjugation, and the statement follows from $\big(f\ast g\big)^\dagger = g^\dagger \ast f^\dagger$. Similarly, $t^{\mu\nu} = \eta^{ab}\mathbf{y}^\mu_a \mathbf{y}^\nu_b - y^\mu y^\nu$ and $\Gamma_{\mu\nu}$ are even under $\pi$, so $\pi$ commutes with $\mathcal{P}$ and preserves $\ker \Gamma_{\alpha\beta}$.
\end{proof}

\subsection{The $\so$ generators}
\label{sec:so-generators}
As a check, we evaluate \eqref{circ-prod} on the $\so$-generators $M_{AB}$ of \eqref{gauge_mdl:cnsq}, which are $AdS$-traceless: $\Gamma_{\mu\nu}M_{AB} = 0$. The Moyal product \eqref{Moyal} gives
\begin{equation}
    \label{MM-star}
    \begin{split}
    M_{AB} \ast M_{CD} = M_{AB} M_{CD} &- \frac{1}{2}\Big( \eta_{AC}M_{BD} - \eta_{AD}M_{BC} - \eta_{BC}M_{AD} + \eta_{BD}M_{AC}\Big) \\
    &+ \frac{1}{2}\Big( \eta_{AC}\eta_{BD} - \eta_{AD}\eta_{BC}\Big) \, ,
    \end{split}
\end{equation}
and the induced product is
\begin{equation}
    \label{MM-circ}
    \boxed{
    \begin{split}
    M_{AB} \circ M_{CD} = \, &M_{AB}M_{CD} - \frac{1}{d-1}\Big( \eta_{AC}\eta^{PQ}M_{PB}M_{QD} - \eta_{AD}\eta^{PQ}M_{PB}M_{QC} - \\
    &- \eta_{BC}\eta^{PQ}M_{PA}M_{QD} + \eta_{BD}\eta^{PQ}M_{PA}M_{QC} \Big) - \\
    &-\frac{t^{\sigma\lambda}t_{\sigma\lambda}}{4d(d-1)}\Big( \eta_{AC}\eta_{BD} - \eta_{AD}\eta_{BC}\Big) - \\
    &-\frac{1}{2}\Big( \eta_{AC}M_{BD} - \eta_{AD}M_{BC} - \eta_{BC}M_{AD} + \eta_{BD}M_{AC}\Big) + \\
    &+\frac{d-3}{2d}\Big( \eta_{AC}\eta_{BD} - \eta_{AD}\eta_{BC}\Big) \, .
    \end{split}}
\end{equation}
The derivation is given in Appendix \ref{app:MM}.
Thus \eqref{MM-circ} is computed directly from \eqref{circ-prod}, without structure constants, and so are its consequences below.
Three consequences follow. First, the $\circ$-commutator coincides with the $\ast$-commutator, so the isometry algebra is undeformed,
\begin{equation}
    \big[ M_{AB} \, , \, M_{CD}\big]_\circ = \big[ M_{AB} \, , \, M_{CD}\big]_\ast = \eta_{BC}M_{AD} - \eta_{AC}M_{BD} - \eta_{BD}M_{AC} + \eta_{AD}M_{BC} \, .
\end{equation}
Second, upon contracting one pair of indices all $Y$-dependent terms cancel,
\begin{equation}
    \label{MM-contr}
    M_{AB} \circ M_{C}{}^{B} = -\frac{d-1}{2}\, M_{AC} + \frac{d-3}{2}\, \eta_{AC} \, ,
\end{equation}
so the traceless symmetric part vanishes, which is one of the relations generating the Joseph ideal.
Third, the quadratic Casimir is a number,
\begin{equation}
    \label{MM-casimir}
    M_{AB} \circ M^{AB} = \frac{(d+1)(d-3)}{2} \, .
\end{equation}
In terms of the hermitian generators $T_{AB} = -i\,M_{AB}$ this reads $\frac{1}{2}T_{AB}T^{AB} = -\frac{(d-1)^2-4}{4}$, the value characteristic of the minimal representation of $\so$. Both cancellations leading to \eqref{MM-contr} and \eqref{MM-casimir} are fixed by the coefficients $c_{(k,m)}(4)$ of \eqref{c}, so they also test the horizontal projector.

It is instructive to contrast \eqref{MM-casimir} with its off-shell counterpart. Contracting \eqref{MM-star} with $\eta^{AC}\eta^{BD}$, all $M$-linear terms drop out, each of them containing either $\eta^{AB}M_{AB}$ or a trace of an antisymmetric tensor, and
\begin{equation}
    M_{AB} \ast M^{AB} = M_{AB}M^{AB} + \frac{d(d+1)}{2} = -\frac{1}{4}\, t^{\sigma\lambda}t_{\sigma\lambda} + \frac{d(d+1)}{2} \, ,
\end{equation}
where the last equality follows from the definition of $M_{AB}$. Unlike \eqref{MM-casimir}, this is not a number: in $\off$ the quadratic Casimir differs from a constant precisely by the invariant that generates $\id$. The factorisation removes that invariant in two steps. The $m=0$ term of \eqref{circ-prod} is $\mathcal{P}\big( \bullet \big)$ itself, which annihilates $t^{\sigma\lambda}t_{\sigma\lambda}$ by \eqref{property-1} and leaves $d(d+1)/2$, while the $m=1$ term contributes, by \eqref{beta-action} and \eqref{r-coef},
\begin{equation}
    r_1(0) \cdot \Big( -\frac{1}{4}\Big) \cdot \beta_{(1,1)}(0) = \frac{1}{4d} \cdot \Big( -\frac{1}{4}\Big) \cdot 24\, d(d+1) = -\frac{3(d+1)}{2} \, ;
\end{equation}
the terms with $m \geqslant 2$ vanish on degree grounds, and the sum reproduces \eqref{MM-casimir}. Here the indices are contracted before \eqref{circ-prod} is applied, whereas \eqref{MM-circ} is contracted afterwards, and the two routes rest on different data: the former only on \eqref{property-1} and \eqref{beta-action}, the latter on the explicit coefficients $c_{(k,m)}(4)$. Their agreement therefore tests Appendix \ref{app:MM} and the horizontal projector against each other.\footnote{It also fixes the ordering convention of Section~\ref{sec:constructing}. Had $r_m$ been evaluated at the degree before the traces are taken, the $m=1$ coefficient would be $r_1(4) = 1/\big(4(d+4)\big)$ and the Casimir would come out as $d(d+1)^2/\big(2(d+4)\big)$, which is not the value characteristic of the minimal representation.}

\subsection{Lorentz-traceless condition}
The explicit splitting \eqref{tA-map} allows a direct proof that \eqref{Lorentz-traceless} also defines a space of representatives with respect to $\id$. Define
\begin{equation}
    \gamma_{\alpha\beta} := \eta_{ab} \frac{\partial^2}{\partial \mathbf{y}^\alpha_a \partial \mathbf{y}^\beta_b} \, ,
\end{equation}
so that \eqref{Lorentz-traceless} takes the form
\begin{equation}
    \gamma_{\alpha\beta} \, f = 0 \, .
\end{equation}
Two statements are needed: $\off = \ker \gamma_{\alpha\beta} + \id$ and $\ker \gamma_{\alpha\beta} \cap \id = \{0\}$.

\paragraph{Existence.} The Lorentz algebra commutes with $\mathfrak{sp}(2)$, so the harmonic decomposition with respect to $\eta_{ab}$ is $\mathfrak{sp}(2)$-covariant and gives, for any $f \in \off$,
\begin{equation}
    \label{Lorentz-split}
    f = f' + \big( \eta^{ab}\, y^\mu_a y^\nu_b \big) X_{\mu\nu} \, , \quad \gamma_{\alpha\beta} f' = 0 \, ,
\end{equation}
with $\mathfrak{sp}(2)$-covariant $X_{\mu\nu}$ satisfying \eqref{sp2-covar}. With $\eta^{ab} y^\mu_a y^\nu_b = t^{\mu\nu} - y^\mu y^\nu$ and \eqref{tA},
\begin{equation}
    f = f' + \underbrace{t^{\mu\nu} \ast X_{\mu\nu}}_{\in \id} + \Gamma^{\mu\nu}X_{\mu\nu} - y^\mu y^\nu X_{\mu\nu} \, .
\end{equation}
Iterating on the two remainders terminates: $\Gamma^{\mu\nu}X_{\mu\nu}$ has degree lower by two, while $y^\mu y^\nu X_{\mu\nu}$ has the same degree but two $y^\alpha_a$'s fewer, and on $\mathfrak{sp}(2)$-singlets the number of $y^\alpha$'s never exceeds that of $y^\alpha_a$'s, as shown below. Ordering the pairs (degree, number of $y^\alpha_a$'s) lexicographically, the iteration stops after finitely many steps, which proves $\off = \ker \gamma_{\alpha\beta} + \id$.

\paragraph{Trivial intersection.} For $g \in \ker \gamma_{\alpha\beta} \, \cap \, \id$,
\begin{equation}
    \label{Lorentz-eq1}
    g + t^{\mu\nu} \ast \mathcal{A}_{\mu\nu}(g) = 0 \,   \Leftrightarrow\,  \sum_{m=0}^{+\infty} \frac{\Gamma\big( \frac{d+\hat{N}}{2} \big)}{\Gamma\big( \frac{d+\hat{N}}{2}+m \big)} \, \frac{1}{m! \cdot 8^m} \, \mathcal{P}\Big( \big(\Gamma_{\alpha\beta}\Gamma^{\alpha\beta} \big)^m g \Big) = 0 \, .
\end{equation}
As $g \in \ker \gamma_{\alpha\beta}$,\footnote{ $A_\alpha A^\alpha \equiv 0$, since $\mathfrak{sp}(2)$ indices are raised and lowered by the symplectic metric $\epsilon_{\alpha\beta}$ (recall \eqref{epsilon}).}
\begin{equation}
    \begin{split}
        \label{Lorentz-eq2}
        \Gamma_{\alpha\beta} \, g = -&\frac{\partial^2}{\partial y^\alpha \partial y^\beta}\, g + \gamma_{\alpha\beta} \, g = -\frac{\partial^2}{\partial y^\alpha \partial y^\beta}\, g \, \\
        & \implies \Gamma^{\alpha\beta}\Gamma_{\alpha\beta} \, g = 0 \, ,
    \end{split}
\end{equation}
so \eqref{Lorentz-eq1} reduces to $\mathcal{P}(\,g\,) = 0$. Since $\big[  \gamma_{\alpha\beta} \, ,\,  \hat{N} \big] = 2 \cdot \gamma_{\alpha\beta}$, decompose
    \begin{equation}
        \begin{split}
            \forall j \in [ 0 \, , \, s]: \hat{N} g_j = & \, 2j \cdot g_j \, , \, \,   g_j \in  \ker \gamma_{\alpha\beta} \, , \\
            & g = \sum_{j=0}^{s} g_j \, \, ,
        \end{split}
    \end{equation}
so that $\mathcal{P}( g_j ) = 0$ for every $j$, $\mathcal{P}\big( \bullet \big)$ being grade preserving. By \eqref{Lorentz-eq2} and \eqref{HorProj},
\begin{equation}
    \label{Lorentz-eq3}
    \sum_{k=0}^{j} c_{(k, 0)}(2j) \, \cdot \big(t^{\mu\nu}\big)^k \Big( - \frac{\partial^2}{\partial y^\mu \partial y^\nu} \Big)^k \, g_j = 0 \, .
\end{equation}
Being an $\mathfrak{sp}(2)$-singlet, $g_j$ has all Greek indices of $y^\alpha$ and $y^\alpha_a$ contracted with one another; since $y_\alpha y^\alpha \equiv 0$, it contains no terms with more $y^\alpha$'s than $y^\alpha_a$'s, so
\begin{equation}
    \exists g^{(n)}_j \in \ker \gamma_{\alpha\beta} \, \, , \, y^\alpha \frac{\partial}{\partial y^\alpha} \, g^{(n)}_j = n \cdot g^{(n)}_j : \quad g_j =  \sum_{n=0}^{j} g^{(n)}_j \, ,
\end{equation}
with $\hat{N}g^{(n)}_j = 2j \cdot g^{(n)}_j$. As $\big(t^{\mu\nu}\big)^k \big( - \partial^2 / \partial y^\mu \partial y^\nu \big)^k$ does not raise the power of $y^\alpha$, the highest power in \eqref{Lorentz-eq3} gives
\begin{equation}
    \Bigg(\sum_{k=0}^{j} c_{(k,0)}(2j) \cdot \frac{j!}{(j-2k)!}\Bigg) \cdot g^{(j)}_j = 0 \, , \quad
    \sum_{k=0}^{j} c_{(k,0)}(2j) \cdot \frac{j!}{(j-2k)!} = \frac{2^{1-j} \cdot \Gamma\big(\frac{d-1}{2} \big) \cdot \Gamma\big(d-3+j \big)}{\Gamma\big(d-2 \big)\cdot\Gamma\big(\frac{d-3}{2}+j \big)} \, ,
\end{equation}
non-zero for $d>3$ and all $j \in \mathds{Z}_{\geqslant 0}$, so $g^{(j)}_{j} = 0$. At the next power,
\begin{equation}
    \begin{split}
    &\Bigg(\sum_{k=0}^{j} c_{(k,0)}(2j) \cdot \frac{(j-1)!}{(j-1-2k)!}\Bigg) \cdot g^{(j-1)}_j = 0 \, , \\
    &\sum_{k=0}^{j} c_{(k,0)}(2j) \cdot \frac{(j-1)!}{(j-1-2k)!}  = \frac{2^{2-j} \cdot \Gamma\big( \frac{d+1}{2} \big) \cdot \Gamma\big( d-2+j \big)}{\Gamma\big(d\big) \cdot \Gamma\big(\frac{d-3}{2} +j \big)} \, ,
    \end{split}
\end{equation}
non-zero for $d>3$ provided $j\geqslant 1$, so $g^{(j-1)}_{j} = 0$. The general step rests on
\begin{equation}
    \Bigg(\sum_{k=0}^{j} c_{(k,0)}(2j) \cdot \frac{(j-m)!}{(j-m-2k)!}\Bigg)  = \frac{2^{1-j+m} \cdot \Gamma\big( \frac{d-1}{2}+m \big) \cdot \Gamma\big( d-3+j+m \big)}{\Gamma\big(d-2+2m\big) \cdot \Gamma\big(\frac{d-3}{2} +j \big)} \, ,
\end{equation}
and gives $g^{(n)}_j = 0$ for all $n$. Hence
\begin{equation}
    g = 0 \Leftrightarrow \begin{cases}
        &\gamma_{\alpha\beta} \, g = 0 \, , \\
        &\mathcal{P}( g ) = 0 \, ,
    \end{cases}
\end{equation}
i.e. $\ker \gamma_{\alpha\beta} \, \cap \, \id = \{0\}$, which together with \eqref{Lorentz-split} establishes the claim.
Only the second step uses $d>3$, through \eqref{tA-map}; the harmonic decomposition in the first step holds for every $d$, by the positivity argument of Lemma~\ref{lem:pos}(a) applied to $\eta_{ab}$ and $\gamma_{\alpha\beta}$.

This implies another way to realise $\onsh$: use $f \in \ker \gamma_{\alpha\beta}$ and a different factorisation map $i_L(\bullet): \off \rightarrow \id $, which we denote as 
\begin{equation}
    i_L(f) = t^{\alpha\beta} \ast L_{\alpha\beta}(f) \, ,
\end{equation}
and the symbol for the product is 
\begin{equation}
    f \circ_L g = f \ast g + t^{\alpha\beta} \ast L_{\alpha\beta}(f\ast g) \, .
\end{equation}
The map $L_{\alpha\beta}(\bullet)$ is not constructed here: we only use its existence, which follows from $\off = \ker \gamma_{\alpha\beta} \oplus \id$, established above, and from the definition of $\id$.

\subsection{The case $d=3$}
\label{sec:d3}
At $d=3$ the family \eqref{quasiprojector} degenerates,
\begin{equation}
    \label{qsprjctr:d3}
    \forall s\in \mathds{N} \, : \Delta\big( t^{\mu\nu}t_{\mu\nu} ;s\big) \Big|_{d=3} \equiv 0 \, ,
\end{equation}
so the proof of Section~\ref{sec:uniqueness} does not apply.
By Lemma~\ref{lem:pos}(a), the decomposition into $\ker\Gamma$ and the image of multiplication by $t^{\mu\nu}$, and hence the projector, exist also at $d = 3$. What requires $d>3$ are the explicit coefficients \eqref{c}, which have poles at $d=3$ for $k+2m=s$, and the quasiprojector argument of Section~\ref{sec:uniqueness}.
The splitting \eqref{4.6} is still valid\cite{vasiliev2004higher}, and \eqref{qsprjctr:nice} gives
\begin{equation}
    \forall f \in \mathfrak{hsgl}_p(2,2|2): \Delta\big(t^{\mu\nu}t_{\mu\nu} ;0\big) \ast f = \Delta\big(t^{\mu\nu}t_{\mu\nu} ;0\big) \cdot \mathfrak{C}(f) \, ,
\end{equation}
where $\mathfrak{C}(f)$ is the degree-zero part of a traceless representative of $f$. It is a character of $\mathfrak{hsgl}_p(2,2|2)$: by associativity,
\begin{equation}
    \mathfrak{C}(f\ast g) \cdot \Delta = \Delta \ast \big( f \ast g\big) = \big(\Delta \ast f \big) \ast g = \mathfrak{C}(f) \cdot \mathfrak{C}(g) \cdot \Delta \, , \quad \mathfrak{C}(1) = 1 \, .
\end{equation}
Hence the equivalence relation defined by the quasiprojector, $f \sim g \Leftrightarrow \Delta \ast (f-g) = 0$, differs from the one defined by the ideal. Its kernel is the set $K$ of Proposition~\ref{prop:K}, a two-sided ideal, $\Delta$ being central; for $d>3$, $K = \id$ by Proposition~\ref{prop:K}, while at $d=3$ one has $K = \ker \mathfrak{C}$, of codimension one.
Every element $t^{\alpha\beta} \ast A_{\alpha\beta}$ whose leading part $t^{\alpha\beta} \cdot A_{\alpha\beta}$ is non-zero has degree at least four, since $\deg A_{\alpha\beta} \geqslant 2$ by \eqref{sp2-covar:cnsq}. We expect the bound for all non-zero elements of $\id$, which would give $\id \subsetneq K$ at $d=3$; we do not prove it. The quasiprojector retains only the scalar $\mathfrak{C}(f)$. The uniqueness of the splitting at $d=3$ remains open.

The same degeneration occurs for the $\mathfrak{so}(N)$ higher-spin structure constants of Joung and Mkrtchyan\cite{joung2014notes}: their bilinear and trilinear forms are hypergeometric expressions with parameters $\frac{N-4}{2}$, $\frac{N-1}{2}$ and argument scaled by $1/8$, which for $N = d+1$ are the parameters $\frac{d-3}{2}$ and $\frac{d}{2}$ of \eqref{quasiprojector}. At $N=4$ the Pochhammer symbols $\big(\frac{N-4}{2}\big)_k$ vanish for $k \geqslant 1$ and both forms collapse to a constant, as $\Delta(t^{\mu\nu}t_{\mu\nu};s)$ does for $s>0$. The correspondence is at the level of parameters; we do not construct the isomorphism explicitly.

The origin of the degeneracy is $\mathfrak{so}(2,2) \cong \mathfrak{sl}(2) \oplus \mathfrak{sl}(2)$: the $AdS$-traceless condition alone does not specify an irreducible $\mathfrak{so}(2,2)$-module associated with a two-row Young diagram, since the traceless module decomposes further into self-dual and anti-self-dual sectors,
\begin{equation}
\frac{1}{2}\varepsilon^{ABCD}
T_{C}\,^{A(s-2) \,,  \,B(s-2)}\,_{D}
=
\pm
T^{A(s-1)\, , \,B(s-1)} .
\end{equation}
This splitting is invisible in the ambient oscillator realisation: any $\varepsilon$-invariant built from the generating variables is of the form $\varepsilon_{ABCD}\, Y^{\alpha A}Y^{\beta B}Y^{\mu C}Y^{\nu D}$, which is totally antisymmetric in four $\mathfrak{sp}(2)$ indices and hence vanishes, these taking only two values; equivalently, $\varepsilon_{ABCD}$ requires four independent vectors, while only $Y^1_A$ and $Y^2_A$ are available. All $\mathfrak{so}(2,2)\oplus\mathfrak{sp}(2)$-invariants are therefore functions of $t^{\mu\nu}t_{\mu\nu}$ and central, \eqref{qsprjctr} is the unique solution of $t_{\alpha\beta} \ast \Delta = 0$ up to normalisation also at $d=3$, and the Pfaffian invariant vanishes,
\begin{equation}
    \label{Pfaff}
    \varepsilon^{ABCD} M_{AB} \ast M_{CD} = 0 \, ,
\end{equation}
i.e. the two $\mathfrak{sl}(2)$ Casimirs coincide on the module realised by the oscillators.
At $d = 3$ the construction contains no free parameter, so unlike the one-parameter family $\mathfrak{hs}[\lambda]$ of three-dimensional higher-spin algebras\cite{joung2014notes} it does not give a family; we do not identify the corresponding value of $\lambda$ here. Recovering the whole family would require
enlarging the set of generating variables, e.g. by deformed oscillators\cite{Vasiliev:1989qh}, which lies beyond the scope of the present work. An explicit analogue of the Moyal star product for the deformed oscillator algebra $\mathsf{Aq}(2,\nu)$, involving additional integration parameters in the multiplication kernel, was constructed in \cite{Korybut:2020vmm}. How the family is reproduced in the ambient-oscillator language, and how the factorisation used here is related to the standard three-dimensional construction at the undeformed point, remains an open question.
\newpage

\section{Towards the $A_\infty$ level: a toy model}
\label{sec:ainfty}

In this section, we recall the established description of HS interactions in terms of minimal $A_\infty$-algebras (see \cite{sharapov2019formal} for a review) and examine the factorisation of Section~\ref{sec:factor-algebra} in this framework. The on-shell equations \cite{Vasiliev:2003ev} are consistent (see \cite{Vasiliev:2025hfh} for their BRST formulation), so an on-shell $A_\infty$-algebra exists, and solving them would produce its maps (the vertices, in HS terms). Since all-order off-shell vertices were obtained in \cite{Didenko:2024zpd}, one may try to obtain these maps by factorising the off-shell minimal $A_\infty$-algebra.

We first recall how the unfolded equations of interacting HS fields are encoded by multilinear maps, called higher products, subject to quadratic relations that generalise associativity. We then study a toy model whose ideal is generated by a single central element $\tau$, realise the factorisation at the level of $m_2$, and formulate conditions under which it extends to $m_3$. These factorisation results do not cover the $\mathfrak{sp}(2)$ ideal; the requirements for extending the construction to that ideal are discussed in Section~\ref{sec:lift}.

\subsection{HS interactions as a minimal $A_\infty$-algebra}

\begin{definition}[$\big($Minimal $A_\infty$-algebra$\big)$]
    Let $W = \bigoplus_{n \in \mathds{Z}} W_n$ be a $\mathds{Z}$-graded vector space, and let $|w| = n$ denote the degree of $w \in W_n$. A minimal $A_\infty$-algebra on $W$ is a collection of multilinear maps, i.e. linear maps from the tensor products of $W$,
    \begin{equation}
        m_n : W^{\otimes n} \rightarrow W \, , \qquad n \geqslant 2 \, ,
    \end{equation}
    called \textit{higher products}, which are of degree $-1$,
    \begin{equation}
        |m_n(w_1, \dots, w_n)| = |w_1| + \dots + |w_n| - 1 \, ,
    \end{equation}
    and subject to the \textit{Stasheff relations}\cite{stasheff1963homotopy,keller2001introduction}
    \begin{equation}
        \label{stasheff}
        \begin{split}
           \sum_{n+p=N} \, \sum_{k=0}^{n-1} (-1)^{\sigma_k} \, m_n\big(w_1, \dots, w_k,\, &m_p(w_{k+1}, \dots, w_{k+p}), w_{k+p+1}, \dots, w_{N-1}\big) = 0 \, , \\
           &\sigma_k = \sum_{j=1}^{k} |w_j| \, , \quad \sigma_0 := 0 \, , 
        \end{split}
    \end{equation}
    for every $N \geqslant 4$ and all $w_1, \dots, w_{N-1} \in W$.
\end{definition}

The Stasheff relations generalise associativity: for $N = 4$ only $m_2$ enters, and \eqref{stasheff} states that $m_2$ is associative with the signs dictated by the grading.
The qualifier \textit{minimal} refers to the absence of a differential $m_1: W \rightarrow W$ of degree $-1$;
over a field, every $A_\infty$-algebra is quasi-isomorphic to a minimal one\cite{kadeishvili1980homology}. The HS setting requires only minimal $A_\infty$-algebras on a space with two non-trivial components.

Let $\mathfrak{A} = (A, \ast)$ be a unital associative algebra equipped with an involution $\pi: A \rightarrow A$, $\pi^2 = 1$; that $\pi$ is an automorphism of $\mathfrak{A}$ will follow below. We take
\begin{equation}
    W_1 = A_1 \cong A \, , \qquad W_0 = A_0 \cong A \, , \qquad W_n = 0 \quad \text{for } n \neq 0, 1 \, ,
\end{equation}
so that $W = A_1 \oplus A_0$. Elements of $A_1$ are the values of the one-form $\omega$, and elements of $A_0$ those of the zero-form $C$. Throughout this section $v, v_i \in A_1$ and $u, u_i \in A_0$. The algebra $\mathfrak{A}$ acts on $A_0$ from the left by $v \ast u$ and from the right by $u \ast \pi(v)$, in accordance with the twisted-adjoint action in \eqref{linear}.\footnote{Only this bimodule structure of $A_0$ enters the construction, and $A_0$ need not be isomorphic to $A$. For a general $\mathfrak{A}$-bimodule $A_0$ with the right action written as $u \ast \pi(v)$, the relations \eqref{stasheff} imply only $u \ast \big(\pi(v_1 \ast v_2) - \pi(v_1) \ast \pi(v_2)\big) = 0$ for all $u \in A_0$; $\pi$ is then an automorphism of $\mathfrak{A}$ provided the right action on $A_0$ is faithful.}

Since $W$ has non-trivial components in degrees $0$ and $1$ only, a higher product vanishes identically whenever none of its arguments, or more than two of them, belong to $A_1$: its value would have degree $-1$ or at least $2$, respectively. The remaining components have either one or two arguments in $A_1$. We denote them by $U$ and $V$, respectively. The components of $m_2$ are fixed by the product $\ast$ and the involution $\pi$, while those of the higher products with $n \geqslant 3$ are the unknowns constrained by \eqref{stasheff}:
\begin{subequations}
    \label{m-to-VU}
    \begin{align}
        & m_2(v_1, v_2) = - v_1 \ast v_2 \, , \qquad m_2(v, u) = - v \ast u \, , \qquad m_2(u, v) = u \ast \pi(v) \, , \\
        & V_n^{(k_1, k_2)}\big(\dots | u_{k_1} | \underline{v_1} | \dots | u_{k_2} | \underline{v_2} | \dots \big) = m_{n+2}\big(\dots , u_{k_1}, v_1, \dots, u_{k_2}, v_2, \dots \big) \, , \\
        & U_n^{(k)}\big(\dots | u_k | \underline{v} | \dots \big) = m_{n+1}\big(\dots , u_k, v, \dots \big) \, ,
    \end{align}
\end{subequations}
with $0 \leqslant k_1 \leqslant k_2 \leqslant n$ and $0 \leqslant k \leqslant n$. Here the dots stand for the arguments $u_1, \dots, u_n$ taken in consecutive order, so that $V_n^{(k_1,k_2)}$ has $k_1$ arguments $u_i$ before $v_1$, $k_2 - k_1$ between $v_1$ and $v_2$, and $n - k_2$ after $v_2$, and similarly for $U_n^{(k)}$; empty ranges are omitted. In the $m_n$ the underline, which marks the arguments taking values in $A_1$, is redundant, since the degree of an argument fixes its role; we keep it in $V$ and $U$ for readability.

For $N = 4$ the relations \eqref{stasheff} involve $m_2$ only. With three arguments in $A_1$ they reproduce the associativity of $\ast$.
For the arguments $(u, v_1, v_2)$ they give $u \ast \pi(v_1) \ast \pi(v_2) = u \ast \pi(v_1 \ast v_2)$, so that, setting $u = 1$, $\pi$ is an automorphism of $\mathfrak{A}$.
The remaining combinations, $(v_1, v_2, u)$ and $(v_1, u, v_2)$, express the compatibility of the left and right actions on $A_0$ with the product $\ast$.

Given such an $A_\infty$-algebra, consider the system
\begin{subequations}
    \label{non-linear-HS}
    \begin{align}
        &d_x\omega + \omega \ast \omega = \mathcal{V}(\omega, \omega, C) + \mathcal{V}(\omega, \omega, C, C) + \dots \, , \\
        &d_xC + \omega \ast C - C \ast \pi(\omega) = \Upsilon(\omega, C, C) + \dots \, ,
    \end{align}
\end{subequations}
where $\omega = \omega_{\underline{m}}(x)\, dx^{\underline{m}} = \omega^{\mathscr{a}}\, T_{\mathscr{a}}$ is a one-form and $C = C(x) = C^{\mathscr{a}}\, T_{\mathscr{a}}$ a zero-form, $\{T_{\mathscr{a}}\}$ being a basis of $A$. The vertices $\mathcal{V}$ and $\Upsilon$ collect all orderings of the arguments,
\begin{equation}
    \begin{split}
        \mathcal{V}(\omega, \omega, C^n) &= \sum_{0\leqslant k_1 \leqslant k_2 \leqslant n} \mathcal{V}\big(\,\underbrace{ \dots | \, C  }_{k_1}\,| \, \underline{\omega} \, | \, \underbrace{C \, | \dots |\, C}_{k_2-k_1} \, | \,\underline{\omega} \, | \underbrace{\,C \, | \dots}_{n-k_2}\,\big) \, , \\
        \Upsilon(\omega, C^{n}) &= \sum_{0\leqslant k \leqslant n} \Upsilon\big(\,\underbrace{ \dots | \, C  }_{k}\,| \, \underline{\omega} \, | \underbrace{\,C \, | \dots}_{n-k}\,\big) \, ,
    \end{split}
\end{equation}
and are defined in terms of $V$ and $U$ as
\begin{equation}
    \begin{split}
        & \mathcal{V}\big(\,\underbrace{ \dots | \, C  }_{k_1}\,| \, \underline{\omega} \, | \, \underbrace{C \, | \dots |\, C}_{k_2-k_1} \, | \,\underline{\omega} \, | \underbrace{\,C \, | \dots}_{n-k_2}\,\big) = \\
        &\qquad = V\big(T_{\mathscr{c}_1} | \dots | \, T_{\mathscr{c}_{k_1}} | \, \underline{T_{\mathscr{a}_1}} \, | \,  T_{\mathscr{c}_{k_1+1}} | \dots | \, T_{\mathscr{c}_{k_2}} \, | \, \underline{T_{\mathscr{a}_2}} \, | \dots \, | \, T_{\mathscr{c}_n} \big)\, \omega^{\mathscr{a}_1}\omega^{\mathscr{a}_2} C^{\mathscr{c}_1} \dots C^{\mathscr{c}_n} \, , \\
        & \Upsilon\big(\,\underbrace{ \dots | \, C  }_{k}\,| \, \underline{\omega} \, | \underbrace{\,C \, | \dots}_{n-k}\,\big) = U\big(T_{\mathscr{c}_1} | \dots | \, T_{\mathscr{c}_{k}} | \, \underline{T_{\mathscr{a}}} \, | \,  T_{\mathscr{c}_{k+1}} | \dots \, | \, T_{\mathscr{c}_n} \big)\, \omega^{\mathscr{a}} C^{\mathscr{c}_1} \dots C^{\mathscr{c}_n} \, .
    \end{split}
\end{equation}
In these formulas the maps are evaluated on basis elements, and the form coefficients are then written in the fixed order $\omega^{\mathscr{a}_1}\omega^{\mathscr{a}_2} C^{\mathscr{c}_1} \dots C^{\mathscr{c}_n}$, irrespective of the positions of the corresponding arguments. We extend this rule to the master field $\Phi = \omega + C$: $m_n(\Phi, \dots, \Phi)$ denotes the sum over all distributions of $\omega$ and $C$ among the arguments, each term being evaluated in the way just described. With the identification \eqref{m-to-VU} the system \eqref{non-linear-HS} takes the compact form
\begin{equation}
    \label{master-eq}
    d_x \Phi = \sum_{n = 2}^{+\infty} m_n(\Phi, \dots, \Phi) \, ,
\end{equation}
whose one-form and zero-form components reproduce the two equations of \eqref{non-linear-HS}. A direct computation shows that, with this convention, the integrability condition $d_x^2 = 0$ of \eqref{master-eq} is equivalent to the Stasheff relations \eqref{stasheff}.

The lowest relations involving the higher products beyond $m_2$ correspond to $N = 5$:
\begin{equation}
    \label{lowest-stasheff}
    \begin{split}
        & m_2\big(m_3(w_1, w_2, w_3), w_4\big) + (-1)^{|w_1|}\, m_2\big(w_1, m_3(w_2, w_3, w_4)\big) + m_3\big(m_2(w_1, w_2), w_3, w_4\big) + \\
        & + (-1)^{|w_1|}\, m_3\big(w_1, m_2(w_2, w_3), w_4\big) + (-1)^{|w_1| + |w_2|}\, m_3\big(w_1, w_2, m_2(w_3, w_4)\big) = 0 \, .
    \end{split}
\end{equation}
Non-trivial components arise when three or two of the arguments belong to $A_1$. The former constrain the maps $V$ entering $\mathcal{V}(\omega, \omega, C)$ and ensure the consistency of the one-form equation at first order in $C$; the latter involve both the maps $V$ and the maps $U$ entering $\Upsilon(\omega, C, C)$ and ensure the consistency of the zero-form equation at second order in $C$. For instance, for $(w_1, w_2, w_3, w_4) = (u_1, v_1, u_2, v_2)$ the relation \eqref{lowest-stasheff} combined with \eqref{m-to-VU} gives\begin{equation}
    \begin{split}
        & U_2^{(1)}(u_1 | \underline{v_1} | u_2) \ast \pi(v_2) - u_1 \ast V^{(0,1)}_1(\underline{v_1}|u_2|\underline{v_2}) + U_2^{(2)}\big(u_1 \ast \pi(v_1)\big| u_2 \big| \underline{v_2}\big) + \\
        & - U_2^{(2)}\big(u_1 \big| v_1 \ast u_2 \big| \underline{v_2}\big) - U^{(1)}_{2}\big(u_1 \big| \underline{v_1} \big| U^{(1)}_1(u_2 \ast \pi(v_2) \big) = 0 \, .
    \end{split}
\end{equation}
The remaining components are obtained in the same way.

The Stasheff relations do not determine the higher products. In HS theory the guiding principle for fixing them is spin-locality\cite{Didenko:2019xzz,Didenko:2020bxd,Didenko:2024zpd,Kirakosiants:2025gpd,Valerev:2026wos,Sharapov:2022faa}. In practice, the vertices $\mathcal{V}(\omega, \omega, C^n)$ and $\Upsilon(\omega, C^{n+1})$ are derived by means of the generating-system approach\cite{Vasiliev:1990en,Vasiliev:2003ev,Didenko:2022qga,Didenko:2023vna,Sharapov:2022faa}, which also makes it possible to control properties of the maps $V$ and $U$ that are not constrained by \eqref{stasheff}.

\subsection{Factorisation at the level of the product}

We now consider the simplest setting in which the factorisation of Section~\ref{sec:factor-algebra} can be lifted to the higher products: the ideal is generated by a single element $\tau \in A$. When inserted into the higher products, $\tau$ is placed in $A_1$. We require
\begin{equation}
    \label{tau-central}
    m_2(\tau, w) + (-1)^{|w|}\, m_2(w, \tau) = 0 \qquad \forall w \in W \, .
\end{equation}
For $w = v \in A_1$ this reads $v \ast \tau - \tau \ast v = 0$, i.e. $\tau$ is central in $\mathfrak{A}$. For $w = u \in A_0$ it reads $u \ast \pi(\tau) - \tau \ast u = 0$; given centrality, setting $u = 1$ yields $\pi(\tau) = \tau$. Conversely, every central $\pi$-invariant element satisfies \eqref{tau-central}.

The element $\tau$ generates the two-sided ideal
\begin{equation}
    \mathcal{I}_\tau = \big\{ \tau \ast a \, : \, a \in A \big\} \, ,
\end{equation}
which is closed under $\pi$. In $W$ it is contained both in $A_1$ and in $A_0$, and in the latter it is invariant under the left and right actions of $\mathfrak{A}$. Let $\mathcal{R} \subset A$ be a space of representatives \eqref{AA:decomp} with respect to $\mathcal{I}_\tau$, closed under $\pi$, and set $\mathcal{R}(W) := \mathcal{R} \oplus \mathcal{R} \subset W$ and $\mathcal{I}_\tau(W) := \mathcal{I}_\tau \oplus \mathcal{I}_\tau \subset W$, the two copies lying in $A_1$ and $A_0$, respectively. Denote by $\mathrm{Ann}(\tau) = \{ a \in A \, : \, \tau \ast a = 0 \}$ the annihilator of $\tau$.

\begin{lemma}
    \label{lem:factor-m2}
    Under the above assumptions:
    \begin{enumerate}
        \item The map $i(\bullet)$ of \eqref{AA:decomp} can be written as $i(f) = \tau \ast \mathcal{A}(f) = \mathcal{A}(f) \ast \tau$ with a linear map $\mathcal{A}: A \rightarrow A$, which is unique up to linear maps into $\mathrm{Ann}(\tau)$. Extending $\mathcal{A}$ and $i$ to $W$ degree-wise, one has
        \begin{equation}
            \label{i-map}
            i(w) = \tau \ast \mathcal{A}(w) = - m_2\big(\tau, \mathcal{A}(w)\big) \, , \qquad |\mathcal{A}(w)| = |w| \, .
        \end{equation}
        \item Every such $\mathcal{A}$ satisfies $\tau \ast \pi\big(\mathcal{A}(f)\big) = \tau \ast \mathcal{A}\big(\pi(f)\big)$ for all $f \in A$.
        \item The map
        \begin{equation}
            \label{p-proj}
            p(w) := w + i(w) = w - m_2\big(\tau, \mathcal{A}(w)\big)
        \end{equation}
        is the projection of $W$ onto $\mathcal{R}(W)$ along $\mathcal{I}_\tau(W)$, and the bilinear map
        \begin{equation}
            \tilde{m}_2(w_1, w_2) := p\big(m_2(w_1, w_2)\big) \, , \qquad w_1, w_2 \in \mathcal{R}(W) \, ,
        \end{equation}
        realises the factor-algebra: on $\mathcal{R} \subset A_1$ it gives $\tilde{m}_2(v_1, v_2) = - v_1 \circ v_2$ with the product $\circ$ of \eqref{AA:circ-prod}, while on $\mathcal{R} \subset A_0$ it gives the induced left and right actions of $(\mathcal{R}, \circ)$.
        \item The relations \eqref{stasheff} with $N = 4$ hold for $\tilde{m}_2$ on $\mathcal{R}(W)$; in particular, $\pi$ is an automorphism of $(\mathcal{R}, \circ)$.
    \end{enumerate}
\end{lemma}
\begin{proof}
    Since $i(A) \subset \mathcal{I}_\tau = \tau \ast A$, for every $f$ there exists $a_f$ with $i(f) = \tau \ast a_f$; choosing $a_f$ on a basis defines a linear $\mathcal{A}$, two such choices differing by a map into $\mathrm{Ann}(\tau)$, and $\tau \ast \mathcal{A}(f) = \mathcal{A}(f) \ast \tau$ by centrality. Since $\pi(\mathcal{R}) = \mathcal{R}$ and $\pi(\mathcal{I}_\tau) = \mathcal{I}_\tau$, uniqueness of the decomposition \eqref{AA:decomp} gives $\pi\big(i(\pi(f))\big) = i(f)$ for all $f \in A$, which together with $\pi(\tau) = \tau$ is item 2. Items 3 and 4 are the statement \eqref{AA:circ-prod} applied to $\mathcal{I}_\tau$, together with $\pi(\mathcal{R}) = \mathcal{R}$.
\end{proof}

The ideal $\id$ of Section~\ref{sec:hs-algebras} is not of this form: it is generated by the $\mathfrak{sp}(2)$-triplet $t^{\alpha\beta}$, which does not belong to $\off$. Nevertheless, $t^{\alpha\beta}$ is invariant under the automorphism \eqref{twist}, and the traceless conditions \eqref{AdS-traceless}, \eqref{Twist-traceless} and \eqref{Lorentz-traceless} define linear spaces closed under $\pi$. The present setting should therefore be regarded as a simplified model, in which the interplay between the ideal and the higher products can be analysed explicitly.

\subsection{Lifting the factorisation to the lowest higher product}
\label{sec:lift}

The factorisation of the product does not by itself induce higher products on $\mathcal{R}(W)$. In general, $m_3$ does not map arguments containing elements of $\mathcal{I}_\tau(W)$ into $\mathcal{I}_\tau(W)$, and the map $(w_1, w_2, w_3) \mapsto p\big(m_3(w_1, w_2, w_3)\big)$ restricted to $\mathcal{R}(W)$ need not satisfy \eqref{lowest-stasheff} together with $\tilde{m}_2$. Additional conditions are required, both on $m_3$ and on the map $\mathcal{A}$.

\begin{theorem}
    \label{thm:lift}
    Let $\mathfrak{A}$, $\pi$, $\tau$, $\mathcal{R}$ and $\mathcal{A}$ be as in Lemma \ref{lem:factor-m2}, and let $m_2$, $m_3$ satisfy \eqref{stasheff} with $N = 4$ and $N = 5$. Assume, in addition, that
    \begin{enumerate}
        \item[(i)] for all $w_1, w_2 \in W$,
        \begin{equation}
            \label{commutation-generalisation}
            0 = m_3(w_1, \tau, w_2) + (-1)^{|w_1|}\, m_3(\tau, w_1, w_2) + (-1)^{|w_2|}\, m_3(w_1, w_2, \tau) \, ;
        \end{equation}
        \item[(ii)] for all $w \in \mathcal{R}(W)$ and $g \in W$,
        \begin{equation}
            \label{do-not-divide-0}
            \mathcal{A}\Big( m_2\big(w, i(g)\big) \Big) = - m_2\big(w, \mathcal{A}(g)\big) \, , \qquad \mathcal{A}\Big( m_2\big(i(g), w\big) \Big) = - m_2\big(\mathcal{A}(g), w\big) \, .
        \end{equation}
    \end{enumerate}
    Define, for $w_1, w_2, w_3 \in W$,
    \begin{equation}
        \label{m3-prime}
        \begin{split}
            m'_3(w_1, w_2, w_3) = \; & m_3(w_1, w_2, w_3) - m_3\Big(\tau, \mathcal{A}\big(m_2(w_1, w_2)\big), w_3\Big) \\
            & - (-1)^{|w_2| + |w_3|}\, m_3\Big(w_1, \mathcal{A}\big(m_2(w_2, w_3)\big), \tau\Big) \, ,
        \end{split}
    \end{equation}
    and
    \begin{equation}
        \tilde{m}_3(w_1, w_2, w_3) := p\big(m'_3(w_1, w_2, w_3)\big) \, , \qquad w_1, w_2, w_3 \in \mathcal{R}(W) \, .
    \end{equation}
    Then $\tilde{m}_3$ maps $\mathcal{R}(W)^{\otimes 3}$ into $\mathcal{R}(W)$, and $(\tilde{m}_2, \tilde{m}_3)$ satisfy \eqref{lowest-stasheff} for all $w_1, \dots, w_4 \in \mathcal{R}(W)$.
\end{theorem}
The proof is given in Appendix \ref{app:lift-proof}.

Through \eqref{m-to-VU}, $\tilde{m}_3$ defines maps $\tilde{V}$ and $\tilde{U}$ on $\mathcal{R}(W)$. The components of \eqref{lowest-stasheff} with three arguments in $\mathcal{R} \subset A_1$ are the lowest Stasheff relations for $\tilde{V}$ with $\circ$ in place of $\ast$, while those with two arguments in $\mathcal{R} \subset A_1$ involve both $\tilde{V}$ and $\tilde{U}$.
In the toy model, the factorisation therefore extends to the maps $\tilde{V}$ and $\tilde{U}$.

Both conditions of Theorem \ref{thm:lift} admit a simple interpretation. Condition \eqref{commutation-generalisation} is the counterpart of \eqref{tau-central} for $m_3$: in both cases $\tau$ is inserted into each argument slot in turn, and the resulting terms add up to zero. Both relations are the cases $n = 1$ and $n = 2$ of
\begin{equation}
    \label{tau-insertion}
    \sum_{k=0}^{n} (-1)^{\sigma_k}\, m_{n+1}(w_1, \dots, w_k, \tau, w_{k+1}, \dots, w_n) = 0 \qquad \forall w_1, \dots, w_n \in W \, ,
\end{equation}
with $\sigma_k$ as in \eqref{stasheff}: for $n = 1$ this is \eqref{tau-central}, and for $n = 2$ it is \eqref{commutation-generalisation} multiplied by $(-1)^{|w_1|}$. In this sense \eqref{commutation-generalisation} generalises the centrality of $\tau$ to the higher product. Its components with $(|w_1|, |w_2|) = (1, 0)$ and $(0, 1)$ constrain the maps $V$, those with $|w_1| = |w_2| = 0$ constrain the maps $U$, and those with $|w_1| = |w_2| = 1$ vanish identically. Note also that for $w_1 = \tau$ the first two terms of \eqref{commutation-generalisation} cancel, so that
\begin{equation}
    \label{tau-w-tau}
    m_3(\tau, w, \tau) = 0 \qquad \forall w \in W \, .
\end{equation}

To understand condition \eqref{do-not-divide-0}, recall that $i(\bullet)$ acts as $-\mathds{1}$ on the ideal \eqref{AA:lemma}. Since $m_2\big(w, i(g)\big)$ and $m_2\big(i(g), w\big)$ belong to $\mathcal{I}_\tau(W)$, the relations \eqref{do-not-divide-0} hold automatically after multiplication by $\tau$:
\begin{equation}
    \tau \ast \mathcal{A}\Big( m_2\big(w, i(g)\big) \Big) = - m_2\big(w, i(g)\big) = - \tau \ast m_2\big(w, \mathcal{A}(g)\big) \, ,
\end{equation}
and similarly for the second relation. Condition \eqref{do-not-divide-0} is thus a stricter version of this identity. It is automatically satisfied if $\tau$ is not a zero divisor, $\mathrm{Ann}(\tau) = \{0\}$; otherwise it restricts the choice of $\mathcal{A}$ among the maps allowed by Lemma \ref{lem:factor-m2}. If $1 \in \mathcal{R}$, setting $w$ equal to the unit in $\mathcal{R} \subset A_1$ and in $\mathcal{R} \subset A_0$ and using item 2 of Lemma \ref{lem:factor-m2}, one finds that \eqref{do-not-divide-0} implies $\pi\big(\mathcal{A}(f)\big) = \mathcal{A}\big(\pi(f)\big)$ for all $f \in A$. Conversely, the components of \eqref{do-not-divide-0} with $w$ in $\mathcal{R} \subset A_0$ follow from those with $w$ in $\mathcal{R} \subset A_1$ once $\pi\big(\mathcal{A}(f)\big) = \mathcal{A}\big(\pi(f)\big)$ for all $f \in A$. Hence, for $1 \in \mathcal{R}$, condition \eqref{do-not-divide-0} is equivalent to
\begin{equation}
    \begin{gathered}
        \mathcal{A}\big(v \ast \tau \ast \mathcal{A}(f)\big) = - v \ast \mathcal{A}(f) \, , \qquad \mathcal{A}\big(\tau \ast \mathcal{A}(f) \ast v\big) = - \mathcal{A}(f) \ast v \qquad \forall v \in \mathcal{R} \, , \, \forall f \in A \, , \\
        \pi\big(\mathcal{A}(f)\big) = \mathcal{A}\big(\pi(f)\big) \qquad \forall f \in A \, .
    \end{gathered}
\end{equation}

The expression \eqref{m3-prime} for $m'_3$ is not unique: by means of \eqref{commutation-generalisation}, each correction term can be rewritten as a combination of terms with $\tau$ in the two other argument slots, and all expressions obtained in this way coincide. We stress that Theorem \ref{thm:lift} concerns the lowest higher product only; the extension of the construction to $m_n$ with $n > 3$, where we expect \eqref{tau-insertion} with $n \geqslant 3$ to play the role of \eqref{commutation-generalisation}, is left for future work.

\paragraph{Towards the ideal $\id$.}
We conclude with remarks on the case $\mathfrak{A} = \off$, with $\pi$ the automorphism \eqref{twist} and the ideal $\id$, which, as noted above, is not of the form considered in Theorem \ref{thm:lift}. Since $m_2$ is expressed through $\ast$ and $\pi$ by \eqref{m-to-VU}, it extends to arguments outside $\off$, in particular to the generators $t^{\alpha\beta}$ placed in $A_1$. By \eqref{sp2-inv:ast} and the $\pi$-invariance of $t^{\alpha\beta}$, each component $t^{\alpha\beta}$ then satisfies \eqref{tau-central} for all $w \in W$. For the arguments $A_{\alpha\beta}$ satisfying \eqref{sp2-covar}, which carry $\mathfrak{sp}(2)$ indices, only the contraction of \eqref{tau-central} over $\alpha\beta$ holds, by \eqref{tA}. At the level of the product, the factorisation is carried out in Sections~\ref{sec:hs-algebras}--\ref{sec:factoring}, the map $i(\bullet) = t^{\alpha\beta} \ast \mathcal{A}_{\alpha\beta}(\bullet)$ being given by \eqref{tA-map}. A concrete choice of $\mathcal{A}_{\alpha\beta}(\bullet)$ realising this map will be given in \cite{TatarenkoPREP}. The additional requirements for lifting the factorisation to $m_3$ remain to be established. Along the lines of Theorem \ref{thm:lift}, these include at least the following. First, $m_3$ must be defined on arguments that are not $\mathfrak{sp}(2)$-singlets, namely on $t^{\alpha\beta}$ and on $\mathcal{A}_{\alpha\beta}(f)$, whereas the vertices in \eqref{non-linear-HS} are defined on $\off$ only; we expect that such an extension can be obtained within the generating-system approach\cite{Didenko:2023vna,Didenko:2024zpd}.
Second, condition \eqref{commutation-generalisation} must be replaced by a suitable counterpart, in the same way as \eqref{tau-central} above. Moreover, the proof in Appendix \ref{app:lift-proof} uses \eqref{tau-w-tau}, whose derivation relies on the same element $\tau$ occupying two slots of $m_3$. For $\id$ these slots are occupied by $t^{\alpha\beta}$ and $t^{\gamma\delta}$, contracted with different arguments, and the counterpart of \eqref{tau-w-tau} does not follow in the same way. Third, condition \eqref{do-not-divide-0} involves the map $\mathcal{A}$ itself rather than $i(\bullet)$.
Zero divisors play no role here: $\off$ is a subalgebra of the Weyl algebra of polynomials in $Y^\alpha_A$ with the product \eqref{Moyal}, which has no zero divisors, so for a non-zero central $\tau \in \off$ one has $\mathrm{Ann}(\tau) = \{0\}$ and condition \eqref{do-not-divide-0} holds automatically. The non-uniqueness relevant for $\id$ has a different origin: the counterpart $\mathcal{A}_{\alpha\beta}(\bullet)$ of $\mathcal{A}$ is fixed by \eqref{tA-map} only up to the freedom \eqref{divisor-of-0}, which is always present by the identity following \eqref{divisor-of-0}; this freedom is a relation among the three generators $t^{\alpha\beta}$ (a syzygy), not a zero divisor. Hence the counterpart of \eqref{do-not-divide-0} is not automatic and restricts the choice of $\mathcal{A}_{\alpha\beta}(\bullet)$. Finally, as for $\mathcal{I}_\tau$, we do not expect $m_3$ to map arguments containing elements of $\id$ into $\id$, so that correction terms analogous to those in \eqref{m3-prime} are required. These questions are left for future work.

\newpage
\section{Conclusion}
We have constructed, for $d>3$, an explicit realisation of the polynomial on-shell HS algebra $\onsh=\off/\id$ on the space of AdS-traceless $\mathfrak{sp}(2)$-singlet polynomials, with the induced product \eqref{circ-prod}. The construction rests on the horizontal projector $\mathcal{P}(\bullet)$ onto $\ker\Gamma_{\alpha\beta}$, which we obtained in closed form as a series of trace-producing operators $\mathcal{T}_{(k,m)}$ with the coefficients \eqref{c}. Its structure follows the restriction \eqref{GLtoSO} of $\mathfrak{gl}(d+1)$ to $\mathfrak{so}(d+1)$: the trace-producing operators are linearly independent up to the multiplicity of the branching, and the remaining ones obey the linear relations \eqref{reductionA} with the coefficients \eqref{B-coef}. These properties are proved in Appendix~\ref{app:B2}, with the independence statements for $d>3$. 

Using the quasiprojector and its action \eqref{qsprjctr:nice} on traceless polynomials, we proved that the splitting \eqref{4.6} of a polynomial into a traceless representative and an element of $\id$ is unique. The map $t^{\alpha\beta}\ast\mathcal{A}_{\alpha\beta}(\bullet)$ and the induced product were then derived explicitly, the former being fixed by representation theory up to the coefficients \eqref{r-coef}. A concrete choice of $\mathcal{A}_{\alpha\beta}(\bullet)$ realising the factorisation map $t^{\alpha\beta}\ast\mathcal{A}_{\alpha\beta}(\bullet)$ of \eqref{tA-map} will be given in \cite{TatarenkoPREP}.

On the $\so$-generators the induced product leaves the isometry algebra undeformed and reproduces one of the relations generating the Joseph ideal and the quadratic Casimir of the minimal representation. The Lorentz-traceless polynomials form a second space of representatives with respect to $\id$. At $d=3$ the quasiprojector family degenerates, so our proof of uniqueness does not apply, and the oscillator realisation reproduces a single member of the one-parameter family of three-dimensional HS algebras. 

In Section~\ref{sec:ainfty} the unfolded interaction equations were written in terms of a minimal $A_\infty$-algebra on a space with two non-trivial components, the Stasheff relations \eqref{stasheff} expressing their integrability. For an ideal generated by a single central $\pi$-invariant element $\tau$, which is simpler than the $\mathfrak{sp}(2)$-ideal, the factorisation was realised at the level of the product and lifted to the lowest higher product $m_3$, under the conditions \eqref{commutation-generalisation} and \eqref{do-not-divide-0}. This is a toy model; it does not cover the ideal $\id$.

For $\id$, the lift to $m_3$ requires higher products on arguments that are not $\mathfrak{sp}(2)$-singlets, a contracted counterpart of \eqref{commutation-generalisation}, a replacement for \eqref{tau-w-tau}, and a choice of $\mathcal{A}_{\alpha\beta}(\bullet)$ compatible with the counterpart of \eqref{do-not-divide-0}. The extension of the toy model to the higher products $m_n$ with $n>3$, where we expect the identities \eqref{tau-insertion} to replace \eqref{commutation-generalisation}, remains to be worked out.

At $d=3$ the uniqueness of the splitting remains open in our approach, as does the relation of the factorisation to the standard three-dimensional construction. Finally, the present analysis is restricted to polynomials, whereas nontrivial interaction equations require a functional class closed under the induced product; this question will be addressed in \cite{TatarenkoPREP}.

Since $\mathcal{P}(\bullet)$ is an explicit differential operator in the generating variables, the factor-algebra can be handled directly in terms of representatives, which may be convenient within the sources formalism of the spin-locality analysis.
The on-shell equations \cite{Vasiliev:2003ev} are consistent (see \cite{Vasiliev:2025hfh} for their BRST formulation) and therefore define an on-shell $A_\infty$-algebra.
Comparing its maps with those obtained by factorising the off-shell minimal $A_\infty$-algebra along the lines of Section~\ref{sec:lift} is a natural next step.

\newpage
\section*{Methods}
Parts of this work were carried out with the assistance of a large language model (LLM) used between September 2026 and October 2026 for checking derivations and editing the manuscript. All results were verified by the author, who takes full responsibility for the content. The LLM was not used to generate scientific results or conclusions.

\section*{Acknowledgments}
The author is grateful to Vyacheslav Didenko, Yura Tatarenko and Danila Valeriev for careful reading of the manuscript and stimulating discussions. The author also thanks Karapet Mkrtchyan for correspondence concerning the case $d=3$. Special thanks go to Gleb Klepov.

\newpage
    \appendix

\renewcommand{\thesection}{\Alph{section}}
\renewcommand{\thesubsection}{\thesection.\arabic{subsection}}

\section{Proof of Theorem \ref{thm:lift}}
\label{app:lift-proof}

In this appendix we prove Theorem \ref{thm:lift}. We first collect the relations used in the proof and derive a few elementary consequences of them. We then show that it suffices to verify the Stasheff relation for $(\tilde{m}_2, \tilde{m}_3)$ modulo the ideal, expand it in terms of $m_2$, $m_3$ and $\mathcal{A}$, and split the resulting terms into seven groups, each of which vanishes separately modulo the ideal. To make the bookkeeping transparent, every term of the expansion carries a label $\mathrm{I}_k, \dots, \mathrm{VII}_k$ indicating its group, and the groups are distinguished by colour.

\subsection*{Notation and relations used}

Throughout this appendix $w, w_1, \dots, w_4 \in W$, and $|\tau| = 1$. The Stasheff relations \eqref{stasheff} with $N = 4$ read
\begin{equation}
    \label{A:S4}
    m_2\big(m_2(x, y), z\big) + (-1)^{|x|}\, m_2\big(x, m_2(y, z)\big) = 0 \qquad \forall x, y, z \in W \, ,
\end{equation}
and those with $N = 5$ are given by \eqref{lowest-stasheff}; we shall apply the latter to various quadruples of arguments, not all of them in $\mathcal{R}(W)$. Condition \eqref{tau-central} will be used in the form
\begin{equation}
    \label{A:T}
    m_2(x, \tau) = - (-1)^{|x|}\, m_2(\tau, x) \qquad \forall x \in W \, .
\end{equation}

Recall that $\mathcal{I}_\tau(W) = \mathcal{I}_\tau \oplus \mathcal{I}_\tau \subset W$ consists of the two copies of the ideal lying in $A_1$ and $A_0$. Since $m_2(\tau, a) = - \tau \ast a$ both for $a \in A_1$ and for $a \in A_0$, by \eqref{m-to-VU},
\begin{equation}
    \label{A:ideal-W}
    \mathcal{I}_\tau(W) = \big\{ m_2(\tau, x) \, : \, x \in W \big\} \, .
\end{equation}
For $x, y \in W$ we write
\begin{equation}
    x \thicksim y \quad \Longleftrightarrow \quad x - y \in \mathcal{I}_\tau(W) \, .
\end{equation}
By Lemma \ref{lem:factor-m2}, the map $p$ of \eqref{p-proj} is the projection of $W$ onto $\mathcal{R}(W)$ along $\mathcal{I}_\tau(W)$; in particular, $p(x) \in \mathcal{R}(W)$, $p(x) \thicksim x$ for all $x \in W$, and $\mathcal{R}(W) \cap \mathcal{I}_\tau(W) = \{0\}$. Moreover, $p$ preserves the degree, since $|\mathcal{A}(x)| = |x|$ and $|\tau| = 1$. Finally, for $w_1, \dots, w_4$ we use the abbreviation
\begin{equation}
    \label{A:Aij}
    \begin{gathered}
        \mathcal{A}_{(ij)} := \mathcal{A}\big(m_2(w_i, w_j)\big) \, , \qquad |\mathcal{A}_{(ij)}| = |w_i| + |w_j| - 1 \, , \\
        p\big(m_2(w_i, w_j)\big) = m_2(w_i, w_j) - m_2\big(\tau, \mathcal{A}_{(ij)}\big) \, .
    \end{gathered}
\end{equation}

\subsection*{Preliminary identities}

\begin{lemma}
    \label{lem:A-prelim}
    For all $a, w, x \in W$:
    \begin{align}
        \label{A:P1}
        & m_2\big(m_2(\tau, a), w\big) = m_2\big(\tau, m_2(a, w)\big) \, , \\
        \label{A:P2}
        & m_2\big(w, m_2(\tau, a)\big) = m_2\big(\tau, m_2(w, a)\big) \, ;
    \end{align}
    $m_2(\tau, x) \thicksim 0$ and $m_2(x, \tau) \thicksim 0$; if $x \thicksim 0$, then $m_2(x, w) \thicksim 0$ and $m_2(w, x) \thicksim 0$. Furthermore, for all $w \in \mathcal{R}(W)$ and $g \in W$, condition \eqref{do-not-divide-0} is equivalent to
    \begin{equation}
        \label{A:D}
        \mathcal{A}\Big( m_2\big(w, m_2(\tau, \mathcal{A}(g))\big) \Big) = m_2\big(w, \mathcal{A}(g)\big) \, , \qquad \mathcal{A}\Big( m_2\big(m_2(\tau, \mathcal{A}(g)), w\big) \Big) = m_2\big(\mathcal{A}(g), w\big) \, .
    \end{equation}
\end{lemma}
\begin{proof}
    Relation \eqref{A:P1} is \eqref{A:S4} for $(x, y, z) = (\tau, a, w)$, since $(-1)^{|\tau|} = -1$. Relation \eqref{A:S4} for $(w, \tau, a)$ reads $m_2\big(m_2(w, \tau), a\big) + (-1)^{|w|}\, m_2\big(w, m_2(\tau, a)\big) = 0$; by \eqref{A:T} and \eqref{A:P1}, its first term equals $-(-1)^{|w|}\, m_2\big(m_2(\tau, w), a\big) = -(-1)^{|w|}\, m_2\big(\tau, m_2(w, a)\big)$, which gives \eqref{A:P2}. The relation $m_2(\tau, x) \thicksim 0$ is \eqref{A:ideal-W}, and $m_2(x, \tau) \thicksim 0$ then follows from \eqref{A:T}. If $x \thicksim 0$, then $x = m_2(\tau, a)$ for some $a \in W$ by \eqref{A:ideal-W}, and \eqref{A:P1}, \eqref{A:P2} show that $m_2(x, w)$ and $m_2(w, x)$ are of the same form. Finally, \eqref{A:D} is \eqref{do-not-divide-0} with $i(g) = - m_2\big(\tau, \mathcal{A}(g)\big)$ substituted from \eqref{i-map}.
\end{proof}

\subsection*{Reduction modulo the ideal}

By construction, $\tilde{m}_2$ and $\tilde{m}_3$ take values in the image of $p$, i.e. in $\mathcal{R}(W)$; this proves the first statement of the theorem. Let $w_1, \dots, w_4 \in \mathcal{R}(W)$. Then all terms of
\begin{equation}
    \label{A:Sigma}
    \begin{split}
        \Sigma := \; & \tilde{m}_2\big(\tilde{m}_3(w_1, w_2, w_3), w_4\big) + (-1)^{|w_1|}\, \tilde{m}_2\big(w_1, \tilde{m}_3(w_2, w_3, w_4)\big) + \tilde{m}_3\big(\tilde{m}_2(w_1, w_2), w_3, w_4\big) \\
        & + (-1)^{|w_1|}\, \tilde{m}_3\big(w_1, \tilde{m}_2(w_2, w_3), w_4\big) + (-1)^{|w_1| + |w_2|}\, \tilde{m}_3\big(w_1, w_2, \tilde{m}_2(w_3, w_4)\big)
    \end{split}
\end{equation}
are well defined, and $\Sigma \in \mathcal{R}(W)$. We have to show $\Sigma = 0$. Since $\mathcal{R}(W) \cap \mathcal{I}_\tau(W) = \{0\}$, it suffices to show $\Sigma \thicksim 0$.

The outer projections in \eqref{A:Sigma} can be dropped modulo the ideal: $p(x) \thicksim x$ for every $x \in W$, and by Lemma \ref{lem:A-prelim} also $m_2\big(p(x), w\big) \thicksim m_2(x, w)$ and $m_2\big(w, p(x)\big) \thicksim m_2(w, x)$. The inner projections in the last three terms cannot be dropped, since $m'_3$ need not map arguments in $\mathcal{I}_\tau(W)$ into $\mathcal{I}_\tau(W)$. Hence
\begin{equation}
    \label{A:Sigma-reduced}
    \begin{split}
        \Sigma \thicksim \; & \underbrace{m_2\big(m'_3(w_1, w_2, w_3), w_4\big)}_{\text{(a)}} + \underbrace{(-1)^{|w_1|}\, m_2\big(w_1, m'_3(w_2, w_3, w_4)\big)}_{\text{(b)}} + \underbrace{m'_3\Big(p\big(m_2(w_1, w_2)\big), w_3, w_4\Big)}_{\text{(c)}} \\
        & + \underbrace{(-1)^{|w_1|}\, m'_3\Big(w_1, p\big(m_2(w_2, w_3)\big), w_4\Big)}_{\text{(d)}} + \underbrace{(-1)^{|w_1| + |w_2|}\, m'_3\Big(w_1, w_2, p\big(m_2(w_3, w_4)\big)\Big)}_{\text{(e)}} \, .
    \end{split}
\end{equation}

\subsection*{Expansion}

In the terms (c), (d) and (e), the map $\mathcal{A}$ in \eqref{m3-prime} acts on products involving a projection. Using \eqref{A:Aij} and \eqref{A:D} with $g = m_2(w_i, w_j)$, we obtain
\begin{equation}
    \label{A:resolve}
    \begin{split}
        \mathcal{A}\Big( m_2\big(p(m_2(w_1, w_2)), w_3\big) \Big) &= \mathcal{A}\Big( m_2\big(m_2(w_1, w_2), w_3\big) \Big) - m_2\big(\mathcal{A}_{(12)}, w_3\big) \, , \\
        \mathcal{A}\Big( m_2\big(w_1, p(m_2(w_2, w_3))\big) \Big) &= \mathcal{A}\Big( m_2\big(w_1, m_2(w_2, w_3)\big) \Big) - m_2\big(w_1, \mathcal{A}_{(23)}\big) \, , \\
        \mathcal{A}\Big( m_2\big(p(m_2(w_2, w_3)), w_4\big) \Big) &= \mathcal{A}\Big( m_2\big(m_2(w_2, w_3), w_4\big) \Big) - m_2\big(\mathcal{A}_{(23)}, w_4\big) \, , \\
        \mathcal{A}\Big( m_2\big(w_2, p(m_2(w_3, w_4))\big) \Big) &= \mathcal{A}\Big( m_2\big(w_2, m_2(w_3, w_4)\big) \Big) - m_2\big(w_2, \mathcal{A}_{(34)}\big) \, .
    \end{split}
\end{equation}
The sign of the last term in \eqref{m3-prime} depends on the degrees of the second and third arguments. For the terms (a), (b) and (c) it equals $-(-1)^{|w_2| + |w_3|}$, $-(-1)^{|w_3| + |w_4|}$ and $-(-1)^{|w_3| + |w_4|}$, respectively. In (d) and (e) one of these arguments is $p\big(m_2(w_i, w_j)\big)$, of degree $|w_i| + |w_j| - 1$, and the sign equals $+(-1)^{|w_2| + |w_3| + |w_4|}$ in both cases. Substituting \eqref{m3-prime}, \eqref{A:Aij} and \eqref{A:resolve} into \eqref{A:Sigma-reduced}, we find
\begin{align}
    \begin{split}
        \text{(a)} = \; & \textcolor{blue}{\underbrace{m_2\big(m_3(w_1, w_2, w_3), w_4\big)}_{\mathrm{I}_1}} \; \textcolor{brown}{\underbrace{- \, m_2\big(m_3(\tau, \mathcal{A}_{(12)}, w_3), w_4\big)}_{\mathrm{IV}_1}} \\
        & \textcolor{red}{\underbrace{- (-1)^{|w_2| + |w_3|}\, m_2\big(m_3(w_1, \mathcal{A}_{(23)}, \tau), w_4\big)}_{\mathrm{VI}_1}} \, ,
    \end{split} \\[6pt]
    \begin{split}
        \text{(b)} = \; & \textcolor{blue}{\underbrace{(-1)^{|w_1|}\, m_2\big(w_1, m_3(w_2, w_3, w_4)\big)}_{\mathrm{I}_2}} \; \textcolor{red}{\underbrace{- (-1)^{|w_1|}\, m_2\big(w_1, m_3(\tau, \mathcal{A}_{(23)}, w_4)\big)}_{\mathrm{VI}_2}} \\
        & \textcolor{teal}{\underbrace{- (-1)^{|w_1| + |w_3| + |w_4|}\, m_2\big(w_1, m_3(w_2, \mathcal{A}_{(34)}, \tau)\big)}_{\mathrm{V}_1}} \, ,
    \end{split} \\[6pt]
    \begin{split}
        \text{(c)} = \; & \textcolor{blue}{\underbrace{m_3\big(m_2(w_1, w_2), w_3, w_4\big)}_{\mathrm{I}_3}} \; \textcolor{brown}{\underbrace{- \, m_3\big(m_2(\tau, \mathcal{A}_{(12)}), w_3, w_4\big)}_{\mathrm{IV}_2}} \; \textcolor{olive}{\underbrace{- \, m_3\Big(\tau, \mathcal{A}\big(m_2(m_2(w_1, w_2), w_3)\big), w_4\Big)}_{\mathrm{II}_1}} \\
        & \textcolor{brown}{\underbrace{+ \, m_3\big(\tau, m_2(\mathcal{A}_{(12)}, w_3), w_4\big)}_{\mathrm{IV}_3}} \; \textcolor{teal}{\underbrace{- (-1)^{|w_3| + |w_4|}\, m_3\big(m_2(w_1, w_2), \mathcal{A}_{(34)}, \tau\big)}_{\mathrm{V}_2}} \\
        & \textcolor{Orange}{\underbrace{+ (-1)^{|w_3| + |w_4|}\, m_3\big(m_2(\tau, \mathcal{A}_{(12)}), \mathcal{A}_{(34)}, \tau\big)}_{\mathrm{VII}_1}} \, ,
    \end{split} \\[6pt]
    \begin{split}
        \text{(d)} = \; & \textcolor{blue}{\underbrace{(-1)^{|w_1|}\, m_3\big(w_1, m_2(w_2, w_3), w_4\big)}_{\mathrm{I}_4}} \; \textcolor{red}{\underbrace{- (-1)^{|w_1|}\, m_3\big(w_1, m_2(\tau, \mathcal{A}_{(23)}), w_4\big)}_{\mathrm{VI}_3}} \\
        & \textcolor{olive}{\underbrace{- (-1)^{|w_1|}\, m_3\Big(\tau, \mathcal{A}\big(m_2(w_1, m_2(w_2, w_3))\big), w_4\Big)}_{\mathrm{II}_2}} \; \textcolor{red}{\underbrace{+ (-1)^{|w_1|}\, m_3\big(\tau, m_2(w_1, \mathcal{A}_{(23)}), w_4\big)}_{\mathrm{VI}_4}} \\
        & \textcolor{purple}{\underbrace{+ (-1)^{|w_1| + |w_2| + |w_3| + |w_4|}\, m_3\Big(w_1, \mathcal{A}\big(m_2(m_2(w_2, w_3), w_4)\big), \tau\Big)}_{\mathrm{III}_1}} \\
        & \textcolor{red}{\underbrace{- (-1)^{|w_1| + |w_2| + |w_3| + |w_4|}\, m_3\big(w_1, m_2(\mathcal{A}_{(23)}, w_4), \tau\big)}_{\mathrm{VI}_5}} \, ,
    \end{split} \\[6pt]
    \begin{split}
        \text{(e)} = \; & \textcolor{blue}{\underbrace{(-1)^{|w_1| + |w_2|}\, m_3\big(w_1, w_2, m_2(w_3, w_4)\big)}_{\mathrm{I}_5}} \; \textcolor{teal}{\underbrace{- (-1)^{|w_1| + |w_2|}\, m_3\big(w_1, w_2, m_2(\tau, \mathcal{A}_{(34)})\big)}_{\mathrm{V}_3}} \\
        & \textcolor{brown}{\underbrace{- (-1)^{|w_1| + |w_2|}\, m_3\big(\tau, \mathcal{A}_{(12)}, m_2(w_3, w_4)\big)}_{\mathrm{IV}_4}} \; \textcolor{Orange}{\underbrace{+ (-1)^{|w_1| + |w_2|}\, m_3\big(\tau, \mathcal{A}_{(12)}, m_2(\tau, \mathcal{A}_{(34)})\big)}_{\mathrm{VII}_2}} \\
        & \textcolor{purple}{\underbrace{+ (-1)^{|w_1| + |w_3| + |w_4|}\, m_3\Big(w_1, \mathcal{A}\big(m_2(w_2, m_2(w_3, w_4))\big), \tau\Big)}_{\mathrm{III}_2}} \\
        & \textcolor{teal}{\underbrace{- (-1)^{|w_1| + |w_3| + |w_4|}\, m_3\big(w_1, m_2(w_2, \mathcal{A}_{(34)}), \tau\big)}_{\mathrm{V}_4}} \, .
    \end{split}
\end{align}
Hence $\Sigma$ is equivalent to the sum of the $24$ labelled terms: five in group $\mathrm{I}$, two in each of the groups $\mathrm{II}$, $\mathrm{III}$ and $\mathrm{VII}$, four in each of the groups $\mathrm{IV}$ and $\mathrm{V}$, and five in group $\mathrm{VI}$. Groups $\mathrm{I}$--$\mathrm{III}$ do not involve the maps $\mathcal{A}_{(ij)}$ explicitly, groups $\mathrm{IV}$, $\mathrm{VI}$ and $\mathrm{V}$ contain $\mathcal{A}_{(12)}$, $\mathcal{A}_{(23)}$ and $\mathcal{A}_{(34)}$, respectively, and group $\mathrm{VII}$ contains both $\mathcal{A}_{(12)}$ and $\mathcal{A}_{(34)}$.

\subsection*{Summation by groups}

In each of the auxiliary relations below, the terms coinciding with the terms of the group under consideration are coloured and labelled accordingly; the remaining (black) terms give the sum of the group.

\begin{enumerate}
    \item \textcolor{blue}{Group $\mathrm{I}$}. The terms $\mathrm{I}_1, \dots, \mathrm{I}_5$ constitute the left-hand side of \eqref{lowest-stasheff} for $(w_1, w_2, w_3, w_4)$; hence
    \begin{equation*}
        \textcolor{blue}{\mathrm{I}_1 + \mathrm{I}_2 + \mathrm{I}_3 + \mathrm{I}_4 + \mathrm{I}_5} = 0 \, .
    \end{equation*}

    \item \textcolor{olive}{Group $\mathrm{II}$}. By linearity of $\mathcal{A}$ and $m_3$,
    \begin{equation*}
        \textcolor{olive}{\mathrm{II}_1 + \mathrm{II}_2} = - m_3\bigg(\tau, \mathcal{A}\Big( m_2\big(m_2(w_1, w_2), w_3\big) + (-1)^{|w_1|}\, m_2\big(w_1, m_2(w_2, w_3)\big) \Big), w_4\bigg) = 0 \, ,
    \end{equation*}
    by \eqref{A:S4} for $(w_1, w_2, w_3)$.

    \item \textcolor{purple}{Group $\mathrm{III}$}. Similarly,
    \begin{equation*}
        \begin{split}
            \textcolor{purple}{\mathrm{III}_1 + \mathrm{III}_2} = \; & (-1)^{|w_1| + |w_2| + |w_3| + |w_4|} \\
            & \times m_3\bigg(w_1, \mathcal{A}\Big( m_2\big(m_2(w_2, w_3), w_4\big) + (-1)^{|w_2|}\, m_2\big(w_2, m_2(w_3, w_4)\big) \Big), \tau\bigg) = 0 \, ,
        \end{split}
    \end{equation*}
    by \eqref{A:S4} for $(w_2, w_3, w_4)$.

    \item \textcolor{brown}{Group $\mathrm{IV}$}. Consider \eqref{lowest-stasheff} for $(\tau, \mathcal{A}_{(12)}, w_3, w_4)$. Since $(-1)^{|\tau|} = -1$ and $(-1)^{|\tau| + |\mathcal{A}_{(12)}|} = (-1)^{|w_1| + |w_2|}$, after multiplication by $-1$ it reads
    \begin{equation*}
        \begin{split}
            & \textcolor{brown}{\underbrace{- \, m_2\big(m_3(\tau, \mathcal{A}_{(12)}, w_3), w_4\big)}_{\mathrm{IV}_1}} + m_2\big(\tau, m_3(\mathcal{A}_{(12)}, w_3, w_4)\big) \; \textcolor{brown}{\underbrace{- \, m_3\big(m_2(\tau, \mathcal{A}_{(12)}), w_3, w_4\big)}_{\mathrm{IV}_2}} \\
            & \textcolor{brown}{\underbrace{+ \, m_3\big(\tau, m_2(\mathcal{A}_{(12)}, w_3), w_4\big)}_{\mathrm{IV}_3}} \; \textcolor{brown}{\underbrace{- (-1)^{|w_1| + |w_2|}\, m_3\big(\tau, \mathcal{A}_{(12)}, m_2(w_3, w_4)\big)}_{\mathrm{IV}_4}} = 0 \, .
        \end{split}
    \end{equation*}
    Hence, by Lemma \ref{lem:A-prelim},
    \begin{equation*}
        \textcolor{brown}{\mathrm{IV}_1 + \mathrm{IV}_2 + \mathrm{IV}_3 + \mathrm{IV}_4} = - m_2\big(\tau, m_3(\mathcal{A}_{(12)}, w_3, w_4)\big) \thicksim 0 \, .
    \end{equation*}

    \item \textcolor{teal}{Group $\mathrm{V}$}. Consider \eqref{lowest-stasheff} for $(w_1, w_2, \mathcal{A}_{(34)}, \tau)$. By \eqref{A:T},
    \begin{equation}
        \label{A:A34tau}
        m_2(\mathcal{A}_{(34)}, \tau) = (-1)^{|w_3| + |w_4|}\, m_2(\tau, \mathcal{A}_{(34)}) \, .
    \end{equation}
    Using this and multiplying by $-(-1)^{|w_3| + |w_4|}$, we obtain
    \begin{equation*}
        \begin{split}
            & - (-1)^{|w_3| + |w_4|}\, m_2\big(m_3(w_1, w_2, \mathcal{A}_{(34)}), \tau\big) \; \textcolor{teal}{\underbrace{- (-1)^{|w_1| + |w_3| + |w_4|}\, m_2\big(w_1, m_3(w_2, \mathcal{A}_{(34)}, \tau)\big)}_{\mathrm{V}_1}} \\
            & \textcolor{teal}{\underbrace{- (-1)^{|w_3| + |w_4|}\, m_3\big(m_2(w_1, w_2), \mathcal{A}_{(34)}, \tau\big)}_{\mathrm{V}_2}} \; \textcolor{teal}{\underbrace{- (-1)^{|w_1| + |w_3| + |w_4|}\, m_3\big(w_1, m_2(w_2, \mathcal{A}_{(34)}), \tau\big)}_{\mathrm{V}_4}} \\
            & \textcolor{teal}{\underbrace{- (-1)^{|w_1| + |w_2|}\, m_3\big(w_1, w_2, m_2(\tau, \mathcal{A}_{(34)})\big)}_{\mathrm{V}_3}} = 0 \, .
        \end{split}
    \end{equation*}
    Hence, by Lemma \ref{lem:A-prelim},
    \begin{equation*}
        \textcolor{teal}{\mathrm{V}_1 + \mathrm{V}_2 + \mathrm{V}_3 + \mathrm{V}_4} = (-1)^{|w_3| + |w_4|}\, m_2\big(m_3(w_1, w_2, \mathcal{A}_{(34)}), \tau\big) \thicksim 0 \, .
    \end{equation*}

    \item \textcolor{red}{Group $\mathrm{VI}$}. Consider \eqref{lowest-stasheff} for $(w_1, \tau, \mathcal{A}_{(23)}, w_4)$, multiplied by $-1$:
    \begin{equation*}
        \begin{split}
            & - m_2\big(m_3(w_1, \tau, \mathcal{A}_{(23)}), w_4\big) \; \textcolor{red}{\underbrace{- (-1)^{|w_1|}\, m_2\big(w_1, m_3(\tau, \mathcal{A}_{(23)}, w_4)\big)}_{\mathrm{VI}_2}} \\
            & - m_3\big(m_2(w_1, \tau), \mathcal{A}_{(23)}, w_4\big) \; \textcolor{red}{\underbrace{- (-1)^{|w_1|}\, m_3\big(w_1, m_2(\tau, \mathcal{A}_{(23)}), w_4\big)}_{\mathrm{VI}_3}} \\
            & + (-1)^{|w_1|}\, m_3\big(w_1, \tau, m_2(\mathcal{A}_{(23)}, w_4)\big) = 0 \, ,
        \end{split}
    \end{equation*}
    and \eqref{lowest-stasheff} for $(\tau, w_1, \mathcal{A}_{(23)}, w_4)$, multiplied by $-(-1)^{|w_1|}$:
    \begin{equation*}
        \begin{split}
            & - (-1)^{|w_1|}\, m_2\big(m_3(\tau, w_1, \mathcal{A}_{(23)}), w_4\big) + (-1)^{|w_1|}\, m_2\big(\tau, m_3(w_1, \mathcal{A}_{(23)}, w_4)\big) \\
            & - (-1)^{|w_1|}\, m_3\big(m_2(\tau, w_1), \mathcal{A}_{(23)}, w_4\big) \; \textcolor{red}{\underbrace{+ (-1)^{|w_1|}\, m_3\big(\tau, m_2(w_1, \mathcal{A}_{(23)}), w_4\big)}_{\mathrm{VI}_4}} \\
            & + m_3\big(\tau, w_1, m_2(\mathcal{A}_{(23)}, w_4)\big) = 0 \, .
        \end{split}
    \end{equation*}
    In the sum of these two relations, the terms containing $m_2(w_1, \tau)$ and $m_2(\tau, w_1)$ combine into
    \begin{equation*}
        - m_3\big(m_2(w_1, \tau) + (-1)^{|w_1|}\, m_2(\tau, w_1), \mathcal{A}_{(23)}, w_4\big) \, ,
    \end{equation*}
    which vanishes by \eqref{A:T}. The remaining two terms of group $\mathrm{VI}$ can be written as
    \begin{equation*}
        \begin{gathered}
            \textcolor{red}{\mathrm{VI}_1} = (-1)^{|\mathcal{A}_{(23)}|}\, m_2\big(m_3(w_1, \mathcal{A}_{(23)}, \tau), w_4\big) \, , \\
            \textcolor{red}{\mathrm{VI}_5} = - (-1)^{|w_1|}\, (-1)^{|m_2(\mathcal{A}_{(23)}, w_4)|}\, m_3\big(w_1, m_2(\mathcal{A}_{(23)}, w_4), \tau\big) \, ,
        \end{gathered}
    \end{equation*}
    since $(-1)^{|\mathcal{A}_{(23)}|} = -(-1)^{|w_2| + |w_3|}$ and $(-1)^{|m_2(\mathcal{A}_{(23)}, w_4)|} = (-1)^{|w_2| + |w_3| + |w_4|}$. Collecting all terms, we obtain
    \begin{equation*}
        \begin{split}
            \textcolor{red}{\mathrm{VI}_1 + \dots + \mathrm{VI}_5} = \; & m_2\Big( m_3(w_1, \tau, \mathcal{A}_{(23)}) + (-1)^{|w_1|}\, m_3(\tau, w_1, \mathcal{A}_{(23)}) \\
            & \hspace{1.2cm} + (-1)^{|\mathcal{A}_{(23)}|}\, m_3(w_1, \mathcal{A}_{(23)}, \tau), \, w_4 \Big) \\
            & - (-1)^{|w_1|} \Big[ m_3\big(w_1, \tau, m_2(\mathcal{A}_{(23)}, w_4)\big) \\
            & \hspace{2.2cm} + (-1)^{|w_1|}\, m_3\big(\tau, w_1, m_2(\mathcal{A}_{(23)}, w_4)\big) \\
            & \hspace{2.2cm} + (-1)^{|m_2(\mathcal{A}_{(23)}, w_4)|}\, m_3\big(w_1, m_2(\mathcal{A}_{(23)}, w_4), \tau\big) \Big] \\
            & - (-1)^{|w_1|}\, m_2\big(\tau, m_3(w_1, \mathcal{A}_{(23)}, w_4)\big) \, .
        \end{split}
    \end{equation*}
    The first argument of $m_2$ in the first two lines and the expression in square brackets vanish by \eqref{commutation-generalisation}, applied with $w_2 = \mathcal{A}_{(23)}$ and with $w_2 = m_2(\mathcal{A}_{(23)}, w_4)$, respectively, and the last line is equivalent to zero by Lemma \ref{lem:A-prelim}. Hence $\textcolor{red}{\mathrm{VI}_1 + \dots + \mathrm{VI}_5} \thicksim 0$.

    \item \textcolor{Orange}{Group $\mathrm{VII}$}. Consider \eqref{lowest-stasheff} for $(\tau, \mathcal{A}_{(12)}, \mathcal{A}_{(34)}, \tau)$. Using \eqref{A:A34tau} and multiplying by $(-1)^{|w_3| + |w_4|}$, we obtain
    \begin{equation*}
        \begin{split}
            & (-1)^{|w_3| + |w_4|}\, m_2\big(m_3(\tau, \mathcal{A}_{(12)}, \mathcal{A}_{(34)}), \tau\big) - (-1)^{|w_3| + |w_4|}\, m_2\big(\tau, m_3(\mathcal{A}_{(12)}, \mathcal{A}_{(34)}, \tau)\big) \\
            & \textcolor{Orange}{\underbrace{+ (-1)^{|w_3| + |w_4|}\, m_3\big(m_2(\tau, \mathcal{A}_{(12)}), \mathcal{A}_{(34)}, \tau\big)}_{\mathrm{VII}_1}} \\
            & - (-1)^{|w_3| + |w_4|}\, m_3\big(\tau, m_2(\mathcal{A}_{(12)}, \mathcal{A}_{(34)}), \tau\big) \; \textcolor{Orange}{\underbrace{+ (-1)^{|w_1| + |w_2|}\, m_3\big(\tau, \mathcal{A}_{(12)}, m_2(\tau, \mathcal{A}_{(34)})\big)}_{\mathrm{VII}_2}} = 0 \, .
        \end{split}
    \end{equation*}
    Hence, by Lemma \ref{lem:A-prelim},
    \begin{equation*}
        \textcolor{Orange}{\mathrm{VII}_1 + \mathrm{VII}_2} \thicksim (-1)^{|w_3| + |w_4|}\, m_3\big(\tau, m_2(\mathcal{A}_{(12)}, \mathcal{A}_{(34)}), \tau\big) \, .
    \end{equation*}
    The right-hand side vanishes by \eqref{tau-w-tau}. Hence $\textcolor{Orange}{\mathrm{VII}_1 + \mathrm{VII}_2} \thicksim 0$.
\end{enumerate}

\subsection*{Conclusion}

All seven groups are equivalent to zero, and therefore $\Sigma \thicksim 0$. Since $\Sigma \in \mathcal{R}(W)$ and $\mathcal{R}(W) \cap \mathcal{I}_\tau(W) = \{0\}$, we conclude that $\Sigma = 0$, i.e. $(\tilde{m}_2, \tilde{m}_3)$ satisfy \eqref{lowest-stasheff} for all $w_1, \dots, w_4 \in \mathcal{R}(W)$. This completes the proof of Theorem \ref{thm:lift}.

\section{Deriving the projector}
\label{app:B}
In this part of the appendix, we first collect formulas used in most calculations. In all presented calculations, $f(Y)$ is supposed to be an $\mathfrak{sp}(2)$-singlet. In \eqref{B4}--\eqref{B6} and in Appendix~\ref{app:B3}, $\hat{N}$ denotes the degree of the expression it multiplies, i.e. of the result.

\subsection{Schouten identities and corollary of $\mathfrak{sp}(2)$-invariance }
Working with $\mathfrak{sp}(2)$-tensors allows us to use some of the identities specific to tensors in $2$ dimensions. Namely, for any symmetric $\mathfrak{sp}(2)$-tensors  $A_{\alpha\beta} = A_{\beta\alpha}$ and $B_{\alpha\beta} = B_{\beta\alpha}$ the following identities hold true 
\begin{subequations}
        \label{B1}
        \begin{align}
            &A_{\alpha \lambda} B^{\lambda\sigma} A_{\sigma \beta} = A_{\alpha\beta} \cdot \big[ A_{\mu\nu} B^{\mu\nu} \big] - \frac{B_{\alpha\beta}}{2} \Big[ A_{\mu\nu} A^{\mu\nu} \Big] \, , \\
            &A_{\alpha\lambda} B^{\lambda\sigma} B_{\sigma\beta} + (\alpha \leftrightarrow \beta) = A_{\alpha\beta} \cdot \Big[ B_{\mu\nu} B^{\mu\nu} \Big] \, .
        \end{align}
    \end{subequations}
Since \eqref{sp2-inv} is represented as a differential equation, the natural thing to do is to explore various differential consequences of it. Here we provide some of them, relevant to the upcoming derivations
\begin{equation}
        \label{B2}
        \begin{split}
        &Y_\alpha \cdot \frac{\partial}{\partial Y^\beta}\bigg( \Big[ \prod_{j=1}^{n} \Gamma_{\mu_j \nu_j} \Big] f \bigg) = \\
        &=-\epsilon_{\alpha\beta} \frac{\hat{N}+2n}{2}  \Big[ \prod_{j=1}^{n} \Gamma_{\mu_j \nu_j} \Big] f - \sum_{i=1}^{n}\big( \epsilon_{\mu_i \alpha}\Gamma_{\beta \nu_i} + (\mu \leftrightarrow \nu) \big) \Big[ \prod_{j\neq i}^{n} \Gamma_{\mu_j \nu_j} \Big] f \, .
        \end{split}
    \end{equation}
    It also implies 
    \begin{equation}
        \label{B3}
        \begin{split}
            &Y^\sigma \cdot \frac{\partial}{\partial Y^\beta}\bigg( \Big[ \prod_{j=1}^{n-1} \Gamma_{\mu_j \nu_j} \Big] \Gamma_{\sigma\alpha}f \bigg) = \\
            &= \frac{\hat{N}+2(n-3)}{2} \Big[ \prod_{j=1}^{n-1} \Gamma_{\mu_j \nu_j} \Big] \Gamma_{\alpha\beta}f - \sum_{i=1}^{n-1}\Big(\Gamma_{\alpha \mu_i}\Gamma_{\nu_i \beta} + (\alpha \leftrightarrow \beta) \Big) \Big[ \prod_{j\neq i}^{n-1} \Gamma_{\mu_j \nu_j} \Big] f \, ,
        \end{split}
    \end{equation}
Using formulas \eqref{B1}, \eqref{B2} and \eqref{B3} it is possible to straightforwardly arrive at 
\begin{equation}
    \label{B4}
    \begin{split}
        &\Gamma_{\alpha\beta}\bigg( \mathcal{T}_{(k,m)} f\bigg)=2k(d+\hat{N}-2k-4m-1) \, \mathcal{T}_{(k-1,m)}\big(\Gamma_{\alpha\beta}f\big) + \\
        &+4m(d+\hat{N}-2m) \, t_{\alpha\beta} \, \mathcal{T}_{(k,m-1)}\big(\Gamma^{\sigma\lambda}\Gamma_{\sigma\lambda} f\big)+2k(k-1) \, t_{\alpha\beta} \, \mathcal{T}_{(k-2,m)}\big(\Gamma^{\sigma\lambda}\Gamma_{\sigma\lambda} f\big) + \\
        &+ \mathcal{T}_{(k,m)}\big(\Gamma_{\alpha\beta}f\big) \, ,
    \end{split}
\end{equation}
which directly implies 
\begin{equation}
    \begin{split}
        &t^{\alpha\beta} \Gamma_{\alpha\beta}\bigg( \mathcal{T}_{(k,m)} f\bigg)= \Big\{2(k+2m)(d+\hat{N}-2k)-6k-8m(m+1) \Big\} \mathcal{T}_{(k,m)} f + \\
        &+2k(k-1) \mathcal{T}_{(k-2,m+1)} f+\mathcal{T}_{(k+1,m)} f \, .
    \end{split}
\end{equation}
There are other consequences that might be extracted from \eqref{B1}-\eqref{B4}. However, the most important one for our tasks is
\begin{equation}
    \label{B6}
    \begin{split}
        &\Gamma^{\alpha\beta}\Big[ t_{\alpha\beta} \, \mathcal{T}_{(k,m-1)}\big( (\Gamma^{\sigma\lambda}\Gamma_{\sigma\lambda})^{n+1} f\big)\Big]= \\
        &=\Big\{2(k+2m+1)(d+\hat{N}-2k)+6(k+1)-8m(m-1) \Big\} \times\mathcal{T}_{(k,m-1)}\big( (\Gamma^{\sigma\lambda}\Gamma_{\sigma\lambda})^{n+1} f\big) +\\
        &+2k(k-1)\,\mathcal{T}_{(k-2,m)}\big( (\Gamma^{\sigma\lambda}\Gamma_{\sigma\lambda})^{n+1} f\big)+\mathcal{T}_{(k+1,m-1)}\big( (\Gamma^{\sigma\lambda}\Gamma_{\sigma\lambda})^{n+1} f\big) \, .
    \end{split}
\end{equation}
\subsection{TPO (in)dependence}
\label{app:B2}

In this part of the appendix we prove properties 3 and 4 of Section~\ref{sec:horproj}, i.e.\ \eqref{indpnd}--\eqref{B-coef}. Throughout, $V_s$ denotes the space of $\mathfrak{sp}(2)$-singlet polynomials of degree $2s$ and $n:=\lfloor s/2\rfloor$. In the shorthand \eqref{TPO-shorthand}, the TPOs \eqref{TPO} are $\mathcal{T}_{(k,m)}$, with
\begin{equation}
Q:=k+2m ,
\end{equation}
and $\mathcal{T}_{(k,m)}$ acts on $f$ through its $Q$-fold trace
\begin{equation}
\Phi_Q(f):=\big\{\Gamma_{\mu_1\nu_1}\cdots\Gamma_{\mu_Q\nu_Q}f\big\} .
\end{equation}
We call $\mathcal{T}_{(k,m)}$ with $k+2m=Q$ a $Q$-fold TPO.

\subsubsection*{Conventions}
\paragraph{Polynomials.} $\mathrm{Pol}_{2s}$ denotes \emph{all} homogeneous polynomials of degree $2s$ in $Y^\alpha_A$, not only $\mathfrak{sp}(2)$-singlets, and $\mathrm{Pol}_{2s}\cap\ker\Gamma_{\mu\nu}$ the subspace annihilated by all $\Gamma_{\mu\nu}$. In contrast to $\ker\Gamma_{\alpha\beta}$ in the main text, the latter is not restricted to $\off$. Only even degrees occur below. In Lemmas \ref{lem:twoRow} and \ref{lem:pos}, $s$ is an arbitrary non-negative integer; they are applied with $s$ replaced by $s-Q$ or $s-j$.

We work with complex coefficients. Since all TPOs have real coefficients, their linear (in)dependence over $\mathbb C$ and over $\mathbb R$ coincide. We fix a basis in which $\eta_{AB}=\mathrm{diag}(\pm1)$, so that $\eta^{AB}$ and $\eta_{AB}$ are the same matrix.

\paragraph{$\mathfrak{sp}(2)$-spin.} By \eqref{t-star},
\begin{equation}
[t^{\alpha\beta},f]_\ast=-2i\Big(Y^\alpha\!\cdot\!\frac{\partial}{\partial Y_\beta}+Y^\beta\!\cdot\!\frac{\partial}{\partial Y_\alpha}\Big)f ,
\end{equation}
so $\mathfrak{sp}(2)$ acts on polynomials by rotations of the index $\alpha$. Using $\mathfrak{sp}(2)\simeq\mathfrak{so}(3)$, we call the spin of an $\mathfrak{sp}(2)$-module its \emph{$\mathfrak{sp}(2)$-spin}. It is not to be confused with the physical spin, which is encoded in the length of the diagram.

\paragraph{Diagrams.} For a two-row Young diagram $\mu=(\mu_1,\mu_2)$ we write $[\mu]_{\mathfrak{gl}(d+1)}$ and $[\mu]_{\mathfrak{so}(d+1)}$ for the corresponding modules, as in \eqref{GLtoSO}. A subspace of $\mathrm{Pol}_{2s}$ \emph{transforms as} $[\lambda]_{\mathfrak{so}(d+1)}$ if it is a sum of $\mathfrak{so}(d+1)$-submodules isomorphic to $[\lambda]_{\mathfrak{so}(d+1)}$; \emph{the part} of $\mathrm{Pol}_{2s}$ that transforms as $[\lambda]_{\mathfrak{so}(d+1)}$ is the sum of all such submodules. By \eqref{GLtoSO},
\begin{equation}
V_s=\bigoplus_{0\leqslant2a\leqslant2b\leqslant s}W_{a,b},\qquad \lambda(a,b):=(s-2a,\,s-2b),
\label{Wab}
\end{equation}
where $W_{a,b}$ is the part of $V_s$ that transforms as $[\lambda(a,b)]_{\mathfrak{so}(d+1)}$; it is a single copy of $[\lambda(a,b)]_{\mathfrak{so}(d+1)}$.

\subsubsection*{Representation-theoretic input}
\begin{lemma}\label{lem:twoRow}
Let $s\geqslant0$.
\begin{enumerate}
\item[(a)] The $\mathfrak{sp}(2)$-spins occurring in $\mathrm{Pol}_{2s}$ are integers $j$ with $0\leqslant j\leqslant s$.
\item[(b)] The spin-$j$ part of $\mathrm{Pol}_{2s}$ is (spin $j$)$\,\otimes\,[s+j,\,s-j]_{\mathfrak{gl}(d+1)}$, i.e.\ a polynomial of degree $2s$ and $\mathfrak{sp}(2)$-spin $j$ is a $\mathfrak{gl}(d+1)$-tensor with the two-row Young symmetry $(s+j,\,s-j)$, and each such diagram occurs once.
\item[(c)] Let $d\geqslant4$. The spin-$j$ part of $\mathrm{Pol}_{2s}\cap\ker\Gamma_{\mu\nu}$ is (spin $j$)$\,\otimes\,[s+j,\,s-j]_{\mathfrak{so}(d+1)}$. The modules $[\mu]_{\mathfrak{so}(d+1)}$ of two-row diagrams are irreducible and pairwise inequivalent. In particular, $\mathrm{Pol}_{2s}\cap\ker\Gamma_{\mu\nu}$ is a sum of submodules isomorphic to $[\mu]_{\mathfrak{so}(d+1)}$ with $|\mu|=2s$.
\end{enumerate}
\end{lemma}
\begin{proof}
(a) With $\partial/\partial Y_2=\partial/\partial Y^1$ and $\partial/\partial Y_1=-\partial/\partial Y^2$ ($\epsilon^{12}=+1$),
\[
[t^{12},\bullet]_\ast=-4i\,j_3 ,\qquad j_3:=\frac12\Big(Y^1\!\cdot\!\frac{\partial}{\partial Y^1}-Y^2\!\cdot\!\frac{\partial}{\partial Y^2}\Big),
\]
while $[t^{11},\bullet]_\ast$ and $[t^{22},\bullet]_\ast$ are the ladder operators. Only the eigenvalues of $j_3$ are used below. A monomial of degree $2s$ with $i$ factors $Y^1$ and $2s-i$ factors $Y^2$ has $j_3=i-s$, an integer in $\{-s,\dots,s\}$.

(b) This is the realisation of $\mathfrak{gl}(d+1)$-modules by highest-weight vectors of the rotations of the index $\alpha$ \cite[eq.~(33)]{bekaert2021unitary}. At fixed degree $2s$, a two-row diagram $\mu$ with $|\mu|=2s$ is determined by $\tfrac12(\mu_1-\mu_2)$, which is the $\mathfrak{sp}(2)$-spin. For $j=0$ this is the statement of Section~\ref{sec:gl-so} that $V_s$ is the rectangular diagram.

(c) The three $\Gamma_{\mu\nu}$ form an $\mathfrak{sp}(2)$-triplet, so $\mathrm{Pol}_{2s}\cap\ker\Gamma_{\mu\nu}$ is $\mathfrak{sp}(2)$-stable and splits into spin-$j$ parts. Their highest-weight vectors are the traceless solutions of \cite[eq.~(33)]{bekaert2021unitary}, which realise $[\mu]_{\mathfrak{so}(d+1)}$ \cite[eq.~(34) and Section~4.3]{bekaert2021unitary}.
\end{proof}
At $d=3$ the irreducibility statement in (c) fails: for $\mu_2>0$ the module $[\mu]_{\mathfrak{so}(4)}$ splits into self-dual and anti-self-dual parts (cf. Section~\ref{sec:d3}). Lemma~\ref{lem:pos}(b) and Lemma~\ref{lem:G1}(ii)--(iv) use (c), directly or through Lemma~\ref{lem:pos}(b), and are therefore stated for $d\geqslant4$ only; Lemma~\ref{lem:pos}(a) and Lemma~\ref{lem:G1}(i) do not use (c) and hold for every $d$.

\begin{lemma}[(positivity)]\label{lem:pos}
For multi-indices $\mathbf a=(a_{\alpha A})$ put $Y^{\mathbf a}:=\prod(Y^\alpha_A)^{a_{\alpha A}}$ and $\mathbf a!:=\prod a_{\alpha A}!$, and for $p=\sum_{\mathbf a}p_{\mathbf a}Y^{\mathbf a}$, $q=\sum_{\mathbf a}q_{\mathbf a}Y^{\mathbf a}$ define
\begin{equation}
\langle p,q\rangle:=\sum_{\mathbf a}\mathbf a!\;p_{\mathbf a}\,\overline{q_{\mathbf a}} .
\end{equation}
This form is positive definite, and for every fixed value of $\mu,\nu$
\begin{equation}
\big\langle t^{\mu\nu}p,q\big\rangle=\big\langle p,\Gamma_{\mu\nu}q\big\rangle .
\label{adj}
\end{equation}
Consequently:
\begin{enumerate}
\item[(a)] $\displaystyle \mathrm{Pol}_{2s}=(\mathrm{Pol}_{2s}\cap\ker\Gamma_{\mu\nu})\oplus\sum_{\mu\nu}t^{\mu\nu}\mathrm{Pol}_{2s-2}$, and hence
\begin{equation}
\mathrm{Pol}_{2s}=\sum_{j=0}^{s}\mathbb C[t^{\mu\nu}]_j\cdot\big(\mathrm{Pol}_{2(s-j)}\cap\ker\Gamma_{\mu\nu}\big),
\label{harmdec}
\end{equation}
where $\mathbb C[t^{\mu\nu}]_j$ denotes the polynomials of degree $j$ in the $t^{\mu\nu}$;
\item[(b)] ($d\geqslant4$) the part of $\mathrm{Pol}_{2s}$ that transforms as $[\lambda]_{\mathfrak{so}(d+1)}$ lies in $\mathbb C[t^{\mu\nu}]_Q\cdot\big(\mathrm{Pol}_{2(s-Q)}\cap\ker\Gamma_{\mu\nu}\big)$, where $|\lambda|=2(s-Q)$. In particular, $\mathrm{Pol}_{2s}$ contains only $\mathfrak{so}(d+1)$-types $\lambda$ with $|\lambda|\leqslant2s$.
\end{enumerate}
\end{lemma}
\begin{proof}
The form does not involve $\eta$ and is positive definite in any signature. It is defined component by component in the fixed coordinates $Y^\alpha_A$ and is invariant under neither $\mathfrak{sp}(2)$ nor $\mathfrak{so}(d-1,2)$, so \eqref{adj} is not a tensor equation: it holds for each fixed numerical value of the indices. A check on monomials gives $\langle Y^\alpha_Ap,q\rangle=\langle p,\partial q/\partial Y^\alpha_A\rangle$, hence
\[
\big\langle t^{\mu\nu}p,q\big\rangle=\Big\langle p,\ \eta^{AB}\frac{\partial^2q}{\partial Y^\mu_A\partial Y^\nu_B}\Big\rangle=\big\langle p,\Gamma_{\mu\nu}q\big\rangle ,
\]
where $\eta^{AB}=\eta_{AB}$ was used. The lower position of $\mu,\nu$ on $\Gamma_{\mu\nu}$ is the natural position of the derivative $\partial/\partial Y^\mu_A$; no index is lowered with $\epsilon$. For example, the adjoint of $t^{11}$ is $\Gamma_{11}$, whereas $\Gamma^{11}=\Gamma_{22}$.

(a) By \eqref{adj}, $q\perp t^{\mu\nu}\mathrm{Pol}_{2s-2}$ for all $\mu,\nu$ if and only if $\Gamma_{\mu\nu}q=0$ for all $\mu,\nu$. So $\mathrm{Pol}_{2s}\cap\ker\Gamma_{\mu\nu}$ is the orthogonal complement of $\sum_{\mu\nu}t^{\mu\nu}\mathrm{Pol}_{2s-2}$, and positive definiteness gives the direct sum. Iteration gives \eqref{harmdec}.

(b) The $t^{\mu\nu}$ are $\mathfrak{so}(d+1)$-invariant, so by Lemma \ref{lem:twoRow}(c) the $j$-th summand in \eqref{harmdec} is a sum of submodules isomorphic to $[\mu]_{\mathfrak{so}(d+1)}$ with $|\mu|=2(s-j)$. Since $\mathfrak{so}(d+1)$ acts completely reducibly, the part of $\mathrm{Pol}_{2s}$ that transforms as $[\lambda]_{\mathfrak{so}(d+1)}$ is the sum of the corresponding parts of the summands, and by inequivalence only the summand with $2(s-j)=|\lambda|$ contributes.
\end{proof}

\subsubsection*{Symbols of TPOs}
A linear combination of $Q$-fold TPOs is encoded by its \emph{symbol}, the polynomial on $\mathbb C^3\times\mathbb C^3$ obtained by linear extension of
\begin{equation}
\mathcal{T}_{(k,m)}\ \longmapsto\ \langle v,w\rangle^k\big(\langle v,v\rangle\langle w,w\rangle\big)^m ,\qquad \langle X,Z\rangle:=X^{\mu\nu}Z_{\mu\nu} ,
\label{symbol}
\end{equation}
where $v\leftrightarrow t^{\mu\nu}$, $w\leftrightarrow\Gamma^{\mu\nu}$ (indices raised with $\epsilon$), all $t$ standing to the left of all $\Gamma$. This is consistent since the $t^{\mu\nu}$ commute among themselves and so do the $\Gamma_{\mu\nu}$. With
\begin{equation}
r:=\sqrt{\langle v,v\rangle\langle w,w\rangle},\qquad x:=\frac{\langle v,w\rangle}{r},
\end{equation}
the symbol of $\mathcal{T}_{(Q-2m,m)}$ is $r^Qx^{Q-2m}$. For $0\leqslant J\leqslant Q$ with $J-Q$ even, let $E_{Q,J}$ be the combination of $Q$-fold TPOs with symbol
\begin{equation}
E_{Q,J}(v,w):=r^QP_J(x),
\end{equation}
where $P_J$ is the Legendre polynomial. Since $P_J$ has the parity of $J$ and $Q-J$ is even, $E_{Q,J}(v,w)$ is a combination of the monomials \eqref{symbol}.

We use two standard facts about $\mathfrak{so}(3)$, see e.g.\ \cite{landau1958quantum}. The symmetric tensors $X^{\mu\nu}$ form the vector of $\mathfrak{sp}(2)\simeq\mathfrak{so}(3)$. First, the homogeneous polynomials of degree $Q$ on $\mathbb C^3$, i.e.\ $S:=\mathrm{Sym}^Q(\mathbb C^3)$, the one-row diagram $[Q]_{\mathfrak{gl}(3)}$, decompose as $\langle X,X\rangle^k$ times harmonic polynomials of degree $Q-2k$, $k=0,1,\dots$; in diagrams, $[Q]_{\mathfrak{gl}(3)}=[Q]_{\mathfrak{so}(3)}\oplus[Q-2]_{\mathfrak{so}(3)}\oplus\dots$, and harmonic polynomials of degree $J$ carry $\mathfrak{sp}(2)$-spin $J$. Hence $S$ contains the $\mathfrak{sp}(2)$-spins $Q,Q-2,\dots$, each once. Second, for fixed $v$, the polynomial $w\mapsto r^JP_J(x)$ is harmonic of degree $J$ in $w$ (addition theorem for spherical harmonics).

\begin{lemma}\label{lem:E}
\begin{enumerate}
\item[(a)] $x^k=\sum_J\chi_{k,J}P_J(x)$, with the sum over $J\leqslant k$, $J-k$ even, and
\begin{equation}
\chi_{k,J}=\frac{2J+1}{2}\int_{-1}^1x^kP_J(x)\,dx=\frac{2J+1}{2}\cdot\frac{k!}{(k-J)!\,2^JJ!}\int_{-1}^1x^{k-J}(1-x^2)^J\,dx>0 .
\end{equation}
\item[(b)] $\mathcal{T}_{(Q-2m,m)}=\sum_{J\leqslant Q-2m}\chi_{Q-2m,J}\,E_{Q,J}$, and $\mathrm{span}\{\mathcal{T}_{(Q-2m,m)}\}_m=\mathrm{span}\{E_{Q,J}\}_J$.
\item[(c)] $E_{Q,J}f$ depends on $\Phi_Q(f)$ only through its $\mathfrak{sp}(2)$-spin-$J$ part $\pi_J\Phi_Q(f)$.
\end{enumerate}
\end{lemma}
\begin{proof}
(a) Rodrigues' formula and $J$ integrations by parts; the boundary terms vanish, and the last integral is positive since $k-J$ is even.

(b) The first statement is (a) multiplied by $r^Q$. Conversely, expanding $P_J$ in powers $x^{J-2i}$ expresses each $E_{Q,J}$ through the $\mathcal{T}_{(Q-2m,m)}$.

(c) Every combination $F$ of $Q$-fold TPOs acts as $Ff=B_F\big(t^{\otimes Q},\Phi_Q(f)\big)$, where $B_F$ is the $\mathfrak{sp}(2)$-invariant bilinear form on $S$ obtained by polarising the symbol $F(v,w)$. Since $S$ contains each $\mathfrak{sp}(2)$-spin once, $B_F$ is block-diagonal in the spin decomposition by Schur's lemma. For $F=E_{Q,J}$ and fixed $v$,
\[
E_{Q,J}(v,w)=\big(\langle v,v\rangle\langle w,w\rangle\big)^{(Q-J)/2}\cdot r^JP_J(x).
\]
The second factor is harmonic of degree $J$ in $w$, hence of $\mathfrak{sp}(2)$-spin $J$, and the first is $\mathfrak{so}(3)$-invariant in $w$. So $w\mapsto E_{Q,J}(v,w)$ lies in the spin-$J$ part of the degree-$Q$ polynomials in $w$. Since the $v^{\otimes Q}$ span $S$, all blocks of $B_{E_{Q,J}}$ with spin different from $J$ vanish.
\end{proof}

\subsubsection*{The main lemma}
Put
\begin{equation}
s_Q:=\min(Q,\,s-Q),\qquad N_Q:=\#\big\{J\in\mathbb Z:\ 0\leqslant J\leqslant s_Q,\ J-Q\ \text{even}\big\}.
\end{equation}

\begin{lemma}\label{lem:G1}
Let $0\leqslant Q\leqslant s$ and $\zeta_0,\dots,\zeta_{\lfloor Q/2\rfloor}\in\mathbb C$.
\begin{enumerate}
\item[(i)] If
\begin{equation}
\sum_{m}\frac{\zeta_m}{Q-2m+J+1}=0\qquad\text{for all }J\in\{0,\dots,s_Q\}\ \text{with}\ J-Q\ \text{even},
\label{moment}
\end{equation}
then $\sum_m \zeta_m\,\mathcal{T}_{(Q-2m,m)}=0$ on $V_s$.
\item[(ii)] ($d\geqslant4$) Conversely, if $\sum_m \zeta_m\,\mathcal{T}_{(Q-2m,m)}=0$ on $V_s$, then \eqref{moment} holds.
\item[(iii)] ($d\geqslant4$) The $Q$-fold TPOs span a space of operators on $V_s$ of dimension $N_Q$.
\item[(iv)] ($d\geqslant4$) For $f\in W_{a,b}$, $\Phi_Q(f)=0$ if $Q>a+b$. On $W_{a,b}$ with $a+b=Q$, a combination $\sum_J\theta_JE_{Q,J}$ vanishes only if $\theta_J=0$ for $J=b-a$.
\end{enumerate}
\end{lemma}

\begin{proof}
\emph{Step 1: spin bound.} Let $f\in V_s$. Since $f$ is a singlet and $\Gamma_{\mu\nu}$ is $\mathfrak{sp}(2)$-covariant, $\Phi_Q(f)$ is $\mathfrak{sp}(2)$-invariant as a whole: it is annihilated by the total generators, which rotate the free indices and act by $[t^{\alpha\beta},\bullet]_\ast$ on the coefficients. By the Clebsch--Gordan rule, the spin-$J$ part of the index structure pairs only with $\mathfrak{sp}(2)$-spin $J$ in the coefficients. The coefficients lie in $\mathrm{Pol}_{2(s-Q)}$, which by Lemma \ref{lem:twoRow}(a) contains only $\mathfrak{sp}(2)$-spins $\leqslant s-Q$. Hence $\pi_J\Phi_Q(f)=0$ for $J>s-Q$, and by Lemma \ref{lem:E}(c)
\begin{equation}
E_{Q,J}=0\ \ \text{on }V_s\qquad\text{for }J>s_Q .
\label{vanish}
\end{equation}
For $J>Q$ the operator $E_{Q,J}$ is not defined, which explains the minimum in $s_Q$.

\emph{Step 2: moments.} Put $\zeta(x):=\sum_m\zeta_mx^{Q-2m}$. By Lemma \ref{lem:E}(b) and \eqref{vanish}, on $V_s$
\begin{equation}
\sum_m\zeta_m\mathcal{T}_{(Q-2m,m)}=\sum_{J\leqslant s_Q}\theta_J\,E_{Q,J},\qquad \theta_J:=\sum_m\zeta_m\chi_{Q-2m,J}=\frac{2J+1}{2}\int_{-1}^1\zeta(x)P_J(x)\,dx .
\label{theta}
\end{equation}
The polynomial $\zeta(x)$ has the parity of $Q$, and $\{P_J\}$ and $\{x^J\}$ with $J\leqslant s_Q$, $J-Q$ even, span the same space. So all $\theta_J$ with $J\leqslant s_Q$ vanish if and only if
\[
\int_{-1}^1\zeta(x)\,x^J\,dx=\sum_m\frac{2\zeta_m}{Q-2m+J+1}=0\qquad\text{for these }J ,
\]
i.e.\ if and only if \eqref{moment} holds. This proves (i).

From now on $d\geqslant4$. Fix $W_{a,b}$ and put $\lambda:=\lambda(a,b)$, so that $|\lambda|=2(s-a-b)$ and $\tfrac12(\lambda_1-\lambda_2)=b-a$.

\emph{Step 3: vanishing of higher traces and spin selection.} Since each component of $\Phi_Q$ is $\mathfrak{so}(d+1)$-covariant, the span of all components of $\Phi_Q(f)$ as $f$ ranges over $W_{a,b}$ is contained in the subspace $\mathrm{Pol}_{2(s-Q)}$ that transforms as $[\lambda]_{\mathfrak{so}(d+1)}$. If $Q>a+b$, then $|\lambda|>2(s-Q)$, and by Lemma \ref{lem:pos}(b) this subspace is zero: $\Phi_Q(f)=0$. If $Q=a+b$, then $|\lambda|=2(s-Q)$; by Lemma \ref{lem:pos}(b) with $s$ replaced by $s-Q$ the subspace lies in $\mathrm{Pol}_{2(s-Q)}\cap\ker\Gamma_{\mu\nu}$, and by Lemma \ref{lem:twoRow}(c) its $\mathfrak{sp}(2)$-spin is $\tfrac12(\lambda_1-\lambda_2)=b-a$. As in Step 1, invariance of $\Phi_Q(f)$ then forces $\pi_J\Phi_Q(f)=0$ for $J\neq b-a$, and by Lemma \ref{lem:E}(c)
\begin{equation}
E_{Q,J}f=0\qquad\text{for all }f\in W_{a,b},\ a+b=Q,\ J\neq b-a .
\label{select}
\end{equation}

\emph{Step 4: non-vanishing.} Let $a+b=Q$. By Lemma \ref{lem:pos}(b), $W_{a,b}\subset\mathbb C[t^{\mu\nu}]_Q\cdot\big(\mathrm{Pol}_{2(s-Q)}\cap\ker\Gamma_{\mu\nu}\big)$, so every $f\in W_{a,b}$ can be written as
\begin{equation}
f=t^{\mu_1\nu_1}\cdots t^{\mu_Q\nu_Q}\,\tilde f_{\mu_1\nu_1\dots\mu_Q\nu_Q},\qquad \tilde f_{\mu_1\nu_1\dots\mu_Q\nu_Q}\in\mathrm{Pol}_{2(s-Q)}\cap\ker\Gamma_{\mu\nu} ,
\label{layer}
\end{equation}
with summation over all index values. In the shorthand \eqref{TPO-shorthand},\[\mathcal{T}_{(Q,0)}=t^{\mu_1\nu_1}\cdots t^{\mu_Q\nu_Q}\,\Gamma_{\mu_1\nu_1}\cdots\Gamma_{\mu_Q\nu_Q}.\]Each term pairs a product of $t$'s with its adjoint \eqref{adj}, so
\[
\big\langle f,\mathcal{T}_{(Q,0)}f\big\rangle=\sum_{\mu_1,\nu_1,\dots,\mu_Q,\nu_Q}\big\|\Gamma_{\mu_1\nu_1}\cdots\Gamma_{\mu_Q\nu_Q}f\big\|^2 .
\]
Let $0\neq f\in W_{a,b}$. If all $\Gamma_{\mu_1\nu_1}\cdots\Gamma_{\mu_Q\nu_Q}f$ vanished, \eqref{layer} and \eqref{adj} would give
\[
\langle f,f\rangle=\sum\big\langle\tilde f_{\mu_1\nu_1\dots\mu_Q\nu_Q},\,\Gamma_{\mu_1\nu_1}\cdots\Gamma_{\mu_Q\nu_Q}f\big\rangle=0 .
\]
Hence $\langle f,\mathcal{T}_{(Q,0)}f\rangle>0$, and $\mathcal{T}_{(Q,0)}f\neq0$; this conclusion does not depend on the choice of basis. By Lemma \ref{lem:E}(b) and \eqref{select}, $\mathcal{T}_{(Q,0)}f=\chi_{Q,b-a}\,E_{Q,b-a}f$ with $\chi_{Q,b-a}>0$. Therefore
\begin{equation}
E_{Q,b-a}f\neq0\qquad\text{for every }0\neq f\in W_{a,b},\ a+b=Q .
\end{equation}
Together with \eqref{select} this proves (iv).

\emph{Step 5: independence.} Let $0\leqslant J\leqslant s_Q$ with $J-Q$ even. Then $(a,b):=\big(\tfrac{Q-J}{2},\tfrac{Q+J}{2}\big)$ satisfies $0\leqslant2a\leqslant2b\leqslant s$, so $W_{a,b}\neq0$ by \eqref{Wab}, and $b-a=J$. If $\sum_{J\leqslant s_Q}\theta_JE_{Q,J}=0$ on $V_s$, applying it to $0\neq f\in W_{a,b}$ and using (iv) gives $\theta_J=0$. Hence the $E_{Q,J}$ with $J\leqslant s_Q$ are linearly independent on $V_s$. With \eqref{theta} this gives (ii), and with Lemma \ref{lem:E}(b) and \eqref{vanish} it gives (iii).
\end{proof}

\subsubsection*{Consequences}
\begin{corollary}[(property 3, \eqref{indpnd}; $d\geqslant4$)]\label{cor:indep}
On $V_s$, the spans of the $Q$-fold TPOs for different $Q$ form a direct sum. In particular, all TPOs with $k+2m\leqslant n$ are linearly independent on $V_s$, which is \eqref{indpnd}.
\end{corollary}
\begin{proof}
Let $\sum_QF_Q=0$ on $V_s$, where $F_Q$ is a combination of $Q$-fold TPOs, and suppose some $F_Q$ is non-zero on $V_s$. Let $Q_0$ be the smallest such $Q$, and write $F_{Q_0}=\sum_{J\leqslant s_{Q_0}}\theta_JE_{Q_0,J}$ on $V_s$ by Lemma \ref{lem:E}(b) and \eqref{vanish}. On every $W_{a,b}$ with $a+b=Q_0$, the $F_Q$ with $Q>Q_0$ vanish by Lemma \ref{lem:G1}(iv), and those with $Q<Q_0$ vanish by the choice of $Q_0$. So $F_{Q_0}=0$ on these $W_{a,b}$. Every $J\leqslant s_{Q_0}$ with $J-Q_0$ even occurs as $b-a$ for one of them (Step 5), so Lemma \ref{lem:G1}(iv) gives $\theta_J=0$ for all such $J$, i.e.\ $F_{Q_0}=0$ on $V_s$, a contradiction. For $Q\leqslant n$ we have $s-Q\geqslant Q$, hence $s_Q=Q$ and $N_Q=\lfloor Q/2\rfloor+1$, which is the number of $Q$-fold TPOs; so each family is linearly independent by Lemma \ref{lem:G1}(iii), and \eqref{indpnd} follows.
\end{proof}

\begin{proposition}[(property 4, \eqref{reductionA}--\eqref{B-coef})]\label{prop:reduction}
Let $Q=2n-q$ with $q\geqslant0$ and $Q>n$, and put $l:=\lfloor q/2\rfloor$. Then $s_Q=s-Q\in\{q,q+1\}$, the admissible values of $J$ are $J_i:=q-2i$, $i=0,\dots,l$, and $N_Q=l+1$.
\begin{enumerate}
\item[(a)] For every $d$ and every $m>l$ with $Q-2m\geqslant0$, the relation \eqref{reductionA} holds on $V_s$ with the coefficients $B_j(m;n,l)$ of \eqref{B-coef}.
\item[(b)] ($d\geqslant4$) The operators $\mathcal{T}_{(Q-2j,j)}$, $j=0,\dots,l$, on the right-hand side of \eqref{reductionA} are linearly independent on $V_s$.
\end{enumerate}
\end{proposition}
\begin{proof}
Since $Q>n$, $s-Q\leqslant Q$, so $s_Q=s-Q=s-2n+q$. For $m\leqslant l$, \eqref{reductionA} is trivial. Let $m>l$ and put $u_i:=J_i+1$, $z_j:=Q-2j$ and $z_*:=Q-2m$. With $\zeta_m=1$ and $\zeta_j=-B_j$ for $j\leqslant l$, condition \eqref{moment} reads
\begin{equation}
\sum_{j=0}^l\frac{B_j}{u_i+z_j}=\frac{1}{u_i+z_*},\qquad i=0,\dots,l .
\label{cauchy}
\end{equation}
Here $u_i+z_j=2(n-i-j)+1$ and $u_i+z_*=2(n-i-m)+1$ are odd, hence non-zero. The $u_i$ are pairwise distinct, the $z_j$ are pairwise distinct, and $z_*\neq z_j$, so \eqref{cauchy} is a Cauchy system with non-vanishing determinant and has a unique solution. To find it, consider
\[
G(z):=\frac1{z+z_*}-\sum_{j=0}^l\frac{B_j}{z+z_j}.
\]
It vanishes at $z=u_0,\dots,u_l$, decays as $1/z$ and has simple poles at $-z_*$ and $-z_j$, so $G(z)=C\prod_i(z-u_i)\big/\big[(z+z_*)\prod_j(z+z_j)\big]$ with a constant $C$. Comparing the residues ($1$ at $-z_*$, $-B_j$ at $-z_j$) gives
\[
B_j=\prod_{k\neq j}\frac{z_k-z_*}{z_k-z_j}\cdot\prod_{i=0}^{l}\frac{z_j+u_i}{z_*+u_i} .
\]
With $z_k-z_*=2(m-k)$, $z_k-z_j=2(j-k)$, $z_j+u_i=2(n-i-j)+1$ and $\prod_{i=0}^l(X-i)=\Gamma(X+1)/\Gamma(X-l)$, this becomes
\[
B_j=\frac{m!}{j!}\,\frac{(-1)^{j+l}}{(m-j)\,\Gamma(l-j+1)\,\Gamma(m-l)}\cdot\frac{\Gamma\!\big(n+\tfrac32-j\big)}{\Gamma\!\big(n+\tfrac32-m\big)}\cdot\frac{\Gamma\!\big(n-l-m+\tfrac12\big)}{\Gamma\!\big(n-l-j+\tfrac12\big)} ,
\]
which is \eqref{B-coef} for $m>l$. Lemma \ref{lem:G1}(i) now gives (a). For (b), a vanishing combination of the $\mathcal{T}_{(Q-2j,j)}$, $j\leqslant l$, satisfies the homogeneous version of \eqref{cauchy} by Lemma \ref{lem:G1}(ii), so all its coefficients vanish.
\end{proof}

\begin{corollary}
If $s$ is odd, all $s$-fold TPOs vanish on $V_s$:
\[
\mathcal{T}_{(s-2m,m)}\big|_{V_s}=0,
\qquad 0\leqslant m\leqslant \lfloor s/2\rfloor .
\]
\end{corollary}
\begin{proof}
For $Q=s$, one has $s_Q=0$. Since $Q$ is odd, there is no integer $J$ satisfying $0\leqslant J\leqslant s_Q$ and $J-Q$ even. The conditions in Lemma~\ref{lem:G1}(i) are therefore vacuous, so every linear combination of the $s$-fold TPOs vanishes on $V_s$.
\end{proof}

\subsection{Deriving the horizontal projector}
\label{app:B3}
Recall that we study
\begin{equation}
    \label{B26}
        \begin{split}
             &\Gamma_{\alpha\beta}\Big( \mathcal{P}\big( f \big)\Big) = 0 \, , \\
            \forall \tilde{f} \in & \ker \Gamma_{\alpha\beta}: \mathcal{P}\big( \tilde{f} \big) = \tilde{f} \, ,
        \end{split}
\end{equation}
with $\mathcal{P}\big( f \big)$ having the following structure
\begin{equation}
    \mathcal{P}\big( f \big) = \sum_{k=0}^{+\infty}\sum_{m=0}^{+\infty}c_{(k, m)}(\hat{N}) \, \mathcal{T}_{(k,m)} f
\end{equation}
Let us first work out the second requirement in \eqref{B26}. For $\tilde{f} \in \ker \Gamma_{\alpha\beta}$ we have 
\begin{equation}
    \label{B28}
    \mathcal{P} \big( \tilde{f} \big) = c_{(0, 0)}(\hat{N}) \cdot \tilde{f} \implies c_{(0, 0)}(\hat{N}) \overset{!}{=} 1 \, .
\end{equation}
Using \eqref{B4}, it is straightforward to derive that the first equation in \eqref{B26} is actually
    \begin{equation}
            \sum_{k=0}^{+\infty}\sum_{m=0}^{+\infty} \mathscr{A}_{(k, m)}(\hat{N}) \cdot \mathcal{T}_{(k,m)}\big(\Gamma_{\alpha\beta} f\big)+\sum_{k=0}^{+\infty}\sum_{m=0}^{+\infty} \mathscr{B}_{(k, m)}(\hat{N}) \cdot t_{\alpha\beta} \, \mathcal{T}_{(k,m)}\big(\Gamma^{\sigma\lambda} \Gamma_{\sigma\lambda} f\big) = 0 \, .
    \end{equation}
with 
\begin{equation}
    \label{B30}
    \mathscr{A}_{(k, m)}(\hat{N}) = 2(k+1)(d+\hat{N}-2k-4m-3) \cdot c_{(k+1, m)}(\hat{N}+2) + c_{(k, m)}(\hat{N}+2) \, ,
\end{equation}
and
\begin{equation}
    \label{B31}
    \begin{split}
        \mathscr{B}_{(k, m)}(\hat{N}) = 4(m&+1)(d+\hat{N}-2m-2) \cdot c_{(k, m+1)}(\hat{N}+2) + \,  \\
        &+2(k+1)(k+2) \cdot c_{(k+2, m)}(\hat{N}+2) \, .
    \end{split}
\end{equation}
Substituting this into \eqref{rqrmnt} and combining with the initial condition \eqref{B28} yields the following recurrence system
\begin{equation}
    \label{rcr-appendix}
    \scalebox{0.92}{ $
    \begin{split} 
        & c_{(0, 0)}(2s) = 1 \, , \\
        &k+2m\leqslant s - 1:  2(k+1)(d+2s-2k-4m-5) \cdot c_{(k+1, m)}(2s) + c_{(k, m)}(2s) = 0 \, , \\
        &k+2m \leqslant s - 2:  (k+1)(k+2) c_{(k+2, m)}(2s)+2(m+1)(d+2s-2m-4) c_{(k, m+1)}(2s) = 0 \, .
    \end{split}  $}
\end{equation}
We start by analysing the 1$^{\text{st}}$ equation in \eqref{rcr-appendix}. It expresses $c_{(k+1, m)}(2s)$ in terms of $c_{(k, m)}(2s)$, if $d + 2s - 2k - 4m - 5 \neq 0$. However, due to the constraint $k+2m \leqslant s-1$ we have 
\begin{equation}
    d + 2(s-k-2m) - 5 \geqslant d - 3 \, ,
\end{equation}
and so for $d > 3$ it is always non-zero in our equations. Notice also that within the imposed constraints on $(k, m)$ it allows one to find all $c_{(k, m)}(2s)$ for $k+2m \leqslant s$ in terms of $c_{(0, m)}(2s)$ as 
\begin{equation}
    \label{B34}
    c_{(k, m)}(2s) = \frac{(-)^k}{4^k \cdot k!} \cdot \frac{\Gamma\big( \frac{d+2s-3}{2} - k -2m \big)}{\Gamma\big( \frac{d+2s-3}{2} -2m \big)} \cdot c_{(0, m)}(2s) \, .
\end{equation}
Substituting this into the 2$^{\text{nd}}$ equation in \eqref{rcr-appendix} yields 
\begin{equation}
    \Gamma\Big( \frac{d+2s-3}{2} - k - 2 -2m\Big) \cdot \Big( \frac{\hat{c}_m}{4^3} + (m+1) \cdot \Big(\frac{d+2s-4}{2} -m \Big) \cdot \hat{c}_{m+1} \Big) = 0 \, .
\end{equation}
Here we introduced $\hat{c}_m := c_{(0, m)}(2s) / \Gamma\big( \frac{d+2s-3}{2} -2m \big)$ for convenience. Note that since in this equation $k+2m \leqslant s-2$ and we work with $d > 3$, the argument of the gamma function in this equation satisfies
\begin{equation}
    \frac{d+2s-3}{2} - k - 2 -2m > 0 \, ,
\end{equation}
and so it must be that 
\begin{equation}
    \label{B37}
    \frac{\hat{c}_m}{4^3} + (m+1) \cdot \Big(\frac{d+2s-4}{2} -m \Big) \cdot \hat{c}_{m+1} = 0 \, .
\end{equation}
Again, from $k+2m \leqslant s-2$ and $d > 3$ it follows that 
\begin{equation}
    \frac{d+2s-4}{2} -m > 0 \, ,
\end{equation}
and so \eqref{B37} is well-defined as a recurrence equation, with solution 
\begin{equation}
    \label{B39}
    \hat{c}_m = \frac{(-)^m}{4^{3m} \cdot m! } \cdot \frac{\Gamma \big( \frac{d+2s-2}{2} - m \big)}{\Gamma\big(\frac{d+2s-2}{2} \big)} \cdot \hat{c}_0 \, .
\end{equation}
Combining \eqref{B34} with \eqref{B39} and the initial condition $c_{(0, 0)}(2s) = 1$ yields
\begin{equation}
    c_{(k, m)}(2s) = \frac{\Gamma\big( \frac{d+2s-3}{2} - k -2m \big) \Gamma\big(\frac{d+2s-2}{2} -m \big)}{\Gamma\big( \frac{d+2s-3}{2}\big) \Gamma\big(\frac{d+2s-2}{2}\big)} \cdot \frac{(-)^{k+m}}{k! \cdot m! \cdot 2^{2k+6m}} \, .
\end{equation}
\section{Constructing the splitting}
\label{app:C}
In this appendix, we prove the auxiliary results used in Section~\ref{sec:factoring}.

\subsection{Properties used to prove the uniqueness of the splitting}

We first prove some auxiliary technical formulas. Direct calculation gives

\begin{equation}
    \label{C1}
    \Gamma_{\alpha\beta} \Big( (t^{\sigma\lambda}t_{\sigma\lambda})^{m} f_s\Big) = 4m(d+2s + 2m-2) \cdot t_{\alpha\beta} \, (t^{\sigma\lambda}t_{\sigma\lambda})^{m-1} f_s + (t^{\sigma\lambda}t_{\sigma\lambda})^{m} \, \Gamma_{\alpha\beta} f_s \, ,
\end{equation}
in particular implying 
\begin{equation}
    \label{C2}
        \Gamma_{\alpha\beta} \Big( (t^{\sigma\lambda}t_{\sigma\lambda})^{m}\Big) = 4m(d + 2m-2) \cdot t_{\alpha\beta} \, (t^{\sigma\lambda}t_{\sigma\lambda})^{m-1}
\end{equation}
Using \eqref{t-star:left}, we obtain 
\begin{equation}
    \label{C3}
    t^{\alpha\beta} \ast t_{\alpha\beta} = t^{\alpha\beta}t_{\alpha\beta} - \Gamma^{\alpha\beta }t_{\alpha\beta} = t^{\alpha\beta}t_{\alpha\beta} - 6(d+1) \, ,
\end{equation}
which, after simplification using \eqref{C3} and \eqref{B4}--\eqref{B6}, gives 
\begin{equation}
    \label{C4}
    \scalebox{0.92}{$
    \begin{split}
        \big(t^{\alpha\beta}t_{\alpha\beta} \big) \ast \Big[ \big( t^{\sigma\lambda}t_{\sigma\lambda}  & \big)^{m} \cdot \tilde{f}_s \Big] = \big( t^{\sigma\lambda}t_{\sigma\lambda} \big)^{m+1} \cdot \tilde{f}_s-\Big(8m(d+2s+2m+1)+12s \Big) \cdot \big( t^{\sigma\lambda}t_{\sigma\lambda} \big)^{m} \cdot \tilde{f}_s  + \\
        &+8m(2m+1)(d+2m+2s-1)(d+2m+2s-2) \big( t^{\sigma\lambda}t_{\sigma\lambda} \big)^{m-1} \cdot \tilde{f}_s \, .
    \end{split} $}
\end{equation}
for $\tilde{f_s} \in \ker \Gamma_{\alpha\beta}: \hat{N} \tilde{f_s} = 2s \cdot \tilde{f_s}$. We use this identity to establish properties of the functions introduced in \eqref{quasiprojector}, beginning with the quasiprojector itself 
\begin{equation}
    \label{C5}
    \Delta(t^{\mu\nu}t_{\mu\nu}; 0) = \sum_{m=0}^{+\infty} \frac{\Gamma(d/2)}{\Gamma(m+d/2)} \cdot \frac{\big(t^{\mu\nu}t_{\mu\nu} / 8 \big)^m}{m!} \, ,
\end{equation}
and directly calculate its product with $t_{\alpha\beta}$ using \eqref{C2}
\begin{equation}
    \begin{split}
        & t_{\alpha\beta} \ast \Delta(t^{\mu\nu}t_{\mu\nu}; 0) = t_{\alpha\beta} \cdot\Delta(t^{\mu\nu}t_{\mu\nu}; 0) - \Gamma_{\alpha\beta} \,  \Delta(t^{\mu\nu}t_{\mu\nu}; 0) = \\
        & = t_{\alpha\beta} \cdot \Gamma(d/2)  \sum_{m=0}^{+\infty} \bigg(\frac{1}{\Gamma(m+d/2)} - \frac{m+d/2}{\Gamma(m+1+d/2)} \bigg)\cdot \frac{\big(t^{\mu\nu}t_{\mu\nu} / 8 \big)^m}{m!} \equiv 0 \, ,
    \end{split}
\end{equation}
due to $\Gamma(z+1) = z \cdot \Gamma(z)$.

Proving \eqref{qsprjctr:nice} is more involved. First, let us prove the following auxiliary statement 
\begin{lemma}
\begin{equation}
    \label{C7}
    \Big[ \big(t_{\sigma\lambda}t^{\sigma\lambda} \big)^m \Big] \ast  \tilde{f_s} = \sum_{k=0}^{m} (-8)^{m-k} \cdot \frac{s!}{(s-m+k)!} \cdot \frac{m!}{(m-k)! \cdot k!} \cdot \frac{\Gamma\Big(m+\frac{3}{2} \Big)}{\Gamma\Big(k+\frac{3}{2} \Big)} \cdot \big(t_{\sigma\lambda}t^{\sigma\lambda} \big)^k \cdot \tilde{f}_s \, .
\end{equation}
\end{lemma}
\begin{proof}
For $m\geqslant0$ and $k\in\mathbb Z$ put
\begin{equation}
    \label{C8}
    g_k(m;s) := (-8)^{m-k}\,\frac{s!}{(s-m+k)!}\,\binom{m}{k}\,
    \frac{\Gamma\big(m+\frac32\big)}{\Gamma\big(k+\frac32\big)} \, ,
\end{equation}
with $\binom{m}{k}=0$ for $k<0$ or $k>m$, and $g_k(-1;s):=0$. Here
$s!/(s-m+k)! = s(s-1)\cdots(s-m+k+1)$ is a polynomial in $s$ that vanishes
for integer $0\leqslant s<m-k$, so the sum in \eqref{C7} may be taken over
$0\leqslant k\leqslant m$.

We use induction on $m$; for $m=0$, \eqref{C7} is trivial. By \eqref{C4}
with $s=0$ and $\tilde f_0=1$,
\begin{equation}
    \begin{split}
    \big(t_{\sigma\lambda}t^{\sigma\lambda}\big)^{m+1}
    = {}& \big(t_{\alpha\beta}t^{\alpha\beta}\big)\ast\big(t_{\sigma\lambda}t^{\sigma\lambda}\big)^{m}
    + 8m(d+2m+1)\big(t_{\sigma\lambda}t^{\sigma\lambda}\big)^{m} \\
    & - 8m(2m+1)(d+2m-1)(d+2m-2)\big(t_{\sigma\lambda}t^{\sigma\lambda}\big)^{m-1} .
    \end{split}
\end{equation}
Multiply by $\tilde f_s$ from the right with $\ast$, use associativity,
\eqref{C7} for $m$ and $m-1$, and \eqref{C4} for each
$\big(t_{\alpha\beta}t^{\alpha\beta}\big)\ast\big[\big(t_{\sigma\lambda}t^{\sigma\lambda}\big)^k\tilde f_s\big]$.
The coefficient of $\big(t_{\sigma\lambda}t^{\sigma\lambda}\big)^k\tilde f_s$
in $\big(t_{\sigma\lambda}t^{\sigma\lambda}\big)^{m+1}\ast\tilde f_s$ is then
\begin{equation}
    \label{C10}
    \begin{split}
    g'_k = {}& g_{k-1}(m;s) - \big(8k(d+2s+2k+1)+12s\big)\,g_k(m;s) \\
    & + 8(k+1)(2k+3)(d+2s+2k+1)(d+2s+2k)\,g_{k+1}(m;s) \\
    & + 8m(d+2m+1)\,g_k(m;s) - 8m(2m+1)(d+2m-1)(d+2m-2)\,g_k(m-1;s) \, .
    \end{split}
\end{equation}
For $k=m+1$ only the first term survives, and $g'_{m+1}=g_m(m;s)=1=g_{m+1}(m+1;s)$.
For $0\leqslant k\leqslant m$, both $g'_k$ and $g_k(m+1;s)$ are polynomials in
$s$, so it suffices to prove $g'_k=g_k(m+1;s)$ for non-integer $s$, where
$g_k(m;s)\neq0$. From \eqref{C8} and $\Gamma(x+1)=x\,\Gamma(x)$,
\begin{align*}
    g_{k-1}(m;s) &= -\frac{4k(2k+1)(s-m+k)}{m-k+1}\,g_k(m;s) , \\
    8(k+1)(2k+3)\,g_{k+1}(m;s) &= -\frac{2(m-k)}{s-m+k+1}\,g_k(m;s) , \\
    8m(2m+1)\,g_k(m-1;s) &= -\frac{2(m-k)}{s-m+k+1}\,g_k(m;s) , \\
    g_k(m+1;s) &= -\frac{4(s-m+k)(m+1)(2m+3)}{m-k+1}\,g_k(m;s) ,
\end{align*}
where the first three also hold at $k=0$, $k=m$ and $m=0$, both sides vanishing.
Hence the third and fifth terms of \eqref{C10} add up to
\begin{equation*}
    \begin{split}
    &\frac{2(m-k)}{s-m+k+1}\Big[(d+2m-1)(d+2m-2)-(d+2s+2k+1)(d+2s+2k)\Big]g_k(m;s) \\
    &\qquad = 2(m-k)\big(2-4(d+m+s+k)\big)\,g_k(m;s) ,
    \end{split}
\end{equation*}
since the bracket equals $(s-m+k+1)\big(2-4d-4(m+s+k)\big)$. Adding the second
and fourth terms, the $d$-dependence cancels:
\begin{equation*}
    \begin{split}
    &\Big[-8k(d+2s+2k+1)-12s+8m(d+2m+1)+2(m-k)\big(2-4(d+m+s+k)\big)\Big] \\
    &\qquad = -4(s-m+k)(2m+2k+3) .
    \end{split}
\end{equation*}
With the first term,
\begin{equation*}
    \begin{split}
    \frac{g'_k}{g_k(m;s)}
    &= -\frac{4(s-m+k)\big[k(2k+1)+(2m+2k+3)(m-k+1)\big]}{m-k+1} \\
    &= -\frac{4(s-m+k)(m+1)(2m+3)}{m-k+1}
    = \frac{g_k(m+1;s)}{g_k(m;s)} ,
    \end{split}
\end{equation*}
since $k(2k+1)+(2m+2k+3)(m-k+1)=(m+1)(2m+3)$. This proves \eqref{C7} for $m+1$.
\end{proof}

Since \eqref{C7} holds for all $m , s \in \mathds{N} \cup \{ 0 \}$, for any $\mathcal{F}\big(t^{\mu\nu}t_{\mu\nu}\big)$ expandable as a series in $t^{\mu\nu}t_{\mu\nu} $ the following must be true 
\begin{equation}
    \forall \tilde{f}_s \in \ker \Gamma_{\alpha\beta}: \hat{N} \tilde{f}_s = 2s \cdot \tilde{f}_s \hookrightarrow \mathcal{F}\big(t^{\mu\nu}t_{\mu\nu}\big) \ast \tilde{f}_s = \mathcal{F}\big(t^{\mu\nu}t_{\mu\nu} ; s\big) \cdot \tilde{f}_s \, ,
\end{equation}
with some $\mathcal{F}\big(t^{\mu\nu}t_{\mu\nu} ; s\big)$. Therefore there exists a family of functions $\Delta\big( t^{\mu\nu}t_{\mu\nu} ; s\big)$ s.t.
\begin{equation}
    \Delta\big( t^{\mu\nu}t_{\mu\nu} ; 0\big) \ast \tilde{f}_s = \Delta\big( t^{\mu\nu}t_{\mu\nu} ; s\big) \cdot \tilde{f}_s \, .
\end{equation}
These functions are naturally represented as series
\begin{equation}
    \Delta\big( t^{\mu\nu}t_{\mu\nu} ; s\big) = \sum_{m=0}^{+\infty} \xi_m(s) \cdot \big( t^{\mu\nu}t_{\mu\nu}\big)^m \, .
\end{equation}
In order to deduce these coefficients $\xi_m(s)$, we invoke the central property of $\Delta\big( t^{\mu\nu}t_{\mu\nu} ; 0\big)$ given by \eqref{qsprjctr:central}. From that it follows that for any $s \in \mathds{N} \cup \{0\}$ the following must be true 
\begin{equation}
    t_{\alpha\beta} \ast \Big( \Delta\big( t^{\mu\nu}t_{\mu\nu} ; s\big) \cdot \tilde{f}_s \Big) = 0 \, .
\end{equation}
Using \eqref{C1} it translates into the following recurrence system for $\xi_m(s)$
\begin{equation}
    \xi_m(s) - 4(m+1)(d+2s+2m) \cdot \xi_{m+1}(s) = 0 \, .
\end{equation}
Provided $s, d \in \mathds{N} \cup \{0 \}$ it has a unique solution
\begin{equation}
    \label{C16}
    \xi_m(s) = \frac{\xi_0(s) \, \Gamma(s+d/2)}{\Gamma(m+s+d/2)} \cdot \frac{1}{m! \cdot 8^m} \, .
\end{equation}
On the other hand, direct combination of \eqref{C5} with \eqref{C7} yields 
\begin{equation}
    \xi_k(s) = \sum_{m=k}^{+\infty} \frac{\Gamma(d/2)}{\Gamma(m+d/2)} \cdot \frac{g_k(m; s)}{m! \cdot 8^m} \, .
\end{equation}
However, actually all of these series terminate for some $m$. Indeed, due to the definition of $g_k(m; s)$ it follows that for $s-m+k < 0$ all coefficients are just $0$, and therefore 
\begin{equation}
    \label{C18}
    \xi_k(s) = \frac{\Gamma(d/2)}{\Gamma(k+3/2) \cdot 8^k} \cdot \sum_{m=k}^{s+k} \frac{\Gamma(m+3/2)}{\Gamma(m+d/2)} \cdot \frac{s!}{(s-m+k)!} \cdot \frac{(-1)^{m-k}}{(m-k)! \cdot k!} \, .
\end{equation}
To evaluate these sums, we consider the following cases.
\begin{itemize}
    \item \textbf{s = 0}. For $s=0$ the sum above is trivial to calculate: 
    \begin{equation}
        \begin{split}
            \xi_k(0) = \frac{\Gamma(d/2)}{\Gamma(k+3/2) \cdot 8^k } \cdot &\sum_{m=k}^{k} \frac{\Gamma(m+3/2)}{\Gamma(m+d/2)} \cdot \frac{0!}{(0-m+k)!} \cdot \frac{(-1)^{m-k}}{(m-k)! \cdot k!} = \\
            &=\frac{\Gamma(d/2)}{\Gamma(k+d/2)} \cdot \frac{1}{k! \cdot 8^k} \, , 
        \end{split}
        \end{equation}
    as expected, since $s= 0$ corresponds to multiplication of $\Delta\big( t^{\mu\nu}t_{\mu\nu} ; 0\big)$ with some $Y$-independent function, and in this case the star-product reduces to the point-wise product. \\
    \item \textbf{s $>$ 0}.
For every $d$, the sum in \eqref{C18} can be brought to the form
        \begin{equation}
            \xi_k(s) = \frac{\Gamma(d/2)}{\Gamma(k+3/2) \cdot k! \cdot 8^k} \cdot \sum_{m=0}^{s} \frac{\Gamma(m+k+3/2)}{\Gamma(m+k+d/2)} \cdot \frac{s!}{(s-m)!} \cdot \frac{(-1)^m}{m!} \, ,
        \end{equation}
which is a terminating hypergeometric series,
        \begin{equation}
            \xi_k(s) = \frac{\Gamma(d/2)}{k! \cdot 8^k  \cdot \Gamma(k+d/2)} \cdot {}_2F_{1} \bigg( \begin{array}{c}
        {-s \, \, , \, \, k+3/2 } \\
        {k+d/2}\\
    \end{array} \, \Bigg| \, 1\bigg) \, .
        \end{equation}
For a non-negative integer $s$, the Chu--Vandermonde identity ${}_2F_1(-s, b; c; 1) = (c-b)_s/(c)_s$, with $(x)_s := x(x+1)\cdots(x+s-1)$, holds for all $b$ and all $c$ that are not non-positive integers. With $b = k+3/2$ and $c = k+d/2$ it gives
        \begin{equation}
            \xi_k(s) = \frac{\Gamma\big(\frac{d}{2} \big)}{\Gamma\big(k+s+\frac{d}{2}\big)} \cdot \frac{1}{k! \cdot 8^k} \cdot \Big( \frac{d-3}{2} \Big)_s \, .
        \end{equation}
For $d>3$, $\big(\frac{d-3}{2}\big)_s = \Gamma\big(s+\frac{d-3}{2}\big)/\Gamma\big(\frac{d-3}{2}\big)$, which is consistent with \eqref{C16}. For $d=3$, $(0)_s = 0$ for $s>0$, so $\xi_k(s)\big|_{d=3} = 0$, which yields the important result \eqref{qsprjctr:d3}.
\end{itemize}

\subsection{Deriving the coefficients $r_m(\hat{N})$}
Here we provide some auxiliary statements that allow us to find $r_m(\hat{N})$. First, we need
\begin{lemma}
\begin{equation}
    \forall \tilde{f}_s \in \ker \Gamma_{\alpha\beta}: \hat{N} \tilde{f}_s = 2s \cdot \tilde{f}_s \hookrightarrow \big( \Gamma_{\alpha\beta} \Gamma^{\alpha\beta} \big)^m \Big[ \big( t_{\sigma\lambda} t^{\sigma\lambda}  \big)^k \, \tilde{f}_s\Big] = \beta_{(m, k)}(s) \cdot \big( t_{\sigma\lambda} t^{\sigma\lambda}  \big)^{k-m} \, \tilde{f}_s \, ,
\end{equation}
where 
\begin{equation}
    \beta_{(m, k)}(s) = 2^{6m} \cdot \frac{\Gamma(k+1)\Gamma(k+\frac{3}{2})}{\Gamma(k-m+1)\Gamma(k-m+\frac{3}{2})} \cdot \frac{\Gamma(\frac{d+2s}{2}+k) \Gamma(\frac{d+2s+1}{2}+k)}{\Gamma(\frac{d+2s}{2}+k-m)\Gamma(\frac{d+2s+1}{2}+k-m)} \, ,
\end{equation}
in particular implying 
\begin{equation}
    \forall m>k: \,  \beta_{(m, k)}(s) = 0 \, .
\end{equation}
\end{lemma}
\begin{proof}
    In order to prove this, we first notice that from \eqref{C2} and \eqref{B6} there is the following consequence 
    \begin{equation}
        \begin{split}
            &\forall \tilde{f}_s \in \ker \Gamma_{\alpha\beta}: \hat{N} \tilde{f}_s = 2s \cdot \tilde{f}_s \hookrightarrow \big( \Gamma_{\alpha\beta} \Gamma^{\alpha\beta} \big) \Big[ \big( t_{\sigma\lambda} t^{\sigma\lambda}  \big)^k \, \tilde{f}_s\Big] =  \\
            &= 8k(2k+1)(d+2s+2k-2)(d+2k+2s-1) \cdot \big( t_{\sigma\lambda} t^{\sigma\lambda}  \big)^{k-1} \, \tilde{f}_s  =\\
            &=2^{6} \cdot \frac{\Gamma(k+1)\Gamma(k+\frac{3}{2})}{\Gamma(k)\Gamma(k+\frac{1}{2})} \cdot \frac{\Gamma(\frac{d+2s}{2}+k) \Gamma(\frac{d+2s+1}{2}+k)}{\Gamma(\frac{d+2s}{2}+k-1)\Gamma(\frac{d+2s+1}{2}+k-1)} \cdot \big( t_{\sigma\lambda} t^{\sigma\lambda}  \big)^{k-1} \, \tilde{f}_s
        \end{split}
    \end{equation}
This serves as the base of the induction proof. The induction step is straightforward to prove for $m\leqslant k$, using the formula above alongside properties of $\Gamma(z)$. \\
For $m>k$ it is enough to notice that 
\begin{equation}
    \forall \tilde{f}_s \in \ker \Gamma_{\alpha\beta}: \hat{N} \tilde{f}_s = 2s \cdot \tilde{f}_s \hookrightarrow \big( \Gamma_{\alpha\beta} \Gamma^{\alpha\beta} \big)^k \Big[ \big( t_{\sigma\lambda} t^{\sigma\lambda}  \big)^k \, \tilde{f}_s\Big] = \kappa \, \tilde{f}_s \, ,
\end{equation}
with a constant $\kappa$.
\end{proof}
Another very useful property is that 
\begin{lemma}
\begin{equation}
    \label{C29}
    \forall \tilde{f}_s \in \ker \Gamma_{\alpha\beta}: \hat{N} \tilde{f}_s = 2s \cdot \tilde{f}_s \hookrightarrow \Delta\big(t^{\mu\nu}t_{\mu\nu};0 \big) \ast \Big[ (t^{\alpha\beta}t_{\alpha\beta})^m \tilde{f}_s \Big] =  h_m(s)\cdot \Delta\big(t^{\mu\nu}t_{\mu\nu};0 \big) \ast  \tilde{f}_s \, ,
\end{equation}
with 
\begin{equation}
    h_m(s) = 8^m \cdot \frac{\Gamma\big( m + \frac{3}{2} \big)}{\Gamma\big( \frac{3}{2} \big)} \cdot \frac{\Gamma\big( \frac{d+2s+1}{2}+m \big)}{\Gamma\big(\frac{d+2s+1}{2} \big)} \, .
\end{equation}
\end{lemma}
\begin{proof}
    For $m=0$ this relation is obviously true, which we use as a base of our induction proof.
    Now, let us choose any $\tilde{f}_s \in \ker \Gamma_{\alpha\beta}: \hat{N} \tilde{f}_s = 2s \cdot \tilde{f}_s$. Then, from \eqref{C3} and \eqref{C4} we have 
    \begin{equation}
        \label{C31}
        \begin{split}
        &\big(t_{\alpha\beta} \ast t^{\alpha\beta} \big) \ast \Big[ \big(t_{\sigma\lambda}t^{\sigma\lambda}  \big)^m \cdot \tilde{f}_s \Big] = \\
        &=\big(t_{\sigma\lambda}t^{\sigma\lambda}  \big)^{m+1} \cdot \tilde{f}_s - \Big(8m(d+2s+2m+1)+6(d+2s+1) \Big) \cdot \big(t_{\sigma\lambda}t^{\sigma\lambda}  \big)^m \cdot \tilde{f}_s + \, \\
        & + 8m(2m+1)(d+2m+2s-1)(d+2m+2s-2) \cdot \big(t_{\sigma\lambda}t^{\sigma\lambda}  \big)^{m-1} \cdot \tilde{f}_s \, .
        \end{split} 
    \end{equation}
Suppose \eqref{C29} holds for some $m$ and $m-1$. Then multiplying \eqref{C31} by $\Delta\big( t^{\mu\nu}t_{\mu\nu}; 0 \big)$ gives 
\begin{equation}
    \begin{split}
       &0 = \Delta\big( t^{\mu\nu}t_{\mu\nu}; 0 \big) \ast \Big[ \big(t_{\sigma\lambda}t^{\sigma\lambda}  \big)^{m+1} \cdot \tilde{f}_s\Big]  -  \\
       &-\Big(8m(d+2s+2m+1)+6(d+2s+1) \Big) \cdot h_m(s) \cdot \Delta\big( t^{\mu\nu}t_{\mu\nu}; 0 \big) \ast \tilde{f}_s + \\
       &+8m(2m+1)(d+2m+2s-1)(d+2m+2s-2)\cdot h_{m-1}(s) \cdot \Delta\big( t^{\mu\nu}t_{\mu\nu}; 0 \big) \ast \tilde{f}_s \, .
    \end{split}
\end{equation}
Due to properties of $\Gamma(z)$, it directly yields 
\begin{equation}
    \Delta\big( t^{\mu\nu}t_{\mu\nu}; 0 \big) \ast \Big[ \big(t_{\sigma\lambda}t^{\sigma\lambda}  \big)^{m+1} \cdot \tilde{f}_s\Big]  = h_{m+1}(s) \cdot  \Delta\big( t^{\mu\nu}t_{\mu\nu}; 0 \big) \ast \tilde{f}_s \, .
\end{equation}
\end{proof}

\section{Evaluating $M_{AB} \circ M_{CD}$}
\label{app:MM}
Here we derive the products \eqref{MM-star} and \eqref{MM-circ} used in Section~\ref{sec:so-generators}. Throughout, $M_{AB} = \frac{i}{2}\epsilon_{\alpha\beta}Y^\alpha_A Y^\beta_B$.

\subsubsection*{The star product of two generators}
The star product \eqref{Moyal} reads
\begin{equation}
    f \ast g = f \exp\Big( i\,\epsilon^{\delta\gamma}\eta_{CD}\, \frac{\overleftarrow{\partial}}{\partial Y^\delta_C}\,\frac{\overrightarrow{\partial}}{\partial Y^\gamma_D} \Big) g \, ,
\end{equation}
in particular $\big[ Y^\alpha_A \, , \, Y^\beta_B\big]_\ast = 2i\,\epsilon^{\alpha\beta}\eta_{AB}$. Since each power of the operator differentiates each factor once, the expansion of $M_{AB}\ast M_{CD}$ terminates at the second order.

For the first order, using $\partial M_{AB} / \partial Y^\delta_P = \frac{i}{2}\big(\epsilon_{\delta\beta}\delta^P_A Y^\beta_B + \epsilon_{\alpha\delta}\delta^P_B Y^\alpha_A\big)$ together with $\epsilon^{\delta\gamma}\epsilon_{\delta\beta} = \delta^\gamma_\beta$ and $\epsilon_{\alpha\delta}\epsilon^{\delta\gamma} = -\delta^\gamma_\alpha$, one finds
\begin{equation}
    M_{AB}\,\epsilon^{\delta\gamma}\eta_{PQ}\frac{\overleftarrow{\partial}}{\partial Y^\delta_P}\frac{\overrightarrow{\partial}}{\partial Y^\gamma_Q}\,M_{CD} = \frac{i}{2}\big(\eta_{AC}M_{BD} - \eta_{AD}M_{BC} - \eta_{BC}M_{AD} + \eta_{BD}M_{AC}\big) \, .
\end{equation}
\emergencystretch=2em
For the second order, $\partial^2 M_{AB} / \partial Y^\delta_P \partial Y^{\delta'}_{P'} = \frac{i}{2}\epsilon_{\delta\delta'}\big( \delta^P_A \delta^{P'}_B - \delta^{P'}_A\delta^P_B \big)$; the $\mathfrak{sp}(2)$ factor is $\epsilon^{\delta\gamma}\epsilon^{\delta'\gamma'}\epsilon_{\delta\delta'}\epsilon_{\gamma\gamma'} = \epsilon^{\gamma\gamma'}\epsilon_{\gamma\gamma'} = 2$ and the $\so$ factor is $2\big(\eta_{AC}\eta_{BD} - \eta_{AD}\eta_{BC}\big)$, so that
\begin{equation}
    M_{AB}\Big(\epsilon^{\delta\gamma}\eta_{PQ}\frac{\overleftarrow{\partial}}{\partial Y^\delta_P}\frac{\overrightarrow{\partial}}{\partial Y^\gamma_Q}\Big)^2 M_{CD} = -\big(\eta_{AC}\eta_{BD} - \eta_{AD}\eta_{BC}\big) \, .
\end{equation}
\emergencystretch=0pt
Collecting the three terms yields \eqref{MM-star}. Its antisymmetric part reproduces the commutator of Section~\ref{sec:star}, which fixes the relative normalisation.

\subsubsection*{The induced product}
The right-hand side of \eqref{MM-star} consists of pieces of degree $4$, $2$ and $0$ in $Y^\alpha_A$. Since $\Gamma_{\mu\nu}M_{AB} = 0$, the last two are annihilated by $\Gamma_{\alpha\beta}\Gamma^{\alpha\beta}$ and left invariant by $\mathcal{P}\big( \bullet \big)$, so they pass through \eqref{circ-prod} unchanged. It remains to evaluate \eqref{circ-prod} on
\begin{equation}
    X_{AB, CD} := M_{AB}M_{CD} \, .
\end{equation}
One has $\partial M_{AB}/\partial Y^\mu_P = -\frac{i}{2}\big( \delta^P_A Y_{\mu B} - \delta^P_B Y_{\mu A}\big)$, and, since $\Gamma_{\mu\nu}M_{AB}=0$, only the cross terms survive in
\begin{equation}
    \Gamma_{\mu\nu}X_{AB,CD} = -\frac{1}{4}\Big[ \eta_{AC}\big( Y_{\mu B}Y_{\nu D} + Y_{\nu B}Y_{\mu D}\big) - \eta_{AD}\big( \cdots \big) - \eta_{BC}\big( \cdots \big) + \eta_{BD}\big( Y_{\mu A}Y_{\nu C} + Y_{\nu A}Y_{\mu C}\big) \Big] \, .
\end{equation}
Contracting with $t^{\mu\nu}$ and using $t^{\mu\nu}Y_{\mu B}Y_{\nu D} = -4\,\eta^{PQ}M_{PB}M_{QD}$ gives
\begin{equation}
    \begin{split}
    t^{\mu\nu}\Gamma_{\mu\nu} X_{AB,CD} = 2\Big( &\eta_{AC}\eta^{PQ}M_{PB}M_{QD} - \eta_{AD}\eta^{PQ}M_{PB}M_{QC} \\
    &- \eta_{BC}\eta^{PQ}M_{PA}M_{QD} + \eta_{BD}\eta^{PQ}M_{PA}M_{QC} \Big) \, .
    \end{split}
\end{equation}
Applying a second trace operator, with $\Gamma_{\alpha\beta}\big( Y_{\mu B}Y_{\nu D}\big) = \eta_{BD}\big( \epsilon_{\mu\alpha}\epsilon_{\nu\beta} + \epsilon_{\mu\beta}\epsilon_{\nu\alpha}\big)$, one obtains
\begin{equation}
    \Gamma_{\mu\nu}\Gamma_{\alpha\beta}X_{AB,CD} = -\big( \epsilon_{\alpha\mu}\epsilon_{\beta\nu} + \epsilon_{\alpha\nu}\epsilon_{\beta\mu}\big)\big( \eta_{AC}\eta_{BD} - \eta_{AD}\eta_{BC}\big) \, ,
\end{equation}
and hence, using $\epsilon^{\alpha\mu}\epsilon^{\beta\nu}\epsilon_{\alpha\mu}\epsilon_{\beta\nu} = 4$, $\epsilon^{\alpha\mu}\epsilon^{\beta\nu}\epsilon_{\alpha\nu}\epsilon_{\beta\mu} = 2$ and $t^{\alpha\beta}t^{\mu\nu}\epsilon_{\alpha\mu}\epsilon_{\beta\nu} = t^{\alpha\beta}t_{\alpha\beta}$,
\begin{equation}
    \label{D8}
    \Gamma^{\alpha\beta}\Gamma_{\alpha\beta}X_{AB,CD} = -6\big( \eta_{AC}\eta_{BD} - \eta_{AD}\eta_{BC}\big) \, , \quad
    \mathcal{T}_{(2,0)} X_{AB,CD} = -2\,\big(t^{\sigma\lambda}t_{\sigma\lambda}\big)\big( \eta_{AC}\eta_{BD} - \eta_{AD}\eta_{BC}\big) \, .
\end{equation}
The ratio of the two expressions in \eqref{D8} is $3$, in agreement with $B_0(1;1,0)$ of \eqref{B-coef}, as it must be, since for $\hat{N}=4$ one has $n = \lfloor s/2 \rfloor = 1 < k+2m = 2$ and the two double-trace operators are related by \eqref{reductionA}.

The relevant coefficients of \eqref{c} at $\hat{N}=4$ are
\begin{equation}
    c_{(1,0)}(4) = -\frac{1}{2(d-1)} \, , \quad c_{(2,0)}(4) = \frac{1}{8(d-1)(d-3)} \, , \quad c_{(0,1)}(4) = -\frac{1}{8d(d-1)(d-3)} \, ,
\end{equation}
and, after the reduction \eqref{reductionA}, the coefficients enter only through the combination
\begin{equation}
    c_{(2,0)}(4) + 3\,c_{(0,1)}(4) = \frac{1}{8d(d-1)} \, .
\end{equation}
Therefore
\begin{equation}
    \begin{split}
    \mathcal{P}\big( X_{AB,CD}\big) = X_{AB,CD} &- \frac{1}{d-1}\Big( \eta_{AC}\eta^{PQ}M_{PB}M_{QD} - \eta_{AD}\eta^{PQ}M_{PB}M_{QC} \\
    &\hphantom{- \frac{1}{d-1}\Big(} - \eta_{BC}\eta^{PQ}M_{PA}M_{QD} + \eta_{BD}\eta^{PQ}M_{PA}M_{QC}\Big) \\
    &+ \frac{M_{PQ}M^{PQ}}{d(d-1)}\Big( \eta_{AC}\eta_{BD} - \eta_{AD}\eta_{BC}\Big) \, .
    \end{split}
\end{equation}
In \eqref{circ-prod} the operator $\hat{N}$ stands to the left of $\big( \Gamma_{\alpha\beta}\Gamma^{\alpha\beta}\big)^m$, so for the degree-four piece the term with $m=1$ carries the coefficient
\begin{equation}
    \frac{\Gamma\big( \frac{d}{2}\big)}{\Gamma\big( \frac{d}{2}+1\big)} \cdot \frac{1}{8} = \frac{1}{4d} \, ,
\end{equation}
because $\Gamma^{\alpha\beta}\Gamma_{\alpha\beta}X_{AB,CD}$ has degree zero; the terms with $m \geqslant 2$ vanish. Adding its contribution, $-\frac{3}{2d}\big( \eta_{AC}\eta_{BD} - \eta_{AD}\eta_{BC}\big)$, to the constant term of \eqref{MM-star} gives the total constant $\frac{1}{2}-\frac{3}{2d} = \frac{d-3}{2d}$, and collecting all pieces yields \eqref{MM-circ}.

\subsubsection*{Contractions}
One has the identity
\begin{equation}
    \label{D13}
    M_{AB}M^{AB} = -\frac{1}{4}\,t^{\sigma\lambda}t_{\sigma\lambda} \, ,
\end{equation}
since both sides equal $-\frac{1}{4}\epsilon_{\alpha\beta}\epsilon_{\gamma\delta}t^{\alpha\gamma}t^{\beta\delta}$. Contracting \eqref{MM-circ} with $\eta^{BD}$ and using\linebreak{} $\eta^{BD}\eta^{PQ}M_{PB}M_{QD} = M_{AB}M^{AB}$ we obtain
\begin{equation}
    \begin{split}
    \eta^{BD}\big( M_{AB} \circ M_{CD}\big) = {}&\underbrace{\eta^{PQ}M_{PA}M_{QC} - \eta^{PQ}M_{PA}M_{QC}}_{=0} \underbrace{- \frac{\eta_{AC}M_{PQ}M^{PQ}}{d-1} + \frac{\eta_{AC}M_{PQ}M^{PQ}}{d-1}}_{=0} \\
    &- \frac{d-1}{2}M_{AC} + \frac{d-3}{2}\eta_{AC} \, ,
    \end{split}
\end{equation}
which is \eqref{MM-contr}; the terms quadratic in the oscillators cancel by \eqref{D13}. A further contraction with $\eta^{AC}$ gives \eqref{MM-casimir}. Finally, contracting \eqref{MM-star} with $\varepsilon^{ABCD}$, the terms linear in $M_{AB}$ and the constant terms drop out because they contain a symmetric $\eta$ contracted with antisymmetrised indices, while the quadratic term vanishes because $M_{AB} \propto Y^1_{[A}Y^2_{B]}$ is a decomposable bivector, so that $\varepsilon^{ABCD}M_{AB}M_{CD} \propto \varepsilon^{ABCD}Y^1_AY^2_BY^1_CY^2_D = 0$; this proves \eqref{Pfaff}.

    \bibliographystyle{ieeetr}
    \bibliography{refers}          %

\end{document}